\documentclass[aps,english,preprintnumbers,amsmath,amssymb,nofootinbib,superscriptaddress,notitlepage]{revtex4-1}

\pdfoutput=1

\usepackage{graphicx}
\usepackage{bbm}
\usepackage{amssymb}
\usepackage{amsmath}

\usepackage{dsfont}

\def\0#1#2{\frac{#1}{#2}}

\def\s0#1#2{\mbox{\small{$ \frac{#1}{#2} $}}}

\newcommand{\E}{\mathrm{e}}
\newcommand{\I}{\mathrm{i}}
\newcommand{\be}{\begin{eqnarray}}
\newcommand{\ee}{\end{eqnarray}}

\newcommand{\nn}{\nonumber }

\usepackage{babel}
\begin{document}

\title{Formation of Selfbound States in a One-Dimensional Nuclear Model\\ -- A Renormalization Group based Density Functional Study --}

\author{Sandra Kemler}\affiliation{Institut f\"ur Kernphysik (Theoriezentrum), Technische Universit\"at Darmstadt, 
D-64289 Darmstadt, Germany}
\author{Martin Pospiech}\affiliation{Institut f\"ur Kernphysik (Theoriezentrum), Technische Universit\"at Darmstadt, 
D-64289 Darmstadt, Germany}
\author{Jens Braun}\affiliation{Institut f\"ur Kernphysik (Theoriezentrum), Technische Universit\"at Darmstadt, 
D-64289 Darmstadt, Germany}
\affiliation{ExtreMe Matter Institute EMMI, GSI, Planckstra{\ss}e 1, D-64291 Darmstadt, Germany}
  
\begin{abstract}
In nuclear physics, Density Functional Theory (DFT) provides the basis for state-of-the art studies of ground-state properties of heavy nuclei. However, the direct
relation of the density functional underlying these calculations and the microscopic nuclear forces is not yet fully understood. We present a
combination of DFT and Renormalization Group (RG) techniques which allows to study selfbound many-body systems 
from microscopic interactions. We discuss its application with the aid of systems of identical fermions interacting
via a long-range attractive and short-range repulsive two-body {force} in one dimension.
We compute ground-state energies, intrinsic densities, {and density correlation functions of these systems and compare our results to those 
obtained from other methods.} In particular, we show {how energies of excited states as well as
the absolute square of the ground-state wave function} can be extracted from
the correlation functions within our approach. The relation between many-body perturbation theory and our DFT-RG approach is discussed
and illustrated with the aid of the calculation of the second-order energy correction for a system of $N$ identical fermions interacting via a general two-body interaction. Moreover, we 
discuss the control of spuriously emerging fermion self-interactions in DFT studies within our framework. In general, our approach
may help to guide the development of energy functionals for future quantitative DFT studies of heavy nuclei from microscopic interactions.
\end{abstract}

\maketitle

%
\section{Introduction}
It is an exciting era for nuclear physics as we are now starting to understand the formation of light and medium-mass 
nuclei from the microscopic nuclear forces via chiral effective field theory 
{interactions~\cite{Borasoy:2006qn,Epelbaum:2008ga,Navratil:2009ut,Machleidt:2011zz,Quaglioni:2012ue,Hammer:2012id,%
Lahde:2013uqa,Soma:2013ona,Hagen:2013nca,Hergert:2015awm,Hebeler:2015hla,Lynn:2015jua}.}
For heavy nuclei, Density Functional Theory (DFT) 
remains currently to be the only feasible approach for a calculation of ground-state properties~\cite{Bender:2003jk}. 
The application of DFT to the nuclear many-body problem has been indeed very successful in recent years. This includes
conceptional advances aiming at, e.~g., {\it ab initio} studies of heavy nuclei, as well as more phenomenologically guided 
advances underlying an impressive variety of applications and providing us with a universal understanding of properties of nuclei,
see, e.~g., Refs.~\cite{Duguet:2007be,Lesinski:2007ys,Bender:2009ty,Lesinski:2011rn} and also Ref.~\cite{Dobaczewski:2010gr} for a review.
In fact, the nuclear energy density functional approach represents  a very active {research field, as also documented 
by the UNEDF/NUCLEI \mbox{SciDAC} collaboration~\cite{RJF,Nam:2012gy} representing just one example for the impressive efforts 
undertaken in this field.} 

Conventional DFT is based on the famous {\it Hohenberg-Kohn} theorem~\cite{Hohenberg:1964zz,Kohn:1965zzb}. For a given interaction potential, 
this theorem states that there exists a one-to-one correspondence between the ground-state density and a given one-body 
potential which effectively confines the {fermions. In particular, there is a map between the ground-state density and the ground-state wave function.
This implies the existence
of an energy density} functional from which, e.g., the ground-state energy and density can be computed.
Unfortuntately, the {\it Hohenberg-Kohn} theorem 
does {\it not} provide a recipe for the computation of the energy density functional
which usually consists of infinitely many terms. 
Thus, it is in general not possible to write down the exact energy density functional for a given many-body problem
and therefore systematic approximation schemes are ideally required in order to compute the ground-state properties of, e.g., nuclei reliably.
For conventional DFT approaches, the construction of reliable and systematic approximations of the energy density functional
indeed defines a critical point. In fact, 
despite the great success of nuclear DFT studies in particular with respect to precision, our understanding of the direct relation of
the functionals underlying these DFT studies to the microscopic nuclear forces is still incomplete.
Indeed, DFT studies in nuclear physics
are currently mostly based on fitting the parameters of a given ansatz for the density functional such that {it
reproduces a given set of experimentally determined values of ground-state properties} of various (heavy) nuclei~\cite{Dobaczewski:2001ed,Kortelainen:2014iha}. 
{The density functionals resulting from such an approach have then been used to study ground-state properties of nuclei across the nuclear chart~\cite{Kortelainen:2014iha} as well as
reactions~\cite{Dobaczewski:2010gr}.}

The search for a direct and systematic {approximation scheme for the construction} of the energy density functional from the underlying microscopic forces represents 
a very active frontier in recent years. Based on density matrix expansions, for example,
there have been attempts to give microscopic constraints on the nuclear energy density functional~\cite{Carlsson:2008gm,Stoitsov:2010ha}.
Moreover, the density-matrix expansion has been tested against {\it ab initio} calculations of trapped neutron drops~\cite{Bogner:2011kp} and
has been used to derive a nuclear energy density functional from chiral two- and three-nucleon interactions~\cite{Kaiser:2009me,Holt:2011nj}.
A future reliable and systematic microscopic construction of density functionals for studies of finite nuclei will certainly benefit from a 
multi-tier strategy based on various different and complementary approaches, ranging from the above-mentioned approaches and microscopic calculations
of the equation of state of nuclear matter~\cite{Hebeler:2009iv,Hebeler:2010xb,Holt:2011jd,Tews:2012fj} over 
a direct optimization of density functionals~\cite{Schunck:2009jc,Kortelainen:2010hv} and the construction of local functionals~\cite{Dobaczewski:2015eva} 
to density-matrix expansions~\cite{Carlsson:2008gm,Stoitsov:2010ha,Kaiser:2009me,Holt:2011nj,Bogner:2011kp}.

In the present work we discuss a renormalization-group (RG) inspired approach to DFT 
which complements the existing efforts in the construction of a microscopic energy density functional in
nuclear physics and ideally extends the available methods in many-body theory in general.
To this end, we utilize the fact that the {\it Hohenberg-Kohn} energy density functional can be identified with the 
two-particle-point-irreducible ($2$PPI) quantum effective action which 
can be derived directly from the path integral via a Legendre transformation with respect to sources
coupled to the densities,
see, e.g., Refs.~\cite{Furnstahl:2007xm,Drut:2009ce,Braun:2011pp} for reviews. Field-theoretically speaking, this
implies that the densities play the role of effective bosonic degrees of freedom describing the dynamics of fermions.

The possibility to compute the energy density functional from the path integral allows us to combine straightforwardly DFT with existing RG approaches. 
Here, we shall employ a DFT-RG approach put forward in Refs.~\cite{PhysRevB.66.155113,Schwenk:2004hm}
and {developed further in Refs.~\cite{Braun:2011pp,Kemler:2013yka}, where}
also the connection of this approach to conventional perturbation theory as well as the construction of systematic
truncation schemes is discussed. {For a general discussion of the properties of $n$PPI effective actions}, 
we refer the reader to Ref.~\cite{Pawlowski:2005xe}. A discussion of the relation of
our DFT-RG ($2$PPI-RG) approach {to $2$PI-RG equations} can be found in Refs.~\cite{Dupuis:2013vda,Rentrop:2015tia}.

The basic idea of our DFT-RG approach is to study gradually the change of the energy density functional
from a weakly interacting or even non-interacting system, which defines the starting point of the RG flow,
to the fully interacting system, e.g. the nucleus under consideration. A special feature of this approach is that it does not aim
at a computation of {the global density functional but rather relies on an expansion of the density functional about}
its ground state which is followed continuously while the interaction is turned on in the flow.

Based on Ref.~\cite{Kemler:2013yka}, we now extend the conceptional discussion of this approach
and {also} apply it to a system of identical fermions in one dimension interacting via a long-range attractive and
short-range repulsive two-body interaction. At first glance, such systems may be simply considered as 
theoretical toy models. However, systems of this kind can indeed be realized in experiments 
with ultracold fermionic atoms interacting via a dipolar interaction~\cite{Deuretzbacher}.
In nuclear physics, such one-dimensional systems have originally been introduced to 
test Monte Carlo calculations~\cite{Alexandrou:1988jg} and have subsequently also been used
to benchmark other approaches, see, e.g., Ref.~\cite{Jurgenson:2008jp}. In Sec.~\ref{sec:model},
we introduce the one-dimensional nuclear model which will be used for explicit calculations with
our DFT-RG approach in this work. The derivation of the DFT-RG flow equation for general
systems of identical fermions is
given in Sec.~\ref{sec:RGDFT}, including a discussion of the computation of the absolute square of the ground-state wave function as well as the 
spectral function from
the density-density correlation function which gives us access to the energies of excited states. Moreover, we discuss that our approach
reproduces correctly the results from {many-body perturbation theory. For illustration, we} present the
calculation of the second-order energy correction  within our framework for a system with an arbitrary number of fermions
interacting via a general two-body interaction. {Finally, we also discuss the relation of our DFT-RG approach to 
conventional DFT studies (including the local density approximation and the gradient expansion) as well as
the emergence and control of spurious fermion self-interactions within our DFT-RG framework.}
In Sec.~\ref{sec:res}, we apply this DFT-RG approach to the {above-mentioned one-dimensional model.}
Our results are discussed critically and compared to results from other approaches, including the exact solution of the
two-body problem.
Our conclusions and outlook can be found in Sec.~\ref{sec:conc}.


%
\section{One-dimensional Nuclear Model}\label{sec:model}
\begin{figure}[t]
\includegraphics[width=0.5\columnwidth]{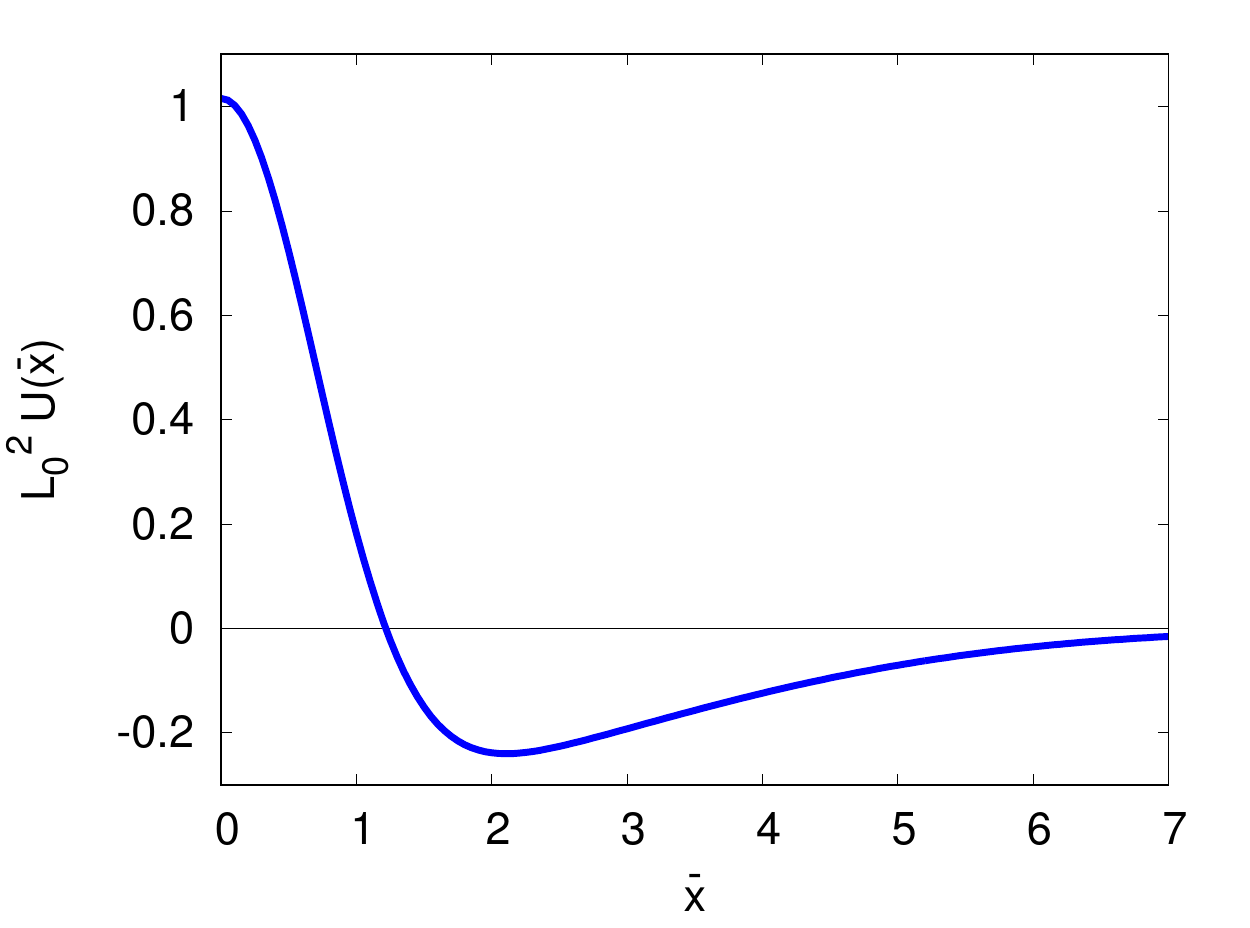}
\caption{\label{fig:pot} Dimensionless two-body interaction potential $L_0^2 {U}$ as a function of the 
dimensionless distance $\bar{x}=x/L_0:=|x_1-x_2|/L_0$
of two toy-nucleons as determined by the set of parameters given in Eq.~\eqref{eq:para}.
}
\end{figure}

In this work, we only consider systems of {\it identical} fermions in one dimension.
For explicit calculations with our DFT-RG approach we adopt a ``bare" 
two-body interaction potential $U$ introduced in Ref.~\cite{Alexandrou:1988jg}.
This interaction potential
is given by a superposition of two Gau\ss ians with opposite signs to simulate repulsive
short-range and attractive long-range nucleon-nucleon interactions:
\be
U(x_1-x_2)=
\frac{g}{\sigma_1 \sqrt{\pi}}\ {\rm e}^{-\frac{(x_1-x_2)^2}{\sigma_1^2}}
- \frac{g}{\sigma_2 \sqrt{\pi}}\ {\rm e}^{-\frac{(x_1-x_2)^2}{\sigma_2^2}}
\,,\label{eq:pot}
\ee
where $g>0$, $\sigma_1>0$, and $\sigma_2>0$ are constant parameters, see Fig.~\ref{fig:pot} for an illustration.
Throughout this work, we set $m=1$
for the mass of our toy-model nucleons. This implies that energy is measured in units of inverse length squared
and $g$ is measured in units of inverse length.
The parameters $\sigma_i$ set the length scale for the short-range repulsion and the range of attraction, respectively.
Following Ref.~\cite{Alexandrou:1988jg}, we {choose 
\be
\bar{g}\equiv g L_0 = 2.4\,\quad\text{and}\quad
\bar{\sigma}_2\equiv \sigma_2 L_0^{-1} = 4.0\,,
\label{eq:para}
\ee
where} the length scale $L_0\equiv \sigma_1=0.2$ is a measure for the extent of our toy-model nucleons and sets 
the scale for all dimensionful quantities in our calculations. On the other hand, the parameter~$\sigma_2$ can be associated with
the range of the interaction. In any case, $\sigma_1$ and~$\sigma_2$ are of the same order of magnitude.
From a phenomenological point of view, the parameter choice~\eqref{eq:para} ensures that the one-dimensional saturation 
properties correspond to empirical three-dimensional properties in nuclear physics~\cite{Alexandrou:1988jg}.


%
\section{Flow Equation for the Energy Density Functional}\label{sec:RGDFT}
In this section we derive the flow equation for the energy density functional as well as the generating functional of the connected density correlation
functions for a system of identical fermions in one dimension and give
a general discussion of field-theoretical aspects of our
DFT-RG approach. In particular, we discuss the connection of our approach
to many-body perturbation theory and show how the absolute square of the wave function and the spectral function 
can be calculated. We would like to add that the generalization of this part of our work 
to higher dimensions is comparatively straightforward. 
\begin{figure}[t]
\includegraphics[width=0.3\columnwidth]{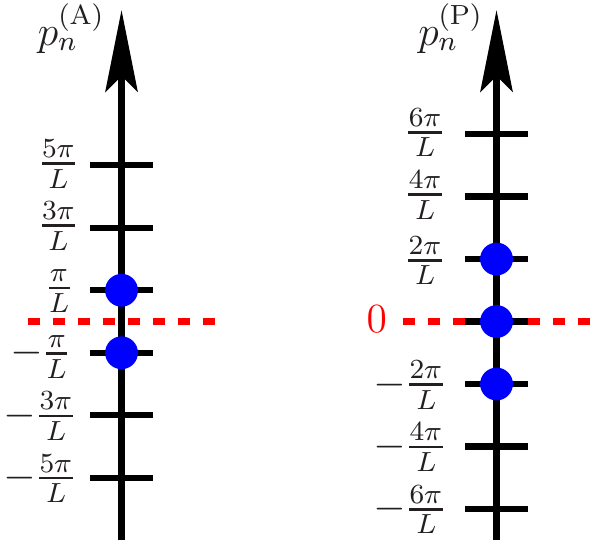}
\caption{\label{fig:momenta} (color online) Ground-state configuration of 
two identical fermions in a box with finite extent~$L$ and antiperiodic boundary conditions (left-hand side)
as well as the ground-state configuration of three identical fermions in a box with
periodic boundary conditions (right-hand side).
}
\end{figure}

%
\subsection{RG Flow Equation for DFT}\label{sec:rgflowdft}

We aim at a study of the formation of selfbound systems of fermions interacting via a 
non-local interaction which is repulsive at short distances and attractive at long range. 
Since the flow equation underlying our present work describes 
the change of the energy density functional under gradual changes of the interaction,
a convenient starting point for the RG flow
is given by a system of confined but non-interacting fermions. Here, we shall confine the fermions
in a box with extent $L$ which keeps the system bound in the RG flow.
The limit $L\to\infty$ is then eventually taken at the end of the flow, i.e. when the interaction has been fully turned on.
Since the box does not have a physical meaning in our case, the choice of the boundary conditions
is at our disposal. In this work, we shall choose periodic boundary conditions for a system with an odd
number of fermions and antiperiodic boundary conditions for a system with an even number of fermions.
This ensures that the ground state of a free gas of $N$ fermions, i.e. the starting point of the RG flow, is 
not degenerate.
The possible fermion momenta are then given by
\be
p_n^{\rm (P)}=\frac{2\pi n}{L}\label{eq:pP}
\ee
for odd numbers of particles and by
\be
p_n^{\rm (A)}=\frac{(2n+1)\pi}{L}\label{eq:pA}
\ee
for even number of particles with $n\in {\mathbb Z}$, see Fig.~\ref{fig:momenta} 
for an illustration. We emphasize that, in both cases, the density correlation functions being built up from one-particle propagators 
obey periodic boundary conditions. This follows from the fact that the density operator is given by $\hat{n}(x)\sim \hat{\psi}^{\dagger}(x)\hat{\psi}(x)$. 
Therefore the associated one-particle wave functions and their complex conjugates 
enter the correlation functions always pairwise, leaving us with periodic boundary conditions 
for these functions in either case. For systems of fermions with an additional internal degree of {freedom (e.g.~spin),} 
the same strategy can in principle be applied. Depending
on the actual particle-number configuration, however, it may then be the case that the two species obey different boundary conditions.
In any case, the fact that the density correlation functions obey periodic boundary conditions is quite convenient since this allows
us to rewrite these functions in terms of a Fourier series straightforwardly and it does
not introduce an additional artificial breaking of translation invariance.

Starting from the so-called classical action and the path integral, we now derive a flow equation for the
energy density functional describing the dynamics of a system of identical fermions interacting via
a general two-body interaction. From this equation, flow equations for the density correlation functions can then be obtained.
The classical action~$S$ defining the
many-body problem under consideration in this work is given by:\footnote{Throughout this work, we employ the imaginary-time formalism.}{
\be
S_{\lambda}[\psi^{\ast},\psi] &=& 
\int_{\tau}\int_x\, \psi^{\ast}(\tau,x) 
\left(\partial_{\tau} - \frac{1}{2}\partial_x^2 
+ V(x)\right) \psi(\tau,x)
\nn\\
&& \qquad+\,\frac{1}{2} 
\int _{\tau}\!\int _x\! \int _{\tau^{\prime}}\!\int _{x^{\prime}} \,
\psi^{\ast}(\tau,x) \psi^{\ast}(\tau^{\prime},{x}^{\prime})
U_{2b}(\tau,x,\tau^{\prime},x^{\prime}){\mathcal{R}_{\lambda}(\tau,x,\tau^{\prime},x^{\prime})}
\psi(\tau^{\prime},{x}^{\prime}) \psi(\tau,{x})\,, 
\label{eq:action}
\ee
where we have set~$m=1$ and
introduced the shorthands $\int_{\tau}=\int _{-\infty} ^{\infty}{\rm d}\tau$, $\int _{x}=\int_{-L/2}^{L/2} {\rm d}x$, $\partial_{\tau}=\frac{\partial}{\partial {\tau}}$, 
and $\partial_x=\frac{\partial}{\partial x}$. 
Here, $U_{2b}(\tau,x,\tau^{\prime},x^{\prime})=\delta(\tau-\tau^{\prime})U(x-x^{\prime})$ describes 
an instantaneous two-body (2b) interaction and~$\mathcal{R}_{\lambda}$ is a (dimensionless) regulator function which obeys
the following conditions:
\be
\lim_{\lambda\to 0}\mathcal{R}_{\lambda}(\tau,x,\tau^{\prime},x^{\prime}) =0\quad\text{and}\quad \lim_{\lambda\to 1}\mathcal{R}_{\lambda}(\tau,x,\tau^{\prime},x^{\prime}) =1\,.\label{eq:regfctcond}
\ee
Here,~$\lambda\in[0,1]$ is a dimensionless control parameter. The first condition
ensures that the two-body potential $U$ is switched off for~$\lambda=0$ and we are left with a system of non-interacting 
fermions in a box. For~$\lambda=1$, the second condition ensures that~$U$ is fully switched on and the action~\eqref{eq:action} describes fermions in a box interacting
via the two-body interaction~$U$. By taking the limit $L\to\infty$ at the point~$\lambda=1$, we then approach the actual many-body problem under consideration.
Other than the two conditions in Eq.~\eqref{eq:regfctcond}, the function~${\mathcal R}_{\lambda}$ is at our disposal. 
For convenience, however, we shall also assume that this function
does not break the symmetries of the theory under consideration. 

For studies in higher dimensions ($d>1$), a suitable choice for~${\mathcal R}_{\lambda}$ may indeed be required to control
ultraviolet divergences, e.g. in systems of fermions with an internal degree of freedom interacting via a contact interaction. In any case,
{we are free to choose ${\mathcal R}_{\lambda}$ such that the change in the parameter~$\lambda$ corresponds to a change of a 
momentum scale or, correspondingly, a length scale.
For example,~$\lambda$ may effectively correspond to an upper bound for the range of the interaction associated
with the two-body potential~$U$. In this case, the regulator function~${\mathcal R}_{\lambda}$ is designed such that an increase of~$\lambda$
corresponds to an increase of this effective upper bound starting from zero at~$\lambda=0$ and approaching infinity 
for~$\lambda\to 1$. Assuming that the problem under consideration is invariant under translations, for example, one may 
choose~${\mathcal R}_{\lambda}(\tau,x,\tau^{\prime},x^{\prime})= \lambda {\rm e}^{-(1-\lambda)(x-x^{\prime})^2/\ell^2}$, where~$\ell$ is a length
scale at our disposal. Clearly, this choice does not only effectively limit the range of the interaction but also alters its
functional form  for~$0<\lambda<1$ within the effective range defined by this function. Alternatively, one may therefore 
choose~${\mathcal R}_{\lambda}(\tau,x,\tau^{\prime},x^{\prime}) =  \lambda \theta( f_{\mathcal R}(\lambda) - (x-x^{\prime})^2)$,
where~$f_{\mathcal R}(\lambda)$ is a (monotonous) function of dimension length squared with the constraints~$f_{\mathcal R}(\lambda)\to 0$ 
for~$\lambda\to 0$, $f_{\mathcal R}(\lambda)\to \infty$ for~$\lambda\to 1$. 
Of course, other choices for the regulator function~${\mathcal R}_{\lambda}$ are also possible. Thus, if we think in terms of 
loop integrals constructed from density correlation functions, a 
suitably chosen regulator function can not only be used to switch on the interaction in the RG flow but also to
specify the details of the momentum integrations in the loop integrals at a given value of~$\lambda$. In all explicit calculations in this work, we shall
always employ~${\mathcal R}_{\lambda}(\tau,x,\tau^{\prime},x^{\prime})=\lambda$ for simplicity. In any case, with respect to the RG equation
for the energy density functional to be discussed below, we note that the result in the physical limit~$\lambda\to 1$
does not depend on our choice for the regulator function, provided that we solve the RG equation for the energy density functional exactly.}

Finally, we note that we have included a potential $V$ in the action~\eqref{eq:action} which describes a box with extent $L$ and suitably chosen boundary
conditions as discussed above. 
In practice, this fermion-confining potential is not included explicitly 
in our calculations but only implicitly by restricting the quantum fields on the 
domain $[-L/2,L/2)$ and imposing appropriate boundary conditions.
We emphasize that we only take into account a two-body interaction potential here, as, e.g., defined in Eq.~\eqref{eq:pot}. 
Higher-order many-body interactions will be dropped but
can in principle be included straightforwardly in our DFT-RG approach.}

Up to an irrelevant normalization factor, the generating functional $Z_{\lambda}$ of the density correlation functions is given by
\be
Z_{\lambda} [J]
\sim\int  \mathcal{D}\psi^{\ast}  \mathcal{D}\psi\, \E ^{-S_{\lambda}[\psi^{\ast},\psi] + 
\int_{\tau}\int_x J(\tau,{x})
( \psi^{\ast}(\tau,{x}) \psi(\tau,{x}))
}
\equiv
\E ^{W_{\lambda}[J]}\,. \label{eq:pathint}
\ee
Here, we have coupled the external source~$J$ to a term which is bilinear in the fermion fields 
and is associated with the density.
For our study of the dynamics of a system of $N$ fermions, we need to fix the particle number in our
calculations. This can be either done by introducing a chemical potential into the action~$S$ or by
choosing appropriate boundary conditions for the equations of motion~\cite{Puglia2003145}. 
In this work, we shall follow the latter approach to fix the particle number which amounts to
fixing the particle number in the initial conditions for the RG flow equations of the density correlation
functions. It can then be shown that the RG flow preserves the particle number as we shall discuss below.

From Eq.~\eqref{eq:pathint}, we obtain the generating functional~$W_{\lambda}$ of the 
{\it connected} density correlation functions:
\be
W_{\lambda}[J]=\ln Z_{\lambda}[J]\,.\label{eq:wz}
\ee
This functional can be expanded in terms of the source~$J$:
\be
W_{\lambda}[J] = G^{(0)}_{\lambda} 
+ \int_{\tau}\int_x G^{(1)}_{\lambda}(\tau,x)J(\tau,{x}) + \frac{1}{2}  \int_{\tau_1}\int_{x_1}\int_{\tau_2}\int_{x_2}G^{(2)}_{\lambda}(\tau_1,x_1,\tau_2,x_2)J(\tau_1,{x_1})J(\tau_2,{x_2})
+\dots\,,\label{eq:wexp}
\ee
where $G^{(0)}_{\lambda}=W_{\lambda}[0]$ is related to the ground-state energy of the system.
The quantity
\be
\rho_{{\rm gs},\lambda}(\tau,x):=G^{(1)}_{\lambda}(\tau,x)=\frac{\delta W_{\lambda}[J]}{\delta J(\tau,x)}\Bigg|_{J=0}\,\label{eq:Gdensity}
\ee
is the time-dependent ground-state (gs) density, and 
\be
G^{(2)}_{\lambda}(\tau_1,x_1,\tau_2,x_2)=\frac{\delta^2 W_{\lambda}[J]}{\delta J(\tau_1,x_1)\delta J(\tau_2,x_2)}\Bigg|_{J=0}\,
\ee
is the fully time-dependent density-density correlation function. The $n$-density correlation functions~$G^{(n)}_{\lambda}$,
\be
G^{(n)}_{\lambda}(\tau_1,x_1,\dots,\tau_n,x_n)=\frac{\delta^n W_{\lambda}[J]}{\delta J(\tau_1,x_1)\cdots \delta J(\tau_n,x_n)}\Bigg|_{J=0}\,,
\,
\ee
can in principle be computed from
the one-particle propagators. We shall exploit this below to compute the initial conditions for the RG flow equations. Note that, even for
the non-interacting system, all density correlation functions~$G^{(n)}_{\lambda}$ are in general finite which reflects the well-known fact that
{even} the energy density functional of a non-interacting system represents a non-local functional.

For convenience, we define the time-independent ground-state density~$n_{{\rm gs},\lambda}({x})$
as follows: 
\be
{n}_{{\rm gs},\lambda}({x}) := \lim_{\beta\to\infty}\frac{1}{\beta}\int _{-\beta/2}^{\beta/2}{\rm d}\tau\, \rho_{{\rm gs},\lambda}(\tau,{x})\,,
\label{eq:ngsrhogs}
\ee
where $\beta$ is an auxiliary parameter introduced to define a finite imaginary time interval, $\tau\in [-\beta/2,\beta/2)$. In our numerical calculations,
we shall always consider the case $\beta\to\infty$, i.e. $\tau \in (-\infty,\infty)$.
In any case, if the density $\rho_{{\rm gs},\lambda}(\tau,{x})$ turns out to be independent of the imaginary time~$\tau$, then
we have ${n}_{\rm gs,\lambda}({x}) \equiv \rho_{{\rm gs},\lambda}(\tau,{x})$. For instantaneous interactions 
as used in the present work, this is indeed the case.

Next, we introduce the so-called  {\it classical} field~$\rho(\tau,{x})$:
\be
\rho (\tau,{x}) = \frac{\delta W_{\lambda}[{J}]}{\delta J(\tau,{x})}\,.\label{eq:rhodef}
\ee
The classical field~$\rho$ is related to the fermion density, see Eq.~\eqref{eq:Gdensity}, and plays the role of a composite bosonic effective degree of freedom in our approach which
is used to describe the dynamics of fermions.
Note that~$\rho$ is a functional of the source $J$, $\rho=\rho[J]$, and that, in our case, it also depends on $\lambda$.
Here, we do not indicate this dependence explicitly {(e.g., by adding an index $\lambda$) to ensure that this quantity is not confused 
with the actual $\lambda$-dependent ground-state}
density. However, we need to keep {the $\lambda$-dependence} in mind for the derivation of the flow equation of the energy density functional below.

The $2$PPI effective action~$\Gamma_{\lambda}[\rho]$ is now obtained from a Legendre transformation of the functional~$W[J]$
with respect to the source~$J$: 
\be
\Gamma_{\lambda}[\rho] 
= \sup_{J} \left\{ -W_{\lambda}[J] + \int _{\tau}  \int _x\, J(\tau,{x}) \rho(\tau,{x}) 
\right\}.
\label{eq:effactdef}
\ee
Below we shall consider an expansion of $\Gamma_{\lambda}$ in terms of a $\lambda$-independent classical field~$\rho$. From
Eq.~\eqref{eq:effactdef}, we then deduce that the source~$J=J_{\rm sup}$, which fulfills the supremum condition, depends on~$\lambda$ and is a functional of~$\rho$, i.e. $J_{\rm sup}=J_{\rm sup}[\rho]$.
Phenomenologically, this transformation to {the $2$PPI effective action} may be viewed as a bosonization of the theory since we have traded in the fermion fields in the classical action~$S$
for the composite bosonic field~$\rho$ in the effective action~$\Gamma_{\lambda}$.

The $2$PPI effective action~$\Gamma$ can be related to the energy-density functional~$E_{\lambda}[\rho]$ mentioned above in the context of conventional {\it Hohenberg-Kohn} 
DFT.\footnote{Here, we consider a generalization of the original {\it Hohenberg-Kohn} formalism with a time-dependent source. 
The exact equivalent of the energy density functional is obtained by employing a
time-independent source~$J({x})$, {see below and, e.g.,} Refs.~\cite{PTP.92.833,1997cond.mat..2247V,PhysRevB.66.155113,Puglia2003145}
and~Refs.~\cite{Eschrig,Kutzelnigg2006163} for a more general discussion on DFT in terms of a Legendre transformation.}
Indeed, we have
\be
E_{\lambda}[\rho]=\lim_{\beta\to\infty}\frac{1}{\beta}\Gamma_{\lambda}[\rho]\,.
\ee
The ground-state energy is given by
\be
E_{\lambda}=\lim_{\beta\to\infty}\frac{1}{\beta}\Gamma_{\lambda}[\rho_{{\rm gs},\lambda}]=-\lim_{\beta\to\infty}\frac{1}{\beta}W_{\lambda}[0]\,,\label{eq:egs}
\ee
which follows from Eq.~\eqref{eq:wz} and the spectral representation of the partition function,~$Z_{\lambda}\sim \sum_n {\rm e}^{-\beta E_{n,\lambda}}$.
Here, the set~$\{ E_{n,\lambda}\}$ (with $E_{\lambda}\equiv E_{0,\lambda} < E_{1,\lambda} < E_{2,\lambda} < \dots$) denotes the $\lambda$-dependent 
energy spectrum associated with the many-body system under consideration.
The {\it universality} of the {\it Hohenberg-Kohn} functional
follows from the fact that the background potential~$V$ can be absorbed
into the source term~$J$ by a simple shift, $J\to J+V$, see Ref.~\cite{Schwenk:2004hm}. 
We also observe that~$\Gamma_{\lambda}$ does not depend on~$J$ as it should be:
\be
\frac{\delta\Gamma_{\lambda}[\rho]}{\delta J}=0\,. \label{eq:stat}
\ee
Finally, the quantum equation of motion of the composite degree of freedom~$\rho$ follows from
\be
\frac{\delta\Gamma_{\lambda}[\rho]}{\delta \rho(\tau,{x})}=J(\tau,{x})\,\label{eq:eom}
\ee
in the limit~$J\to 0$, being the analogue of the {\it Hohenberg-Kohn} variational principle. 
The solution of this equation yields the ground-state density~$\rho_{{\rm gs},\lambda}$ of the
system in the limit~$J\to 0$ which is required to compute the ground-state energy, see Eq.~\eqref{eq:egs}.

In conventional DFT, often a global ansatz for the {\it a priori} unknown energy density functional is made. This ansatz is then minimized
using the famous {\it Kohn-Sham} equations. Here, we refrain from making a global ansatz for the
density functional but follow the approach detailed in~Ref.~\cite{Kemler:2013yka} and expand~$\Gamma_{\lambda}$ about its
ground state~$\rho_{{\rm gs},\lambda}$:
\be
\!\!\!\!\!\!\Gamma_{\lambda}[\rho]=\Gamma_{\lambda}[\rho_{\rm gs,\lambda}] 
+ \frac{1}{2}\int _{\tau_1}\int _{x_1} \int_{\tau_2}\int_{x_2} \left( \rho(\tau_1,{x_1}) 
\!-\! \rho_{\rm gs,\lambda}(\tau_1,{x_1})\right)\Gamma^{(2)}_{\lambda}(\tau_1,x_1,\tau_2,{x}_2) \left( \rho(\tau_2,{x}_2) 
\!-\! \rho_{\rm gs,\lambda}(\tau_2,{x_2})\right)+\dots\,, \label{eq:vertexexp}
\ee
where
\be
\Gamma_{\lambda}^{(2)}(\tau_1,x_1,\tau_2,{x}_2) =\frac{\delta^2 \Gamma_{\lambda}[\rho]}{\delta \rho(\tau_1,x_1)\delta \rho(\tau_2,x_2)}\Bigg|_{\rho=\rho_{\rm gs}}
\ee
is positive definite by construction. 

Apparently, the expansion of~$W_{\lambda}$ in terms of $J$ and the expansion of $\Gamma_{\lambda}$ in terms of $\rho$ are related.
From Eq.~\eqref{eq:Gdensity}, we deduce that the ground-state density~$\rho_{{\rm gs},\lambda}$ and the one-density correlation function~$G^{(1)}$
are equivalent. 
From Eqs.~\eqref{eq:rhodef} and~\eqref{eq:eom}, it moreover follows that the ``curvature"~$\Gamma_{\lambda}^{(2)}$ is related to the density-density correlation function~$G^{(2)}_{\lambda}$
appearing in the expansion~\eqref{eq:wexp}. In fact, we have
\be
\frac{\delta^2 \Gamma_{\lambda}[\rho]}{\delta \rho \delta \rho } = \left( \frac{\delta ^2 W_{\lambda}[J]}{\delta J \delta J}\right)^{-1}
\ee
and therefore it follows that
\be
\Gamma^{(2)}_{\lambda}(\tau_1,x_1,\tau_2,{x}_2) =\left(G^{(2)}_{\lambda}\right)^{-1}(\tau_1,x_1,\tau_2,{x}_2)\,.\label{eq:g2w2}
\ee
Corresponding relations for density correlation functions of higher order can be derived along these lines. For the three-density correlation function, for example,
we find 
\be
\frac{\delta^3 \Gamma_{\lambda}[\rho]}{\delta \rho \delta \rho \delta\rho} =\frac{\delta}{\delta\rho}\left( \frac{\delta ^2 W_{\lambda}[J]}{\delta J \delta J}\right)^{-1}\,,
\ee
which {yields
\be
\Gamma^{(3)}_{\lambda}(\chi_1,\chi_2,\chi_3) 
=-\int_{\chi_4}\int_{\chi_5}\int_{\chi_6}
\left(\Gamma^{(2)}_{\lambda}(\chi_1,\chi_4) G^{(3)}_{\lambda}(\chi_4,\chi_5,\chi_6) \Gamma^{(2)}_{\lambda}(\chi_5,\chi_2) \Gamma^{(2)}_{\lambda}(\chi_6,\chi_3)\right)\,.
\label{eq:g3w3}
\ee
For} convenience, we have introduced $\chi=\{\tau, x\}$, $\int_{\chi}=\int_{-\infty}^{\infty} {\rm d}\tau\int_{-L/2}^{L/2}{\rm d}x$, {and
\be
\Gamma^{(n)}_{\lambda}(\chi_1,\chi_2,\dots,\chi_n)=\frac{\delta^n \Gamma_{\lambda}[\rho]}{\delta \rho(\chi_1) \cdots \delta \rho(\chi_n)}\Bigg|_{\rho=\rho_{\rm gs}}\,. 
\ee
For} later purposes, we finally give the relation {between~$G^{(4)}_{\lambda}$ and $\Gamma^{(4)}_{\lambda}$:
\be
\!\!\!\!\!\Gamma^{(4)}_{\lambda}(\chi_1,\chi_2,\chi_3,\chi_4)
&=&-\int_{\chi_5}\int_{\chi_6}\int_{\chi_7}
\Big( \Gamma^{(3)}_{\lambda}(\chi_1,\chi_5,\chi_4) G^{(3)}_{\lambda}(\chi_5,\chi_6,\chi_7) \Gamma^{(2)}_{\lambda}(\chi_6,\chi_2) \Gamma^{(2)}_{\lambda}(\chi_7,\chi_3) \nn\\
&& \qquad\qquad\quad + \Gamma^{(2)}_{\lambda}(\chi_1,\chi_5) G^{(3)}_{\lambda}(\chi_5,\chi_6,\chi_7) \Gamma^{(3)}_{\lambda}(\chi_6,\chi_2,\chi_4) \Gamma^{(2)}_{\lambda}(\chi_7,\chi_3)\nn\\[0.3cm]
&& \qquad\qquad\quad\quad + \Gamma^{(2)}_{\lambda}(\chi_1,\chi_5) G^{(3)}_{\lambda}(\chi_5,\chi_6,\chi_7) \Gamma^{(2)}_{\lambda}(\chi_6,\chi_2) \Gamma^{(3)}_{\lambda}(\chi_7,\chi_3,\chi_4)\Big)\nn\\
&& \qquad - \int_{\chi_5}\int_{\chi_6}\int_{\chi_7}\int_{\chi_8}
\Gamma^{(2)}_{\lambda}(\chi_1,\chi_5) G^{(4)}_{\lambda}(\chi_5,\chi_6,\chi_7,\chi_8) \Gamma^{(2)}_{\lambda}(\chi_6,\chi_2) \Gamma^{(2)}_{\lambda}(\chi_7,\chi_3)\Gamma^{(2)}_{\lambda}(\chi_8,\chi_4) 
\,.\label{eq:g4w4}
\ee
Thus,} {general correlation functions~$\Gamma^{(n)}_{\lambda}$ 
can} be obtained from a computation of the (connected) density correlation functions~$G^{(n)}_{\lambda}$. The computation of these functions for
an interacting theory (i.e. for~$\lambda >0$) is in general highly non-trivial and will be discussed below within our DFT-RG framework which 
only uses the density correlation functions of the non-interacting theory as an input. The latter 
can be computed
from the one-particle propagator~$\Delta_0$ of the non-interacting theory,
\be
\!\!\!\!\!\!\!\!\!\Delta_{0}(\tau_1,x_1,\tau_2,x_2)&=&-\langle {\mathcal T}\psi(\tau_1,x_1)\psi^{\ast}(\tau_2,x_2) \rangle\nn\\
&=&-\langle \psi(\tau_1,x_1)\psi^{\ast}(\tau_2,x_2)\rangle \theta_{\sigma}(\tau_1-\tau_2) + \langle \psi^{\ast}(\tau_2,x_2)\psi(\tau_1,x_1)\rangle \theta_{\sigma}(\tau_2-\tau_1)\,,
\label{eq:onepart}
\ee
where~${\mathcal T}$ is the time-ordering operator. Here,~$\theta_{\sigma}(\tau)=1$ for~$\tau>0$ as well as for~$\tau\to 0^{+}$ and~$\theta_{\sigma}(\tau)=0$ otherwise.
The non-interacting ground-state density is then obtained from
\be
\rho_{{\rm gs},\lambda=0}(\tau_1,x_1)\equiv G^{(1)}_{\lambda=0}(\tau_1,x_1)=\lim_{\tau_2\to\tau_1^{+}} \Delta_0(\tau_1,x_1,\tau_2,x_1)\,.\label{eq:densdef}
\ee
Higher-order density correlation functions can be written as expectation values of time-ordered products of the fields as well. In fact, the general $n$-density correlation
functions~$Z_{\lambda}^{(n)}$ including both connected and disconnected diagrams are defined as follows:
\be
Z_{\lambda}^{(n)}(\chi_1,\dots,\chi_n) &=& \langle {\mathcal T} \psi^{\ast}(\chi_1)\psi(\chi_1)\cdots\psi^{\ast}(\chi_n)\psi(\chi_n)\rangle_{\lambda}\,,\label{eq:Zn}
\ee
where~$Z^{(1)}_{\lambda}\equiv G^{(1)}_{\lambda}$ and the index `$\lambda$' refers to the fact that the expectation value has to be computed with
respect to the $\lambda$-dependent ground-state.
Using the {\it Wick} theorem, the (connected) density-density correlation functions~$G^{(n)}_{\lambda=0}$ of the non-interacting theory
can be extracted from the correlation functions $Z^{(n)}_{\lambda=0}$ {and can then be} 
written in terms of one-particle propagators. For the connected density-density correlation function, we find
\be
Z^{(2)}_{\lambda}(\chi_1,\chi_2)=\rho_{{\rm gs},\lambda}(\chi_1)\rho_{{\rm gs},\lambda}(\chi_2) + G_{\lambda}^{(2)}(\chi_1,\chi_2) \quad\text{with}\quad
G_{\lambda=0}^{(2)}(\chi_1,\chi_2) 
=-\Delta_{0}(\chi_2,\chi_1)\Delta_{0}(\chi_1,\chi_2)\,.\label{eq:g2opp}
\ee
It follows from Eq.~\eqref{eq:onepart} that~$G_{\lambda=0}^{(2)}$ only depends on $|\tau_1-\tau_2|$.
For the three-density correlation function, we {obtain
\be
G_{\lambda=0}^{(3)}(\chi_1,\chi_2,\chi_3)&=& \Delta_{0}(\chi_1,\chi_2)\Delta_{0}(\chi_2,\chi_3)\Delta_{0}(\chi_3,\chi_1) 
+ \Delta_{0}(\chi_2,\chi_1)\Delta_{0}(\chi_1,\chi_3)\Delta_{0}(\chi_3,\chi_2)\,.
\label{eq:G3}
\ee
The} four-density correlation function~$G^{(4)}_{\lambda=0}$ entering the computation of the expansion coefficient~$\Gamma^{(4)}_{\lambda=0}$ can be written as follows:
\be
\!\!\!\!\!\!\!\!\!\!\!\! G_{\lambda=0}^{(4)}(\chi_1,\dots,\chi_4)&=&-\Delta_{0}(\chi_1,\chi_2)\Delta_{0}(\chi_2,\chi_3)\Delta_{0}(\chi_3,\chi_4)\Delta_{0}(\chi_4,\chi_1)
-\Delta_{0}(\chi_1,\chi_4)\Delta_{0}(\chi_4,\chi_3)\Delta_{0}(\chi_3,\chi_2)\Delta_{0}(\chi_2,\chi_1)\nn\\
&&-\Delta_{0}(\chi_2,\chi_4)\Delta_{0}(\chi_4,\chi_3)\Delta_{0}(\chi_3,\chi_1)\Delta_{0}(\chi_1,\chi_2)
-\Delta_{0}(\chi_1,\chi_4)\Delta_{0}(\chi_4,\chi_2)\Delta_{0}(\chi_2,\chi_3)\Delta_{0}(\chi_3,\chi_1)\nn\\
&&-\Delta_{0}(\chi_2,\chi_4)\Delta_{0}(\chi_4,\chi_1)\Delta_{0}(\chi_1,\chi_3)\Delta_{0}(\chi_3,\chi_2)
-\Delta_{0}(\chi_3,\chi_4)\Delta_{0}(\chi_4,\chi_2)\Delta_{0}(\chi_2,\chi_1)\Delta_{0}(\chi_1,\chi_3)\,.
\label{eq:G4}
\ee
Higher-order correlation functions can be computed along these lines. We {find
\be
G_{\lambda=0}^{(n)}(\chi_1,\dots,\chi_n)&=& \frac{(-1)^{n+1}}{n}\sum_{(i_1,\dots,i_n) \in S_n} \Delta_{0}(\chi_{i_1},\chi_{i_2}) \Delta_{0}(\chi_{i_2},\chi_{i_3})
\cdots\Delta_{0}(\chi_{i_{n-1}},\chi_{i_n})\,,
\ee
where~$S_n$ is} the set of all permutations of the $n$-tuple $(1,2,\dots,n)$.
In our explicit calculations discussed below, however, we shall only take into account correlation functions~$G^{(n)}_{\lambda}$ with $n\leq 4$. 

At this point we would like to make again contact to conventional DFT. In conventional DFT, 
the energy density functional~$E_{\rm HK}$ is considered to be a functional of the time-{\it independent} density field~$n(x)$ rather
than of the time-dependent density field~$\rho(\tau,x)$ as in our case.\footnote{The functional~$E_{\rm HK}$ should not
be confused with the universal {\it Hohenberg-Kohn} (HK) functional~$F_{\rm HK}[n]=E_{\rm HK}[n]-\int_x V(x)n(x)$, where~$V(x)$ is the fermion-confining (external) potential.}
The use of a derivative/gradient expansion
is then often discussed in this context as it potentially allows for a systematic computation of the energy density functional.
{Starting from an expansion of~$E_{\rm HK}$ about
the ground-state density~$n_{\rm gs}$,
\be
E_{\rm HK}[n]=E_{\rm HK}^{(0)}
+ \frac{1}{2}\int _{x_1} \int_{x_2} \left( n({x_1}) 
- n_{\rm gs}\right){E}^{(2)}_{\rm HK}(x_1,{x}_2) \left( n({x}_2) 
- n_{\rm gs}\right)+\dots\,,
\label{eq:vertexexpHK}
\ee
where we have assumed that the many-body system under consideration is invariant under $x_i\to -x_i$ and translations $x_i\to x_i+a$ with $a$ defining an arbitrary shift,\footnote{In 
particular, we 
assume here that the ground-state density is homogeneous.}
the} expansion~\eqref{eq:vertexexpHK} can be recast into a derivative/gradient expansion of the following form~\cite{PhysRevB.54.17402}:\footnote{Note 
that, in general, translation invariance is 
broken in a finite box. We shall come back to this issue below when we discuss {the implementation of the two-body interaction} in a box with (anti)periodic
boundary conditions.}{
\be
E_{\rm HK}[n]=g_{{\rm HK}}^{(0)}[n]+\int_x g_{{\rm HK}}^{(2)}(n)\left(\partial_x n(x)\right)^2 + \dots\,.\label{eq:gradexpHK}
\ee
Here,} we only show the first two terms of this expansion and drop higher-order derivative terms. 
In contradistinction to the vertex expansion~\eqref{eq:vertexexpHK} of the energy density functional~$E_{\rm HK}$,
{the derivative expansion~\eqref{eq:gradexpHK} does not depend on}
the ground-state density~$n_{{\rm gs}}$, i.e. it is a global functional rather than a local approximation.
{The new quantities appearing} in the derivative expansion can in principle be derived from the correlation functions $E_{\rm HK}^{(m)}$. 
The first term on the right-hand side of Eq.~\eqref{eq:gradexpHK}
is associated with the local density approximation which can be obtained from the continuum equation of state or, equivalently,
from~$E_{\rm HK}^{(0)}$ and its derivatives with respect to a
homogeneous ``trial" density.
{Similarly,~$g_{{\rm HK}}^{(2)}$ can be computed from
$E_{\rm HK}^{(2)}$.}  
To be more {specific,~$g_{{\rm HK}}^{(2)}$ is fully} determined by the derivatives of the 
Fourier transform of $E^{(2)}_{\rm HK}$ with respect to a homogeneous ``trial" density.\footnote{Here, we tacitly assume
that the radius of convergence associated with the vertex expansion~\eqref{eq:vertexexpHK} is infinite.}
This relation between the vertex expansion and the gradient expansion is a direct consequence of the fact
that the correlation functions~$E_{\rm HK}^{(m)}$ obey the following relation:{
\be
\frac{\partial E_{\rm HK}^{(m)}(x_1,\dots,x_m)}{\partial n_{\rm gs}} 
= \int_{x_{m+1}} E_{\rm HK}^{(m+1)}(x_1,\dots,x_{m+1})\,.
\ee
In} general, the {sketched} ``translation" of the vertex expansion~\eqref{eq:vertexexpHK} into a derivative expansion 
yields derivative terms of arbitrarily high orders in Eq.~\eqref{eq:gradexpHK}. For example, the computation of 
terms of the order~${\mathcal O}(\partial_x^4)$ requires the four-density correlation function as input. 
The validity and the convergence of such a gradient/derivative expansion is difficult to estimate and the analysis in general requires the knowledge of the
underlying correlation functions~$E_{\rm HK}^{(m)}$. In any case, due to the presence of a Fermi surface, the expansions of these functions 
in terms of momenta, which are associated with the derivatives appearing in Eq.~\eqref{eq:gradexpHK}, 
will in general only have a finite radius of convergence. Depending on the system under consideration, 
the gradient/derivative expansion may therefore be only of limited use.

Within this work, we consider a 
functional~$\Gamma_{\lambda}$ of the time-dependent density field~$\rho(\tau,x)$. Also in this case, the vertex expansion~\eqref{eq:vertexexp}
can be recast into a derivative expansion,{
\be
\Gamma_{\lambda}[\rho] = g^{(0)}_{\lambda}[\rho] +\int_{\tau}\int_{x} g^{(2,0)}_{\lambda}(\rho)\left(\partial_{\tau} \rho(\tau,x)\right)^2
+\int_{\tau}\int_x g^{(2,1)}_{\lambda}(\rho)\left(\partial_x \rho(\tau,x)\right)^2 + \dots\,,\label{eq:gradexp}
\ee
provided} that 
the many-body system under consideration is invariant under $\tau\to -\tau$ and time translations $\tau\to \tau+\tau_0$ as well, where $\tau_0$ defines an arbitrary shift
in time direction. Again, an investigation of the validity of such an expansion is difficult for the same reasons as 
in the case of conventional DFT and its applicability therefore needs to be carefully examined. 

The projection of the functional~$\Gamma_{\lambda}[\rho]$, which 
maps a time- and space-dependent function~$\rho(\tau,x)$ into a real number, onto the functional~$E_{\rm HK}[n]$, which
maps a space-dependent function~$n(x)$ into a real number, appears in general to be highly non-trivial, even if both functionals are associated with the same ground state. However,
a comparison of the definitions of the (ground-state) density~$\rho_{\rm gs}$ (see Eq.~\eqref{eq:ngsrhogs}) 
and the density-density correlation function~$G^{(2)}_{\lambda}$ (see Eq.~\eqref{eq:g2opp})  
with the corresponding quantities
entering the computation of energy density functionals of the type of~$E_{\rm HK}[n]$ (see Ref.~\cite{Puglia2003145})  
suggests the following projection rule for the correlation functions:
\be
G_{\rm HK}^{(m)}(x_1,\dots,x_m)=\lim_{\beta\to\infty}\frac{1}{\beta}\int_{-\beta/2}^{\beta/2}{\rm d}\tau_1 \cdots \int_{-\beta/2}^{\beta/2}{\rm d}\tau_m\, G^{(m)}_{\lambda}(\tau_1,x_1,\dots,\tau_m,x_m)\,.
\label{eq:projrule}
\ee
Assuming that this rule holds for any~$m\in\mathbb{N}$, the correlation functions~$E_{\rm HK}^{(m)}$ appearing in Eq.~\eqref{eq:vertexexpHK} can be computed from the time-dependent correlation 
functions~$G^{(m)}_{\lambda}$. For~$m=2$, for example, we have~$E_{\rm HK}^{(2)}\sim (G_{\rm HK}^{(2)})^{-1}$.

Let us now come back to the derivation of a flow equation for the density {functional~$\Gamma_{\lambda}$, taking into account
the full time-dependence of the correlation functions and going beyond the gradient/derivative expansion.} Our discussion already makes apparent 
that the computation of~$\Gamma_{\lambda}$ can be traced back to the computation of density correlation functions. In the non-interacting
limit, i.e.~$\lambda=0$, these functions can be calculated analytically for a large class of confining geometries. 
This defines the initial point of our RG flow and we will 
discuss the computation of the initial conditions for our system in more detail below. In the fully interacting limit, i.e.~$\lambda=1$,
the correlation functions cannot be computed straightforwardly. However, we can derive exact equations for the
computation of the change of the generating functional~$W_{\lambda}$ and the 
density functional~$\Gamma_{\lambda}$ under a variation of~$\lambda$.
To this end, we consider Eq.~\eqref{eq:wz} and 
take the derivative of~$W_{\lambda}$ with respect to~$\lambda$. This yields {
\be
\partial_{\lambda} W_{\lambda}[J] &=& -\frac{1}{2}\int_{\chi_1}\int_{\chi_2}\frac{\delta W_{\lambda}[J]}{\delta J(\chi_1)}
U_{2b}(\chi_1,\chi_2)\left(\partial_{\lambda}{\mathcal R}_{\lambda}(\chi_1,\chi_2)\right)\frac{\delta W_{\lambda}[J]}{\delta J(\chi_2)}\nn\\
&& \qquad -\frac{1}{2}\int_{\chi_1}\int_{\chi_2}U_{2b}(\chi_1,\chi_2) \left(\partial_{\lambda}{\mathcal R}_{\lambda}(\chi_1,\chi_2)\right)
\left(\frac{\delta^2 W_{\lambda}[J]}{\delta J(\chi_2)\delta J(\chi_1)} 
- \frac{\delta W_{\lambda}[J]}{\delta J(\chi_2)}\delta(\chi_2-\chi_1)
\right)\,,\label{eq:wfloweq}
\ee
where} $\partial_{\lambda}=\frac{\partial}{\partial \lambda}$ and 
$\delta(\chi_1-\chi_2)\equiv\delta(\tau_1-\tau_2)\delta(x_1-x_2)$. This equation is exact on the level of two-body interactions and allows
to compute the change of the functional~$W_{\lambda}[J]$ under a variation of~$\lambda$. The initial condition for this differential equation is given
by the non-interacting system associated with~$\lambda=0$, i.e. non-interacting {fermions in a box in} our studies below. 
The fully interacting many-body problem is obtained in the limit~$\lambda=1$.
While this flow equation is exact, its solution in general requires approximations, as its right-hand
side depends on functional derivatives of~$W_{\lambda}$ with respect to the source~$J$. To solve this equation in a systematic fashion, we insert the expansion~\eqref{eq:wexp}
of the functional~$W_{\lambda}$ into Eq.~\eqref{eq:wfloweq}.
This allows us to derive flow equations for the density correlation functions~$G^{(n)}_{\lambda}$ by comparing 
the same orders in~$J$ on the left- and right-hand side of Eq.~\eqref{eq:wfloweq}. 
In Sec.~\ref{sec:mbpt},
we shall then show that such an expansion is indeed systematic in the sense that it can be systematically related to many-body perturbation theory. Nevertheless,
our approach is not perturbative. The solutions~$G^{(n)}_{\lambda=1}$ of these flow equations rather contain arbitrarily high orders in the coupling constant associated with a perturbative calculation.
We also emphasize that the expansion~\eqref{eq:wexp} yields an infinite tower of coupled flow equations for the correlation functions~$G^{(n)}_{\lambda}$. From Eq.~\eqref{eq:wfloweq}, we deduce that
the flow equation for~$G^{(n)}_{\lambda}$ in general depends on the correlation functions~$G^{(m)}_{\lambda}$ with $1\leq m \leq n+2$. 
This suggests that a truncation of this set of equations is in general required in order to
compute the density correlation functions.
Below, we shall discuss the truncation scheme underlying our present studies in more detail.  
In any case, the correlation functions~$G^{(n)}_{\lambda}$ obtained from a solution of such a (truncated) set of flow equations 
can be used to compute the functions~$\Gamma^{(n)}_{\lambda}$ determining the density functional~$\Gamma_{\lambda}$.

As we have discussed above, the functionals~$W_{\lambda}$ and~$\Gamma_{\lambda}$ are related via a Legendre transformation. Using the flow
equation~\eqref{eq:wfloweq} for the generating functional of the connected correlation functions and the definition~\eqref{eq:effactdef} of the density functional~$\Gamma_{\lambda}$,
we find the following flow equation for~$\Gamma_{\lambda}$: {
\be
\!\!\!\partial_\lambda\Gamma_\lambda [\rho]
&=&\frac{1}{2}\int_{\chi_1}\int_{\chi_2} \rho(\chi_1)U_{2b}(\chi_1,\chi_2)\left(\partial_{\lambda}{\mathcal R}_{\lambda}(\chi_1,\chi_2)\right)\rho(\chi_2)\nn\\
&&\qquad + \frac{1}{2} \int_{\chi_1}\int_{\chi_2} U_{2b}(\chi_1,\chi_2)\left(\partial_{\lambda}{\mathcal R}_{\lambda}(\chi_1,\chi_2)\right) 
\left(  \left( \frac{\delta^2 \Gamma_{\lambda}[\rho]}{\delta \rho \delta \rho } \right)^{-1}\!\!\!\!\!(\chi_2,\chi_1) \!-\! \rho(\chi_2)\delta(\chi_2\!-\!\chi_1)\right)\,.
\label{eq:gfloweq}
\ee
This} functional differential equation is again {\it exact} on the level of two-body interactions and 
describes the flow of the density functional from the non-interacting system at the initial point $\lambda=0$ to the
fully interacting many-body system at~$\lambda=1$.  
The flow equations for~$W_{\lambda}$ and~$\Gamma_{\lambda}$ look indeed quite similar. However, we would like to stress that the fundamental building blocks of these flow equations, i.e.
the correlation functions, are not identical. Whereas the connected density correlation functions enter the flow of the functional~$W_{\lambda}$, $2$PPI correlation functions underly the
flow of the density functional~$\Gamma_{\lambda}$. 
Of course, these correlation functions can be translated into each other, as shown explicitly in Eqs.~\eqref{eq:g2w2},~\eqref{eq:g3w3}, and~\eqref{eq:g4w4}.

We observe that the flow equation~\eqref{eq:gfloweq} for the $2$PPI effective action has a simple so-called one-loop structure as it is the case for the RG flow
equation for {the one-particle irreducible ($1$PI) effective action} derived by Wetterich~\cite{Wetterich:1992yh}. As already indicated in the case of the flow 
equation for~$W_{\lambda}$, however, this does by no means imply that only 
one-loop corrections are taken into account with this flow equation. In fact, by solving the functional differential equation~\eqref{eq:gfloweq}, we automatically include
arbitrarily high orders in a loop expansion. Moreover, we emphasize that 
the derivation of this flow equation does not require that the interaction strength is small.

In terms of the terminology of many-body physics, the first term on the right-hand side of Eq.~\eqref{eq:gfloweq} (and analogously of Eq.~\eqref{eq:wfloweq}) 
can be identified with the {\it Hartree} term. The second
term on the right-hand side contains all other contributions, including {\it Fock} contributions as we shall discuss in more detail in Sec.~\ref{sec:hfa}.
Moreover, we would like to stress that the expansion~\eqref{eq:vertexexp} of the density functional about the ground state 
is an exact expansion and should by no means be confused with the local
density approximation. This is already clear from the fact that the $\Gamma_{\lambda}^{(n)}$~functions, which determine the functional~$\Gamma_{\lambda}$, 
depend on the imaginary time and the spatial coordinates, see also our discussion of Eq.~\eqref{eq:gradexp}.

As in the case of the flow equation for~$W_{\lambda}$, also the flow equation~\eqref{eq:gfloweq} for the density functional can in general not
be solved exactly. To construct systematically the exact solution of this equation, we can make use of the expansion~\eqref{eq:vertexexp} about the ground state~$\rho_{\rm gs}$.
Plugging this expansion into the general flow equation~\eqref{eq:gfloweq} and comparing 
the same orders in~$(\rho-\rho_{\rm gs})$ on the left- and right-hand side, we obtain flow equations for the $2$PPI correlation
functions~$\Gamma_{\lambda}^{(n)}$. In complete analogy to the flow equations for the connected correlation functions, this yields
an infinite tower of coupled differential equations for the $2$PPI correlation
functions. To be more specific, the flow of the~$\Gamma_{\lambda}^{(n)}$ function for $n\geq 2$ in general
depends on~$\Gamma_{\lambda}^{(m)}$ with $2\leq m \leq n+2$. As we expand about the ground state, the flow equation for the~$\Gamma_{\lambda}^{(1)}$ function vanishes
by construction.

{The formally exact flow equation~\eqref{eq:gfloweq} also constrains the potentially possible approximations for~$\Gamma_{\lambda}$. To illustrate this with a simple example, 
consider an ansatz for~$\Gamma_{\lambda}$ which is of the type of a local density approximation (LDA):
\be
\Gamma_{\lambda}[\rho] \approx \gamma_{0,\lambda} +\sum_{m=2}^{\infty} \frac{\gamma_{m,\lambda}}{m!}\int_{\chi}(\rho(\chi)-\rho_{{\rm gs},\lambda}(\chi))^m\,,
\label{eq:tddftlda}
\ee
where the~$\gamma_m$'s are $\lambda$-dependent real-valued numbers. For a general interaction potential~$U_{2b}$, we observe that this ansatz does not allow for a consistent
solution of the flow equation~\eqref{eq:gfloweq}. In fact, according to the flow equation~\eqref{eq:gfloweq}, we have~$\Gamma_{\lambda}^{(2)} \sim U_{2b}$. On the other hand, 
the ansatz~\eqref{eq:tddftlda} yields~$\Gamma_{\lambda}^{(2)} \sim \delta(\chi_1-\chi_2)$ which is in general only consistent with the flow equation~\eqref{eq:gfloweq}
for an interaction potential of the form~$U_{2b}\sim \delta(\chi_1-\chi_2)$. Whereas an interaction potential of this form is meaningful for systems of fermions with an internal
degree of freedom (such as spin), it is not for identical fermions due to the {\it Pauli} exclusion 
principle. In our studies below, we shall therefore always employ the full vertex expansion~\eqref{eq:vertexexp}, i.e. we take into account
the full dependence of the correlation functions on time-like and spatial coordinates.
} 

The flow equations~\eqref{eq:wfloweq} and~\eqref{eq:gfloweq} are equivalent on a formal level. However, the computation of the initial conditions of
the $2$PPI correlation functions require an inversion of the density-density correlation function~$G^{(2)}_{\lambda}$. The latter can be derived conveniently
from the one-particle propagator, see Eq.~\eqref{eq:g2opp}. While the inversion of this correlation function may be simple for theories which are invariant under
translation in time and space, it may turn out to be highly involved in any other case, such as fermions trapped in a harmonic potential.
Moreover, the computation of the initial conditions for the ~$\Gamma_{\lambda}^{(n)}$ functions is tedious in general, as can be seen from Eqs.~\eqref{eq:g2w2},~\eqref{eq:g3w3}, and~\eqref{eq:g4w4},
and always require the connected density correlators~$G^{(n)}_{\lambda}$ as an input. Therefore,
we shall consider the flow equation~\eqref{eq:wfloweq} from here 
on for any explicit calculation. 
As stated above, the resulting connected density correlation functions can in principle be used to construct the density functional~$\Gamma_{\lambda}$, if needed.

We close this section by commenting briefly on the RG philosophy underlying our approach which is of the {\it Callan-Symanzik} type. Originally, {\it Callan} and {\it Symanzik}
aimed at a study of the scaling behavior of correlation functions under a variation of the renormalized mass in relativistic field theories. Indeed, the famous 
{\it Callan-Symanzik} equation follows 
from taking the derivative of the one-particle irreducible $n$-point correlation function with respect to the (renormalized) mass. This yields a set of differential equations 
which allows to study the scaling behavior of the correlation functions, e.g., in the limit of small masses or, equivalently, large momenta. 
In other words, this type of RG equation follows from taking the derivative of correlation functions
with respect to a suitably chosen parameter/coupling associated with a term bilinear in the fields in the underlying action, e.g. the mass parameter in the case of the {\it Callan-Symanzik} equation.
In our case, we rescale the two-body interaction
with a dimensionless control parameter~$\lambda$ and study then the change of the correlation functions under a variation of this 
parameter.\footnote{Here, we assume that we use~${\mathcal R}_{\lambda}=\lambda$ for the regulator function.}
The 
RG equations for these functions then follow from taking the derivative of the generating functional of either the connected correlation functions or the $2$PPI correlation functions with respect
to~$\lambda$ and describe the change of these functions under a (specific) variation of the two-body interaction. Alternatively, if we bosonized
the action~\eqref{eq:action} already on the level of the path integral by employing a {\it Hubbard-Stratonovich} transformation 
to introduce an (auxiliary) composite field~$\rho(\tau,x) \sim \psi^{\ast}(\tau,x)\psi(\tau,x)$, the two-body interaction would appear in a term bilinear in the auxiliary field~$\rho$ and, in the spirit
of the {\it Callan-Symanzik} equation, may be viewed as a $\lambda$-dependent {non-local (mass-like) gap} for the field~$\rho$. The variation of this term with respect
to~$\lambda$ in the {\it Hubbard-Stratonovich}-transformed theory, which is still equivalent to the original theory, then allows us in principle again to compute the 
behavior of the correlation functions as a function of the control parameter~$\lambda$, making the analogy to the {\it Callan-Symanzik} approach even more apparent.

\subsection{Density-Density Correlation Function and Spectral Function}\label{sec:ddcf} 
Let us now discuss some properties of the density correlation functions~$G^{(n)}_{\lambda}$, 
with an emphasis on the density-density correlation function~$G^{(2)}_{\lambda}$. 

Loosely speaking, the density correlation functions~$G_{\lambda}^{(n)}$ are related to the probability to find a fermion at, e.g., a given point~$x_1$ and the other
fermions at~$x_2, \dots, x_N$. For example,
the density is related to the probability to find a fermion at point~$x$:
\be
n_{{\rm gs},\lambda}(x)\equiv \rho_{{\rm gs},\lambda}(0,x)=\langle \psi^{\ast}(0,x) \psi(0,x)\rangle_{\lambda}\,.\label{eq:dens2}
\ee
Here, we have used Eq.~\eqref{eq:densdef}, which is also valid for finite~$\lambda$, and 
assumed that we are only considering theories which are time-translation invariant. This allows us to set~$\tau=0$ without loss
of generality. From an integration of the density with respect to~$x$, we then obtain the total fermion number~$N$.

Correspondingly, the correlation function~$Z^{(2)}$ evaluated at~$\tau_1=\tau_2$,
\be
Z^{(2)}_{\lambda}(0,x_1,0,x_2)=\langle \psi^{\ast}(0,x_1) \psi(0,x_1)\rangle_{\lambda}  \langle \psi^{\ast}(0,x_2) \psi(0,x_2)\rangle_{\lambda} + G^{(2)}_{\lambda}(0,x_1,0,x_2)\,,  
\ee
is related to the probability to find one fermion at point~$x_1$ and another fermion at point~$x_2$.\footnote{Again, we have set~$\tau_1=\tau_2=0$ without loss of generality.}
To see this, we compute~$Z^{(2)}_{\lambda}$ in the operator {formalism:
\be
\!\!\!\!\!\!\!\!\!
Z^{(2)}_{\lambda}(0,x_1,0,x_2)\!=\!\langle \Psi_{{\rm gs},\lambda} | \hat{n}(x_1)\hat{n}(x_2)|\Psi_{{\rm gs},\lambda}\rangle
\!=\! n_{{\rm gs},\lambda}(x_1)\delta(x_1\!-\!x_2) + N(N\!-\!1) \int_{x_3}\cdots\int_{x_N} |\Psi_{{\rm gs},\lambda}(x_1,x_2,\dots,x_N)|^2
\,,\label{eq:Z2wf}
\ee
where} $|\Psi_{{\rm gs},\lambda}\rangle$ (with $\langle  \Psi_{{\rm gs},\lambda}|  \Psi_{{\rm gs},\lambda} \rangle=1$) denotes the ground-state wave function, $\hat{n}(x)=\sum_{n=1}^{N}\delta(x-\hat{x}_i)$ 
is the density operator, {and
\be
n_{{\rm gs},\lambda}(x)=N \int_{x_2}\cdots \int_{x_N}  |\Psi_{{\rm gs},\lambda}(x,x_2,\dots,x_N)|^2
\ee
is} the density. From these expressions, we read off the following exact identities for the density-density correlation functions~$Z^{(2)}_{\lambda}$ and~$G^{(2)}_{\lambda}$, respectively:
\be
\int_{x_1}\int_{x_2}Z^{(2)}_{\lambda}(0,x_1,0,x_2)=N^2\quad\text{and}\quad \int_{x_1}\int_{x_2}G^{(2)}_{\lambda}(0,x_1,0,x_2)=0\,.\label{eq:Z2sumG2sum}
\ee
For convenience, we now {define
\be
\Phi_{N,\lambda}(x_1,x_2)=
\int_{x_3}\cdots\int_{x_N} |\Psi_{{\rm gs},\lambda}(x_1,x_2,x_3,\dots,x_N)|^2\,,
\ee
which is nothing but the so-called two-body density up to a factor~$N(N-1)$. Moreover, up to a suitably chosen normalization, $\Phi_{N,\lambda}(x_1,x_2)$~determines 
the probability to find one fermion at position~$x_1$ and simultaneously another fermion at position~$x_2$. Note that the two-body density is also related to the single-particle density matrix.
The $N$-body density can in principle be extracted from the correlation function~$Z_{\lambda}^{(N)}$.

Apparently, the ground-state density and the quantity $\Phi_{N,\lambda}(x_1,x_2)$ are closely related:}
\be
n_{{\rm gs},\lambda}(x)=N\int_{x_2} \Phi_{N,\lambda}(x,x_2)\,.
\ee
{Neglecting fluctuation effects and assuming that the positions~$x_1$ and~$x_2$ are far away from each other, we {obtain
\be
\Phi_{N,\lambda}(x_1,x_2) \simeq \frac{1}{N(N-1)} n_{{\rm gs},\lambda}(x_1)n_{{\rm gs},\lambda}(x_2)\,.
\ee
Moreover,} we deduce the following exact relation from Eq.~\eqref{eq:Z2wf}:}
\be
\Phi_{N,\lambda}(x_1,x_2)
=\frac{1}{N(N-1)}\left( n_{{\rm gs},\lambda}(x_1)n_{{\rm gs},\lambda}(x_2)+G_{\lambda}^{(2)}(0,x_1,0,x_2)-n_{{\rm gs},\lambda}(x_1)\delta(x_1-x_2)
\right)\,.\label{eq:G2Pauli}
\ee
Due to {\it Pauli} blocking, we have $\Phi_{N,\lambda}(x,x)=0$. For~$x_1\to x_2$, we therefore find
\be
Z_{\lambda}^{(2)}(0,x_1,0,x_2)=G_{\lambda}^{(2)}(0,x_1,0,x_2) + n_{{\rm gs},\lambda}(x_1)n_{{\rm gs},\lambda}(x_2)\sim n_{{\rm gs},\lambda}(x_1)\delta(x_1-x_2)\,,
\ee
which also follows directly from an evaluation of Eq.~\eqref{eq:g2opp}.
Specifically for~$N=2$, we obtain
\be
|\Psi_{{\rm gs},\lambda}(x_1,x_2)|^2 =\frac{1}{2}\left( n_{{\rm gs},\lambda}(x_1)n_{{\rm gs},\lambda}(x_2)+G_{\lambda}^{(2)}(0,x_1,0,x_2)-n_{{\rm gs},\lambda}(x_1)\delta(x_1-x_2)\right)
\,.\label{eq:psi2N2}
\ee
This implies that we can compute the absolute square of the wave function for~$N=2$, i.e. the probability density to find one fermion at point~$x_1$ and the other one at point~$x_2$, 
directly from the density~$n_{{\rm gs},\lambda}$ and the
density-density correlation function~$G_{\lambda}^{(2)}$ as, e.g., computed with our DFT-RG approach. To extract the absolute square
of the wave function associated with the relative motion~$|\varphi_{N,\lambda}|^2$, i.e. the so-called intrinsic density, it is convenient to 
introduce new coordinates. For~$N=2$, we use~$R=\frac{1}{2}(x_1+x_2)$ and~$r=x_1-x_2$. The intrinsic density is then given by
\be
|\varphi_{N=2,\lambda}(r)|^2=2\int_{-\frac{L-|r|}{2}}^{\frac{L-|r|}{2}}{\rm d}R\,|\Psi_{{\rm gs},\lambda}(R+{\tfrac{1}{2}}|r|,R-\tfrac{1}{2}|r|)|^2 \,,\label{eq:iddef}
\ee
where we have used the symmetry under~$r\to -r$ to define~$r$ such that $0\leq r\leq L$. 
Although~$\Gamma_{\lambda}$ is not a functional of the intrinsic density but only of~$\rho$, we can obtain the intrinsic density from
the correlation functions~$\Gamma^{(n)}_{\lambda}$ determining uniquely the density functional~$\Gamma_{\lambda}$. 
We stress that the density~$\rho_{{\rm gs},\lambda}$ minimizing~$\Gamma_{\lambda}$ 
should by no means be confused
with the intrinsic density. In fact, for a given $N$-body system (with~$N>1$), the intrinsic density cannot be extracted from~$\rho_{{\rm gs},\lambda}$.

{For~$N\geq 3$, it} is possible to extract~$|\Psi_{{\rm gs},\lambda}(x_1,\dots,x_N)|^2$ from the $N$-density correlation function~$G^{(N)}_{\lambda}$.
The associated intrinsic 
density is then obtained along the lines of the case~$N=2$ by, e.g., using {\it Jacobi} coordinates.
We add that the computation of the $G^{(N)}_{\lambda}$ functions and therefore 
of~$|\Psi_{{\rm gs},\lambda}(x_1,\dots,x_N)|^2$ in general requires the fully time-dependent density correlation functions~$G^{(m)}_{\lambda}(\tau_1,x_1,\tau_2,x_2)$
{with $1\leq m \leq N+2$, as} we shall see below. 

We now show that {excited states} can be extracted from the time-dependent density-density correlation function~$G^{(2)}_{\lambda}$. To this
end, we first consider the spectral decomposition of~$Z^{(2)}_{\lambda}$ within the operator formalism,
\be
Z^{(2)}_{\lambda}(\tau_1,x_1,\tau_2,x_2)
=\sum_{n=0}^{\infty}\langle \Psi_{{\rm gs},\lambda}| \hat{\psi}^{\dagger}(x_1)\hat{\psi}(x_1) |\Psi_{\lambda}^{(n)}\rangle\langle \Psi_{\lambda}^{(n)}|  \hat{\psi}^{\dagger}(x_2)\hat{\psi}(x_2)
|\Psi_{{\rm gs},\lambda}\rangle{\rm e}^{-(E_{n,\lambda}-E_{0,\lambda})|\tau_1-\tau_2|}\,,\label{eq:z2spec}
\ee
which is obtained by inserting a suitably chosen~${\mathds{1}}$-operator, 
\be
\mathds{1}=\sum_{n=0}^{\infty} |\Psi^{(n)}_{\lambda}\rangle\langle \Psi^{(n)}_{\lambda}|\,,
\ee
into the analogue of Eq.~\eqref{eq:Zn} in the {\it Heisenberg} picture.
Here, $|\Psi^{(n)}_{\lambda}\rangle$ are the eigenstates of the $N$-body problem with energies~$E_{n,\lambda}$, ordered such 
that~$E_{\lambda}\equiv E_{0,\lambda} < E_{1,\lambda} <\dots$ and~$|\Psi_{{\rm gs},\lambda}\rangle\equiv |\Psi_{\lambda}^{(0)}\rangle$.\footnote{Here, we assume that the ground state is not degenerate.}
From the relation~\eqref{eq:g2opp} between~$Z^{(2)}_{\lambda}$ and~$G^{(2)}_{\lambda}$, it then follows that 
\be
G^{(2)}_{\lambda}(\tau_1,x_1,\tau_2,x_2)
=\sum_{n=1}^{\infty}\langle \Psi_{{\rm gs},\lambda}| \hat{\psi}^{\dagger}(x_1)\hat{\psi}(x_1) |\Psi_{\lambda}^{(n)}\rangle\langle \Psi_{\lambda}^{(n)}|  \hat{\psi}^{\dagger}(x_2)\hat{\psi}(x_2)
|\Psi_{{\rm gs},\lambda}\rangle\,{\rm e}^{-(E_{n,\lambda}-E_{0,\lambda})|\tau_1-\tau_2|}\,.\label{eq:g2spec}
\ee
From this expression, we deduce that the first excited state can be extracted directly from~$G^{(2)}_{\lambda}$ in the limit of large time {differences:
\be
(E_{1,\lambda}-E_{0,\lambda})=-\lim_{|\tau|\to\infty}\frac{1}{|\tau|}\ln G^{(2)}_{\lambda}(\tau,0,0,0)\,,
\ee
where} $E_{0,\lambda}$ is the ground-state energy and we have set~$\tau_2=0$ {and~$\tau_1=\tau$ without} loss of generality. In addition, we have 
set~$x_1=x_2=0$ to ensure that the first excited state
is included in the sum~\eqref{eq:g2spec}. In fact, we have
\be
 \hat{\psi}^{\dagger}(x)\hat{\psi}(x)=\sum_{k,l}\,\hat{a}^{\dagger}_k \hat{a}_l\, (\phi^{(I)}_k(x))^{\ast}\phi^{(I)}_l(x)\,,\label{eq:psidpsi}
\ee
where $\hat{a}_k^{\dagger}$ creates and~$\hat{a}_l$ annihilates a fermion with momentum~$p_k^{(I)}$ and~$p_l^{(I)}$, respectively,
where  $I\in\{{\rm P,A}\}$, see Eqs.~\eqref{eq:pP} and~\eqref{eq:pA}. The position-space representations of the one-particle eigenstates 
used in this work are given by
\be
\!\!\!\!\!\!\phi_{n}^{(I)}(x)=\frac{1}{\sqrt{L}}\,{\E}^{-\I p_n^{(I)}x}\quad\text{with}\quad \int_x (\phi_{m}^{(I)}(x))^{\ast}\phi_{n}^{(I)}(x)=\delta_{m,n}
\quad\text{and}\quad \sum_n (\phi_{n}^{(I)}(x_1))^{\ast}\phi_{n}^{(I)}(x_2)=  \sum_n\delta(x_1\!-\!x_2\!+\! nL)\label{eq:basisfcts}
\,,
\ee
where~$n,m\in {\mathbb Z}$.
From Eq.~\eqref{eq:psidpsi}, we now conclude that it is possible to choose~$x_1$ and~$x_2$ in Eq.~\eqref{eq:g2spec} such that,
at least in the non-interacting limit, the matrix element~$\langle \Psi_{{\rm gs},\lambda}| \hat{\psi}^{\dagger}(x_i)\hat{\psi}(x_i) |\Psi_{\lambda}^{(n)}\rangle$
vanishes for specific excited states, depending on the momentum difference of the created and annihilated
{fermion. The} generalization of our considerations to other basis functions, e.g., associated with a harmonic trap potential, 
is straightforward.

From Eqs.~\eqref{eq:g2spec} and~\eqref{eq:psidpsi}, we also obtain the generalizations of the relations in Eq.~\eqref{eq:Z2sumG2sum}
for time-dependent density-density correlation functions:
\be
\int_{x_1}\int_{x_2}Z^{(2)}_{\lambda}(\tau_1,x_1,\tau_2,x_2)=N^2\quad\text{and}\quad 
\int_{x_i}G^{(2)}_{\lambda}(\tau_1,x_1,\tau_2,x_2)=0\,,
\ee
where $i\in\{1,2\}$.
These relations can be further generalized to the case of $n$-density correlation functions. Using~Eq.~\eqref{eq:psidpsi} and
the spectral decomposition of the $n$-density correlation function, it follows {that
\be
\int_{x_1}\cdots \int_{x_n} Z^{(n)}_{\lambda}(\chi_1,\dots,\chi_n)=N^n\quad\text{and}\quad 
\int_{x_i} G^{(n)}_{\lambda}(\chi_1,\dots,\chi_n)=0\,,\label{eq:intzngn}
\ee
where} $i\in\{1,\dots,n\}$.
As we shall see below, these exact relations are useful to analyze the DFT-RG flows on very general grounds.

In addition to the energy of the first {excited state}, it is possible to extract the spectral function~$\Omega_{\lambda}$ from the density-density correlation
function which gives access to the energies of {higher excited states as well}. To see this, we consider the Fourier transformation 
of~$G^{(2)}_{\lambda}$ with respect to the imaginary time~$\tau$:
\be
\tilde{G}^{(2)}_{\lambda}(\omega,x_1,x_2)=\int_{-\infty}^{\infty} \frac{{\rm d}E}{2\pi}\frac{2E}{\omega^2 + E^2}\,\Omega_{\lambda}(E,x_1,x_2)\,,\label{eq:specimp}
\ee
where
\be
G^{(2)}_{\lambda}(\tau_1,x_1,\tau_2,x_2)=\int_{-\infty}^{\infty}\frac{{\rm d}\omega}{2\pi}\, \tilde{G}^{(2)}_{\lambda}(\omega,x_1,x_2) {\rm e}^{-\I\omega(\tau_1-\tau_2)}
\ee
and the spectral function~$\Omega_{\lambda}$ is defined as
\be
\Omega_{\lambda}(E,x_1,x_2)=2\pi \sum_{n=1}^{\infty}\langle \Psi_{{\rm gs},\lambda}| \hat{\psi}^{\dagger}(x_1)\hat{\psi}(x_1) |\Psi_{\lambda}^{(n)}\rangle\langle \Psi_{\lambda}^{(n)}|  \hat{\psi}^{\dagger}(x_2)\hat{\psi}(x_2)
|\Psi_{{\rm gs},\lambda}\rangle\delta(E-(E_{n,\lambda}-E_{0,\lambda}))\,.
\ee
Within our DFT-RG approach, we have direct access to the density-density correlation function~$G^{(2)}_{\lambda}$. To obtain the spectral function,
we therefore need to solve Eq.~\eqref{eq:specimp} for~$\Omega_{\lambda}$. Using
\be
\frac{2E}{\omega^2 + E^2}=\frac{1}{-\I\omega+E}-\frac{1}{-\I\omega-E}\quad\text{and}\quad
\lim_{\eta\to 0^{+}}\frac{1}{\omega \pm \I\eta}= {\rm P}\frac{1}{\omega}\mp\I\pi\delta(\omega)\,,
\ee
we find that Eq.~\eqref{eq:specimp} can indeed be solved for~$\Omega_{\lambda}$:
\be
\Omega_{\lambda}(E,x_1,x_2)=2\lim_{\eta\to 0^{+}}\operatorname{Im}\tilde{G}^{(2)}_{\lambda}(\I E-\eta,x_1,x_2)\,.
\ee
Moreover, we observe that~$\Omega_{\lambda}$ obeys the following ``sum rule":
\be
\int_{-\infty}^{\infty}\frac{{\rm d}E}{2\pi}\Omega_{\lambda}(E,x,y)=G^{(2)}_{\lambda}(0,x,0,y)\,.
\ee
{Thus, the spectral function is obtained from an analytic continuation of the Fourier transform 
of the density-density correlation function, in complete analogy to the case of 1PI correlation functions, see, e.g., Refs.~\cite{Mattuck:1976xt,Altland:2006si}.} 
If an analytic solution of the flow equation for the density-density correlation function
is not available, 
the analytic continuation has to be performed numerically which
is in general expected to be 
a highly non-trivial problem. However, since~$\tilde{G}^{(2)}$ is invariant under~$\omega\to-\omega$ 
and~$\tilde{G}^{(2)}_{\lambda}\sim 1/\omega^2$ in the large~$\omega$-limit, the
computation of~$\Omega_{\lambda}$ may in principle be achieved by fitting the numerical results 
for~$\tilde{G}^{(2)}_{\lambda}$ to Pad\'{e} approximants respecting these constraints. From
the analytic continuation of the Pad\'{e} approximants, the spectral function can then be obtained. In any case, we shall focus
on the computation of ground-state properties in our numerical studies presented below and defer the
computation of excited states within our DFT-RG formalism to future work.

\subsection{DFT \& Many-Body Perturbation Theory}\label{sec:mbpt}

We now show that our DFT-RG approach allows us to recover the perturbative expansion in 
a systematic fashion. Since we consider in this work only systems of $N$ identical fermions interacting via a two-body interaction,
we first show that the flow equation for the ground-state energy vanishes identically if the two-body interaction is simply given by
a contact interaction, as it should be. Since the ground-state energy is essentially proportional to~$W_{\lambda}[0]$, 
see Eq.~\eqref{eq:egs},
it suffices to consider the flow equation for the latter which is obtained by 
setting~$J=0$ in Eq.~\eqref{eq:wfloweq}:
\be
\partial_{\lambda} W_{\lambda}[0] &=& -\frac{1}{2}\int_{\chi_1}\int_{\chi_2}\rho_{{\rm gs},\lambda}(\chi_1)U_{2b}(\chi_1,\chi_2)\rho_{{\rm gs},\lambda}(\chi_2)\nn\\
&& \qquad - \frac{1}{2}\int_{\chi_1}\int_{\chi_2}U_{2b}(\chi_1,\chi_2)\left( G^{(2)}_{\lambda}(\chi_2,\chi_1) - \rho_{{\rm gs},\lambda}(\chi_2)\delta(\chi_2-\chi_1)\right)\,.
\label{eq:egsflow}
\ee
Assuming time-translation invariance and using Eqs.~\eqref{eq:dens2} and~\eqref{eq:G2Pauli},
the right-hand side of Eq.~\eqref{eq:egsflow} indeed vanishes for any value of~$\lambda$
for a contact interaction~$U_{2b}(\tau_1,x_1,\tau_2,x_2)\sim \delta(\tau_1-\tau_2)\delta(x_1-x_2)$:
\be
\partial_{\lambda} E_{\lambda} \sim \partial_{\lambda}W_{\lambda}[0]= 0\,.
\ee
Thus,
the ground-state energy does not change under a variation of~$\lambda$ and therefore it remains identical to
its initial value which is the energy of a system of $N$ non-interacting identical fermions.

Let us now consider a general two-body interaction to discuss the connection of our approach to many-body perturbation theory, i.e.
an expansion of observables, such as the ground-state energy, in powers of some (small) dimensionless parameter~$\bar{g}_N$, where
the subscript~$N$ indicates that this parameter is suitably normalized with the total particle number. For convenience, we shall assume in
the following that the two-body interaction~$U$ can be written as a product of a coupling parameter~$g$ with dimension of inverse length
and an in general space-dependent shape function~$\mathfrak U$ with dimension of inverse length as 
{it is the case for the interaction potential}~\eqref{eq:pot}: 
\be
U(x_1-x_2)=g\,\mathfrak{U}(x_1-x_2)\,.
\ee
For~$N$ fermions in a box with extent~$L$ {and (anti)periodic boundary}
conditions as, e.g., in this work, the ground-state density~$n_{{\rm gs},\lambda}=N/L$ is homogeneous for any value of~$\lambda$ (see our discussion below). Therefore
we define 
\be
\bar{g}_N=\frac{gL}{N}
\ee
as dimensionless coupling parameter.\footnote{This definition is common in studies of one-dimensional ultracold gases interacting via a 
contact interaction~\cite{2004PhRvL..93i0405T,2004PhRvL..93i0408F,Rammelmuller:2015zaa}. Note that other finite (length) scales defining the range of the interaction and the extent of the repulsive core enter our
calculations in this work, see Eq.~\eqref{eq:pot}.
Thus, the coupling~$g$ as {a naive} measure of the potential energy and the density as {a naive} measure of the kinetic energy are not
the only scales in our numerical studies below. Here, we only use this definition for convenience to define an expansion of the density functional.}
For confining geometries other than a box {with (anti)periodic boundary} 
conditions, a dimensionless coupling can be defined accordingly. For fermions in a harmonic trap, for example, we may use the density of the non-interacting system
in the center of the trap to render~$g$ dimensionless. 
The perturbative expansion of the ground-state energy can now be defined as follows:
\be
E_{\lambda}=N\left(E^{(0)} + E_{\lambda}^{(1)} \bar{g}_N + \frac{1}{2} E_{\lambda}^{(2)} \bar{g}_N^2 + \dots\right)\,,
\label{eq:egsexp}
\ee
where~$E_{\lambda}^{(i)}$ is associated with the energy correction of order~${\mathcal O}(\bar{g}_N^i)$ and~$E^{(0)}\equiv (1/N)E_{\lambda=0}$ 
is the ground-state energy of the
non-interacting system, i.e. $N$ non-interacting {fermions in a box with (anti)periodic boundary conditions in this work. In} 
the same way we can expand other quantities, such as the 
ground-state density and the density-density correlation function. We define:
\be
n_{{\rm gs},\lambda}(x)=n_{{\rm gs}}^{(0)}(x) + n_{{\rm gs},\lambda}^{(1)}(x) \bar{g}_N + \frac{1}{2} n_{{\rm gs},\lambda}^{(2)} (x)\bar{g}_N^2 + \dots\,,\label{eq:densexp}
\ee
where~$n_{{\rm gs}}^{(0)}\equiv n_{{\rm gs},\lambda=0}$ is the ground-state density of the non-interacting system, and
\be
G^{(2)}_{\lambda}(\tau_1,x_1,\tau_2,x_2)=
G^{(2,0)}(\tau_1,x_1,\tau_2,x_2) + G^{(2,1)}_{\lambda}(\tau_1,x_1,\tau_2,x_2)\bar{g}_N
+ \frac{1}{2}G^{(2,2)}_{\lambda}(\tau_1,x_1,\tau_2,x_2) \bar{g}_N^2+\dots\,,\label{eq:g2exp}
\ee
where~$G^{(2,0)}\equiv G^{(2)}_{\lambda=0}$ is the non-interacting density-density correlation function. Density correlation functions of higher order
can be expanded accordingly. Thus, the initial conditions for the RG equations for the energy, density and the density correlation functions are 
simply given by the zeroth order of their perturbative expansions. For our $N$-body system in a box, for example, we have
\be
E_{\lambda}^{(0)} = \frac{1}{N} E_{\lambda=0}=\frac{1}{N}\sum_{n}\epsilon_n^{(I)} \theta(- \bar{\epsilon}_n^{\,(I)})
=\frac{\pi^2}{6}\left(\frac{N}{L}\right)^2\left(1-\frac{1}{N^2}\right)
\,,\label{eq:egs0}
\ee
where the one-particle energies are given by
\be
\epsilon_n^{(I)}=\frac{1}{2}(p_n^{(I)})^2\,.
\ee
The auxiliary function~$\theta$ is defined as follows
\be
\theta(\pm\bar{\epsilon}_n^{\,(I)})=\begin{cases}
    1 & \text{for}\quad \pm\bar{\epsilon}^{(I)}_n \mp\eta > 0\,,\\
    0 & \text{otherwise}\,.
  \end{cases}
\ee
Here,~$\eta\to 0^{+}$ is tacitly assumed. Moreover, we have introduced
$\bar{\epsilon}^{(I)}_n=\epsilon^{(I)}_n -\epsilon_{\rm F}^{(I)}$,
where the Fermi 
energies
\be
\epsilon_{\rm F}^{(A)}=\epsilon_{\frac{N}{2}-1}^{(A)} \quad (N\;\text{even})\quad \text{and}
\quad \epsilon_{\rm F}^{(P)}=\epsilon_{\frac{N-1}{2}}^{(P)}\quad (N\;\text{odd})
\ee
are determined by the fermion number~$N$, respectively.
Note that the expression on the right-hand side of Eq.~\eqref{eq:egs0} depends only on~$N$ and~$L$ but not on the choice of the boundary conditions.

For the one-particle propagator as defined in Eq.~\eqref{eq:onepart}, we find the following expression for a non-interacting system in a box with (anti)periodic boundary conditions for the one-particle {states:
\be
\Delta_{\lambda=0}^{(I)}(\tau_1,x_1,\tau_2,x_2)&=&-\sum_n\int_{-\infty}^{\infty} 
\frac{{\rm d}\omega}{2\pi}\frac{{\rm e}^{-\I\omega(\tau_1-\tau_2)}}{-\I\omega + \bar{\epsilon}^{\,(I)}_n}(\phi_n^{(I)}(x_1))^{\ast}\phi_n^{(I)}(x_2)\nn\\
&=& -\sum_n \left\{ \theta(\bar{\epsilon}_n^{\,(I)})\theta_{\sigma}(\tau_1-\tau_2) - \theta(-\bar{\epsilon}_n^{\,(I)})\theta_{\sigma}(\tau_2-\tau_1)
\right\}(\phi_n^{(I)}(x_1))^{\ast}\phi_n^{(I)}(x_2)\,{\rm e}^{-|\bar{\epsilon}_n^{\,(I)}||\tau_1-\tau_2|}\,.
\ee
The} basis functions~$\phi_n^{(I)}$ are defined in Eq.~\eqref{eq:basisfcts}.  From the one-particle propagator, we can now construct the density as well as
the density correlation functions at~$\lambda=0$ corresponding to the zeroth order in the perturbative expansion. For the density, we find from Eq.~\eqref{eq:densdef} {that
\be
n_{{\rm gs}}^{(0)}(x)\equiv n_{{\rm gs},\lambda=0}(x) \equiv \rho_{{\rm gs},\lambda=0}(0,x) 
=\lim_{\tau\to 0^{-}}\Delta_{\lambda=0}^{(I)}(\tau,x,0,x)
=\sum_n (\phi_n^{(I)}(x))^{\ast}\phi_n^{(I)}(x)\theta(- \bar{\epsilon}_n^{\,(I)})=\frac{N}{L}\,.\label{eq:ngsic}
\ee
Thus, the} density is homogeneous, time-independent, and independent of our choice for the boundary conditions associated with odd and even particle numbers, respectively.
To compute the density-density correlation function in the non-interacting limit, we use Eq.~\eqref{eq:g2opp} to obtain
\be
 G^{(2,0)}(\tau_1,x_1,\tau_2,x_2) \equiv G^{(2)}_{\lambda=0}(\tau_1,x_1,\tau_2,x_2) 
 =\frac{1}{L}\sum_{k,l} \E^{-|\epsilon_{k}^{\,(I)}-\epsilon_{l}^{\,(I)}||\tau_1-\tau_2|}\,\theta(\bar{\epsilon}_{k})\theta(-\bar{\epsilon}_{l})\,
(\phi_{k-l}^{(P)}(x_1))^{\ast}\phi_{k-l}^{(P)}(x_2)\,.\label{eq:g2ic}
\ee
As stated above, the density-density correlation function depends only on~$|\tau_1-\tau_2|$. Note that the density-density correlation
function obeys periodic boundary conditions in the spatial direction, independent of our choice for the boundary conditions of the one-particle states, $I\in\{P,A\}$.
This also holds for the interacting system and can be traced back to the fact that~$(\phi^{(I)}_{k}(x))^{\ast}\phi^{(I)}_{l}(x)\sim \phi^{(P)}_{k-l}(x)$. Moreover, we observe
that the density-density correlation function depends only on~$|x_1-x_2|$. For the interacting system, however, this is only the case if the interaction potential
also exhibits translation invariance.
For~$\tau_1=\tau_2$, we can rewrite~$G^{(2,0)}$ as follows:
\be
G^{(2,0)}(0,x_1,0,x_2) 
 = 
 \delta(x_1-x_2)n_{{\rm gs}}^{(0)}(x_1)
 - \frac{1}{L}\sum_{k,l} \theta(-\bar{\epsilon}_{k})\theta(-\bar{\epsilon}_{l})\,
(\phi_{k-l}^{(P)}(x_1))^{\ast}\phi_{k-l}^{(P)}(x_2)\,,\label{eq:g20ic}
\ee
where we have used~$x_i\in[-L/2,L/2)$ and set~$\tau_1=\tau_2=0$ without loss of generality. As it should be, these results 
for the density-density correlation function are in agreement with
our general considerations in Sec.~\ref{sec:ddcf}. In particular, we have
\be
\int_{x_1}\int_{x_2} G^{(2,0)}(0,x_1,0,x_2) = 0\,
\ee
for any fermion number. Specifically for~$N=2$, we find
\be
|\Psi_{{\rm gs},\lambda=0}(x_1,x_2)|^2=\frac{1}{4}\left(\frac{2}{L}\right)^2\left( 1 - \cos\left(\frac{2\pi (x_1-x_2)}{L}\right)\right) 
\quad\text{and}\quad |\varphi_{2,\lambda=0}(r)|^2= \left(\frac{2}{L}\right)^2\left(L-r\right)\sin^2\left(\frac{\pi r}{L}\right)
\label{eq:idl0}
\ee
for the absolute square of the ground-state wave function and the intrinsic density with $r=|x_1-x_2|\leq L$, respectively.

We now turn to the computation of the leading-order correction to the ground-state energy for which we follow the general
discussion in Ref.~\cite{Kemler:2013yka}. Using Eq.~\eqref{eq:egs} and plugging the expansions~\eqref{eq:egsexp},~\eqref{eq:densexp}, and~\eqref{eq:g2exp}
into the flow equation~\eqref{eq:egsflow}, we observe that the computation of the leading-order correction~$E_{\lambda}^{(1)}$ to the ground-state energy
only requires the density and density-density correlation function in their zeroth-order approximations, i.e.~$n^{(0)}_{{\rm gs}}$ and~$G^{(2,0)}$, respectively.
Thus, we have
\be
\partial_{\lambda} E_{\lambda}^{(1)} &=& \frac{1}{2L}\int_{x_1}\int_{x_2} n_{{\rm gs}}^{(0)}(x_1) {\mathfrak U}(x_1-x_2) n_{{\rm gs}}^{(0)}(x_2) \nn\\
&& \qquad + \frac{1}{2L}\int_{x_1}\int_{x_2} {\mathfrak U}(x_1-x_2)\left( G^{(2,0)}(0,x_2,0,x_1) - n_{{\rm gs}}^{(0)}(x_2)\delta(x_2-x_1)
\right)\,.
\label{eq:diffeqegs1}
\ee
Note that the general structure of this equation does not depend on the confining geometry. 
For fermions in a box with (anti)periodic boundary conditions, we obtain
\be
E_{\lambda=1}^{(1)} &=& \int_0^{\lambda=1}\!\!\! {\rm d}\lambda^{\prime} \partial_{\lambda^{\prime}}E_{\lambda^{\prime}}^{(1)} =
\frac{1}{2}\mathfrak{U}_{0,0}\left(\frac{N}{L}\right)^2
- \frac{1}{2L^2}\sum_{k,l} 
\theta(-\bar{\epsilon}_{k}^{(I)})\theta(-\bar{\epsilon}_{l}^{(I)})\,\mathfrak{U}_{k-l,k-l}\,.
\label{eq:egslo}
\ee
Here, we have used Eqs.~\eqref{eq:ngsic} and~\eqref{eq:g20ic}. Moreover, we have introduced the dimensionless coefficients~$\mathfrak{U}_{m,n}$ which essentially define the Fourier-series 
representation of the function~$\mathfrak{U}$:
\be
\mathfrak{U}(x_1-x_2)=\sum_{m,n} \mathfrak{U}_{m,n}\, (\phi_{m}^{(P)}(x_1))^{\ast}\phi_{n}^{(P)}(x_2)\,.\label{eq:Udecomp}
\ee
We observe that only the diagonal elements of the matrix~$\mathfrak{U}_{m,n}$ enter the 
computation of the leading-order correction~$E_{\lambda}^{(1)}$. Note that~$\mathfrak{U}_{m,n}\neq 0$ for~$m\neq n$ in
a finite box, even if the interaction only depends on~$|x_1-x_2|$. Thus, translation invariance is explicitly broken by the presence
of the box. For convenience, however, we shall employ a specific periodic extension of the original interaction potential in the box from now on, which is obtained by only
taking into account the diagonal elements of the matrix~$\mathfrak{U}_{m,n}$:
\be
\mathfrak{U}_{m,n}=\mathfrak{U}_{m,m}\delta_{m,n}\,.\label{eq:Udiag}
\ee
This corresponds to a {\it redefinition} of the interaction potential~$U$ in such a way that, for $|x_1-x_2|\leq L/2$,
$U$ is still identical to the original
interaction potential in the infinite-volume limit but obeys the periodicity condition
\be
{U}(x_1-x_2+mL) = U(x_1-x_2)
\ee
with $m\in\mathbb{Z}$.
The so-defined two-body potential~$U(x)$ depends only on~$|x_1-x_2|$ even for finite~$L$ and
is a continuous periodic function of~$x$ on the interval~$[-L/2,L/2)$,
which is convenient for numerical studies of the DFT-RG flow {equations. 
Apparently, our {\it redefined} interaction potential
is not identical to the original interaction in the presence of the box. However, it
approaches the original interaction potential in the infinite-volume limit which is of most relevance for a study of the formation
of selfbound states in the absence of this auxiliary fermion-confining box.\footnote{By
solving the two-body problem exactly for both definitions of the interaction in the box, we have checked that the correct infinite-volume limit is indeed recovered, see also our discussion 
in Sec.~\ref{sec:res}.} Note that,
although it is convenient, our DFT-RG approach does not require to impose periodicity of the interaction in the presence of the box but works 
with any type of interaction.} 

Let us now discuss the derivation of the second-order correction~$E_{\lambda}^{(2)}$ to the ground-state energy. 
As for the leading-order correction, we use Eq.~\eqref{eq:egs} and plug the expansions~\eqref{eq:egsexp},~\eqref{eq:densexp}, and~\eqref{eq:g2exp}
into the flow equation~\eqref{eq:egsflow}. From this, we deduce that the computation of~$E_{\lambda}^{(2)}$ requires
the computation of the leading-order correction to the density and the density-density correlation function, i.e.~$n_{{\rm gs},\lambda}^{(1)}$
and~$G^{(2,1)}_{\lambda}$, respectively. The flow equation for the density~$n_{{\rm gs},\lambda}$ 
is obtained by taking a functional derivative of Eq.~\eqref{eq:wfloweq} with respect
to the source~$J$ and setting then~$J=0$. It reads:
\be
\partial_{\lambda}n_{{\rm gs},\lambda}(x) &=& - \int_{\tau_1}\int_{x_1}\int_{x_2} n_{{\rm gs},\lambda}(x_1) {U}(x_1-x_2)G^{(2)}_{\lambda}(\tau_1,x_2,0,x)\nn\\
&& \qquad -\frac{1}{2}\int_{\tau_1}\int_{x_1}\int_{x_2} U(x_1-x_2)\left( G^{(3)}_{\lambda}(\tau_1,x_2,\tau_1,x_1,0,x)-\delta(x_2-x_1)G^{(2)}_{\lambda}(\tau_1,x_1,0,x)\right)\,.
\ee
From Eq.~\eqref{eq:intzngn}, it follows immediately that 
\be
\int_x\partial_{\lambda}n_{{\rm gs},\lambda}(x)=0\,.
\ee
Thus, the fermion number remains constant in the RG flow and is therefore indeed fixed by the initial condition. Using Eqs.~\eqref{eq:Udecomp} and~\eqref{eq:Udiag}
together with Eq.~\eqref{eq:intzngn}, we observe that the flow equation for the density {simplifies:
\be
\!\!\!\!\!\partial_{\lambda}n_{{\rm gs},\lambda}(x) =-\! \int_{\tau_1}\int_{x_1}\int_{x_2} n_{{\rm gs},\lambda}(x_1) {U}(x_1\!-\!x_2)G^{(2)}_{\lambda}(\tau_1,x_2,0,x)
\!-\!\frac{1}{2}\int_{\tau_1}\int_{x_1}\int_{x_2} U(x_1\!-\!x_2) G^{(3)}_{\lambda}(\tau_1,x_2,\tau_1,x_1,0,x)\,.\label{eq:densflowpt}
\ee
The} {leading-order correction~$n_{{\rm gs},\lambda}^{(1)}$ to the density is now obtained from this equation by replacing the full density correlation functions~$G^{(2)}_{\lambda}$
and~$G^{(3)}_{\lambda}$ by their initial conditions, i.e. their 
zeroth-order approximations. Plugging Eqs.~\eqref{eq:G3} and~\eqref{eq:g2ic} into the flow equation~\eqref{eq:densflowpt}, we find
that the right-hand side of the latter equation vanishes:} 
\be
\partial_{\lambda} n_{{\rm gs},\lambda}^{(1)} = 0\,.\label{eq:pertdens}
\ee
This implies that the density remains homogeneous, even at leading order. 
{In fact, with our choice for the boundary conditions, it follows from the definition of
the density as the one-density correlation function, $n_{{\rm gs},\lambda}(x)=\langle \psi^{\dagger}{(x)}\psi(x)\rangle_{\lambda}$, that 
$n_{{\rm gs},\lambda}$ is homogeneous
for any value of~$\lambda$, i.e. at any order of the perturbative expansion, at least if the ground state is not degenerate:
\be
n_{{\rm gs},\lambda}(x)\equiv n_{{\rm gs},\lambda=0}(x)=\frac{N}{L}\,.\label{eq:constnexact}
\ee
Therefore we have~$n_{{\rm gs},\lambda}^{(j)}=0$ {for~$j\geq 1$ as} well. Our leading-order result~\eqref{eq:pertdens} 
for the density is indeed in agreement with this general statement.}
At this point, we would like to stress that this does not imply that the intrinsic density is homogeneous which already becomes apparent from our results 
for the non-interacting system, see Eq.~\eqref{eq:idl0}.

In addition to the density, we need the density-density correlation function in leading order to compute the ground-state energy at next-to-leading order in the perturbative
expansion. The flow equation for the density-density correlation function~$G^{(2)}_{\lambda}$ 
can be derived by taking the second functional derivative of Eq.~\eqref{eq:wfloweq} with respect
to the source~$J$ and setting then~$J=0$. We find:
\be
\partial_{\lambda}G^{(2)}_{\lambda}(\tau_1,x_1,\tau_2,x_2)
&=& - \int_{\tau_3}\int_{x_3}\int_{x_4} G^{(2)}_{\lambda}(\tau_1,x_1,\tau_3,x_3)U(x_3-x_4)G^{(2)}_{\lambda}(\tau_3,x_4,\tau_2,x_2)\nn\\
&& \quad -  \int_{\tau_3}\int_{x_3}\int_{x_4} n_{{\rm gs},\lambda}(x_3)U(x_3-x_4)G^{(3)}_{\lambda}(\tau_3,x_4,\tau_1,x_1,\tau_2,x_2)\nn\\
&& \quad\quad -\frac{1}{2} \int_{\tau_3}\int_{x_3}\int_{x_4} U(x_3-x_4)\left(G^{(4)}_{\lambda}(\tau_3,x_4,\tau_3,x_3,\tau_1,x_1,\tau_2,x_2)\right.\nn\\
&& \qquad\qquad\qquad\qquad\qquad\qquad\qquad\qquad \left. -\delta(x_4-x_3)G^{(3)}_{\lambda}(\tau_3,x_3,\tau_1,x_1,\tau_2,x_2)
\right)\,.
\label{eq:G2flow}
\ee
To calculate the ground-state energy, we only need the density-density correlation function for~$\tau_1=\tau_2=0$. However, 
we observe that the
calculation of the latter requires the knowledge of the time dependence of the correlation {functions~$G^{(2)}_{\lambda}$, $G^{(3)}_{\lambda}$, and~$G^{(4)}_{\lambda}$.

Before} we derive the leading-order correction of the density-density correlation function from the flow equation~\eqref{eq:G2flow}, we shall analyze this equation
from a more general point of view. Using Eq.~\eqref{eq:intzngn}, we observe
\be
\int_{x_1}\int_{x_2}\partial_{\lambda} G^{(2)}_{\lambda} (\tau_1,x_1,\tau_2,x_2)= 0\,.
\ee
Loosely speaking, this implies that the normalization of the ground-state wave-function is preserved in the RG flow which is necessary
to assign a physical interpretation to this correlation function, see also our discussion in Sec.~\ref{sec:ddcf}.
{Using Eqs.~\eqref{eq:Udecomp},~\eqref{eq:Udiag} and~\eqref{eq:constnexact}
together} with Eq.~\eqref{eq:intzngn}, the flow equation for~$G^{(2)}_{\lambda}$ simplifies considerably:
\be
\partial_{\lambda}G^{(2)}_{\lambda}(\tau_1,x_1,\tau_2,x_2)
&=& - \int_{\tau_3}\int_{x_3}\int_{x_4} G^{(2)}_{\lambda}(\tau_1,x_1,\tau_3,x_3)U(x_3-x_4)G^{(2)}_{\lambda}(\tau_3,x_4,\tau_2,x_2)\nn\\
&& \quad -\frac{1}{2} \int_{\tau_3}\int_{x_3}\int_{x_4} U(x_3-x_4) G^{(4)}_{\lambda}(\tau_3,x_4,\tau_3,x_3,\tau_1,x_1,\tau_2,x_2)\,.
\label{eq:G2flow2}
\ee
To derive the leading-order correction of the density-density correlation function from this equation, we have to replace the full density
correlation functions by their initial {conditions, i.e.~$G^{(2)}_{\lambda}\to G^{(2)}_{\lambda=0}\equiv G^{(2,0)}$ 
and~$G^{(4)}_{\lambda}\to G^{(4)}_{\lambda=0}\equiv G^{(4,0)}$, respectively.}
The initial conditions of these correlation functions can be obtained from their definitions in terms of one-particle propagators, see Eqs.~\eqref{eq:g2opp} and~\eqref{eq:G4}.
For the leading-order correction~$G^{(2,1)}_{\lambda}$, we then find
\be
&& G^{(2,1)}_{\lambda}(\tau_1,x_1,0,x_2)\nn\\
&&\qquad\qquad = -\lambda\frac{N}{L^3}\sum_{m\neq0}\,\mathfrak{U}_{m,m}(\phi_{m}^{(P)}(x_1))^{\ast} \phi_{m}^{(P)}(x_2)\bigg\{\bigg(
\sum_{k} \theta(-\bar{\epsilon}_{k}^{(I)})\theta(\bar{\epsilon}_{k+m}^{(I)})  |\tau_1|\E^{-|\epsilon_{k+m}^{(I)}-\epsilon_{k}^{(I)}||\tau_1|}
\nn\\
&&\qquad\qquad \qquad  -2\sum_{k\neq l}\, 
\frac{1}{\epsilon_{k+m}^{(I)}-\epsilon_{l+m}^{(I)}+\epsilon_{l}^{(I)}-\epsilon_{k}^{(I)}}  
\theta(\bar{\epsilon}_{k+m}^{(I)})\theta(-\bar{\epsilon}_{k}^{(I)})\left(\theta(-\epsilon_{l}^{(I)})-\theta(-\epsilon_{l+m}^{(I)})\right)\E^{-|\epsilon_{k+m}^{(I)}-\epsilon_{k}^{(I)}||\tau_1|} \bigg\}
\nn\\
&&\qquad\qquad\quad + \lambda\frac{N}{L^3} \sum_{k,l} (\phi_{k-l}^{(P)}(x_1))^{\ast} \phi_{k-l}^{(P)}(x_2)\bigg\{ 
\sum_{m\neq 0} \mathfrak{U}_{m,m}\theta(-\bar{\epsilon}_{k}^{(I)})\left(\theta(-\bar{\epsilon}_{l+m}^{(I)})\!-\!\theta(-\bar{\epsilon}_{k+m}^{(I)}) \right) |\tau_1|\E^{-|\epsilon_{k+m}^{(I)}-\epsilon_{l+m}^{(I)}||\tau_1|}
\nn\\
&&\qquad\qquad \quad\quad +2\sum_{m\neq 0} \mathfrak{U}_{m,m}\frac{1}{\epsilon_{k+m}^{(I)}-\epsilon_{l+m}^{(I)}+\epsilon_{l}^{(I)}-\epsilon_{k}^{(I)}} 
\theta(-\bar{\epsilon}_{k}^{(I)})\theta(\bar{\epsilon}_{l}^{(I)})\left(\theta(-\bar{\epsilon}_{l+m}^{(I)})-\theta(-\bar{\epsilon}_{k+m}^{(I)}) \right)\E^{-|{\epsilon}_{k}^{(I)}-{\epsilon}_{l}^{(I)}||\tau_1|}\bigg\},
\label{eq:G21}
\ee
where we have set~$\tau_2=0$ for convenience. Using Eq.~\eqref{eq:egs} together with the expansions~\eqref{eq:egsexp},~\eqref{eq:densexp}, and~\eqref{eq:g2exp},
we find the following expression for the second-order correction to the ground-state {energy:
\be
\partial_{\lambda}E^{(2)}_{{\rm gs},\lambda}=\frac{1}{L}\int_{x_1}\int_{x_2}\mathfrak{U}(x_1-x_2)G^{(2,1)}_{\lambda}(0,x_1,0,x_2)\,.\label{eq:egs2nd}
\ee
From} this equation,  we eventually obtain:
\be
\!\!\!\!\!\!\! E^{(2)}_{{\rm gs},\lambda=1}
&=& \int _0^{\lambda=1}\!\!\!{\rm d}\lambda^{\prime}\, \partial_{\lambda^{\prime}} E^{(2)}_{{\rm gs},\lambda^{\prime}}
\nn\\
&=&\frac{1}{L^3}\left(\frac{N}{L}\right) \sum_{m\neq0} \, \sum_{k\neq l}\,\mathfrak{U}_{m,m}
\left(\mathfrak{U}_{k-l,k-l}-\mathfrak{U}_{m,m}\right) \frac{1}{\epsilon_{k+m}^{(I)}\!-\!\epsilon_{l+m}^{(I)}\!-\!\epsilon_{k}^{(I)}\!+\!\epsilon_{l}^{(I)}} 
\theta(\bar{\epsilon}_{k+m}^{(I)})\theta(-\bar{\epsilon}_{k}^{(I)})\theta(\bar{\epsilon}_{l}^{(I)})\theta(-\bar{\epsilon}_{l+m}^{(I)}).
\label{eq:egsnlo}
\ee
We emphasize that this result holds for any particle number~$N$. 
For~$N=2$ and~$N=3$, we have checked numerically for the interaction potential~\eqref{eq:pot} that the results from Eq.~\eqref{eq:egsnlo} 
agree
identically with those from conventional second-order {\it Schr\"odinger} perturbation theory.

For~$N=2$, the absolute square of the wave function can be calculated 
from Eq.~\eqref{eq:G21}:
\be
|\Psi_{{\rm gs},\lambda}(x_1,x_2)|^2=|\Psi_{{\rm gs}}^{(0)}(x_1,x_2)|^2 + |\Psi_{{\rm gs},\lambda}^{(1)}(x_1,x_2)|^2\bar{g}_N + \dots\,,
\label{eq:psi2lo}
\ee
where~$|\Psi_{{\rm gs}}^{(0)}(x_1,x_2)|^2=|\Psi_{{\rm gs},\lambda=0}(x_1,x_2)|^2$, see Eq.~\eqref{eq:idl0}, and
\be
|\Psi_{{\rm gs},\lambda}^{(1)}(x_1,x_2)|^2 = \frac{1}{2} G^{(2,1)}_{\lambda}(0,x_1,0,x_2)\,.
\ee
Note that~$G^{(2,1)}_{\lambda}(0,x_1,0,x_2)=0$ for~$x_1=x_2$, in agreement with the {\it Pauli} principle. From Eq.~\eqref{eq:psi2lo},
the intrinsic density can be computed in leading order by plugging this expansion into Eq.~\eqref{eq:iddef}. 

For completeness, we close this subsection by noting that the spectral function~$\Omega_{\lambda}(E,x_1,x_2)$ in leading order in~$\bar{g}_N$ can also be
computed from~$G^{(2,1)}_{\lambda}$, see Sec.~\ref{sec:ddcf}, which gives us access to excited states. To be specific, we have
\be
\Omega_{\lambda}(E,x_1,x_2)=\Omega_{\lambda}^{(0)}(E,x_1,x_2) + \Omega_{\lambda}^{(1)}(E,x_1,x_2)\bar{g}_N+\dots\,,
\ee
where
\be
\Omega_{\lambda}^{(0)}(E,x_1,x_2)=2\lim_{\eta\to 0^{+}}\operatorname{Im}\tilde{G}^{(2,0)}_{\lambda}(\I E-\eta,x_1,x_2)\quad\text{and}\quad
\Omega_{\lambda}^{(1)}(E,x_1,x_2)=2\lim_{\eta\to 0^{+}}\operatorname{Im}\tilde{G}^{(2,1)}_{\lambda}(\I E-\eta,x_1,x_2)\,
\ee
with~$\tilde{G}^{(2,n)}_{\lambda}$ being the Fourier transformations of~${G}^{(2,n)}_{\lambda}$. For~$x_1=x_2=0$, for example, we {have
\be
\tilde{G}^{(2,0)}(\omega,0,0) 
 =\frac{2}{L^2}\sum_{k,l} \theta(\bar{\epsilon}_{k})\theta(-\bar{\epsilon}_{l})\frac{(\epsilon_{k}^{\,(I)}-\epsilon_{l}^{\,(I)})}{\omega^2 + (\epsilon_{k}^{\,(I)}-\epsilon_{l}^{\,(I)})^2}
\ee
and
\be
&&\tilde{G}^{(2,1)}_{\lambda}(\omega,0,0)\nn\\
&&\qquad = -2\lambda\frac{N}{L^4}\sum_{m\neq0}\,\mathfrak{U}_{m,m}\bigg\{\bigg(
\sum_{k} \theta(-\bar{\epsilon}_{k}^{(I)})\theta(\bar{\epsilon}_{k+m}^{(I)}) \frac{|\epsilon_{k+m}^{(I)}-\epsilon_{k}^{(I)}|^2 - \omega^2}{(\omega^2+(\epsilon_{k+m}^{(I)}-\epsilon_{k}^{(I)})^2)^2}
\nn\\
&&\qquad \qquad -2\sum_{k\neq l}\, 
\frac{1}{\epsilon_{k+m}^{(I)}-\epsilon_{l+m}^{(I)}+\epsilon_{l}^{(I)}-\epsilon_{k}^{(I)}}  
\theta(\bar{\epsilon}_{k+m}^{(I)})\theta(-\bar{\epsilon}_{k}^{(I)})\left(\theta(-\epsilon_{l}^{(I)})-\theta(-\epsilon_{l+m}^{(I)})\right)\frac{|\epsilon_{k+m}^{(I)}-\epsilon_{k}^{(I)}|}{\omega^2 + (\epsilon_{k+m}^{(I)}-\epsilon_{k}^{(I)})^2}
\bigg\}
\nn\\
&&\qquad\quad + 2\lambda\frac{N}{L^4} \sum_{k,l}\bigg\{ 
\sum_m \mathfrak{U}_{m,m}\theta(-\bar{\epsilon}_{k}^{(I)})\left(\theta(-\bar{\epsilon}_{l+m}^{(I)})\!-\!\theta(-\bar{\epsilon}_{k+m}^{(I)}) \right) 
\frac{|\epsilon_{k+m}^{(I)}-\epsilon_{l+m}^{(I)}|^2 - \omega^2}{(\omega^2+(\epsilon_{k+m}^{(I)}-\epsilon_{l+m}^{(I)})^2)^2}
\nn\\
&&\qquad \quad\quad +2\sum_{m\neq 0} \mathfrak{U}_{m,m}\frac{1}{\epsilon_{k+m}^{(I)}-\epsilon_{l+m}^{(I)}+\epsilon_{l}^{(I)}-\epsilon_{k}^{(I)}} 
\theta(-\bar{\epsilon}_{k})\theta(\bar{\epsilon}_{l})\left(\theta(-\bar{\epsilon}_{l+m}^{(I)})-\theta(-\bar{\epsilon}_{k+m}^{(I)}) \right)\frac{|\bar{\epsilon}_{k}^{(I)}-\bar{\epsilon}_{l}^{(I)}|}{\omega^2 + (\bar{\epsilon}_{k}^{(I)}-\bar{\epsilon}_{l}^{(I)})^2}
\bigg\}\,.
\label{eq:G21FT}
\ee
An} analytic continuation of these functions then yields the spectral function in leading order from which the excited states can be extracted.

The reconstruction of the perturbative series from our DFT-RG approach 
can be systematically continued to higher orders in the dimensionless coupling parameter~$\bar{g}_N$. For example, to obtain the third-order correction of the
energy, we need to insert the second-order correction of the density-density correlation function into the flow equation for the energy which, in turn, 
requires the computation of the density-density as well as the four-density correlation function at leading order, see Ref.~\cite{Kemler:2013yka} for a more general discussion.
The direct connection of our approach to many-body perturbation theory
is indeed a very useful feature to guide the construction of systematic approximation schemes for our DFT-RG studies. 
However, this does {\it not} imply that this approach is only perturbative. On the contrary, the solution 
for, e.g., the density-density correlation function from the flow equation~\eqref{eq:G2flow}
includes arbitrarily high orders
in the dimensionless parameter~$\bar{g}_N$.

\subsection{Hartree-Fock Approximation}\label{sec:hfa}

\begin{figure}[t]
\includegraphics[width=0.6\columnwidth]{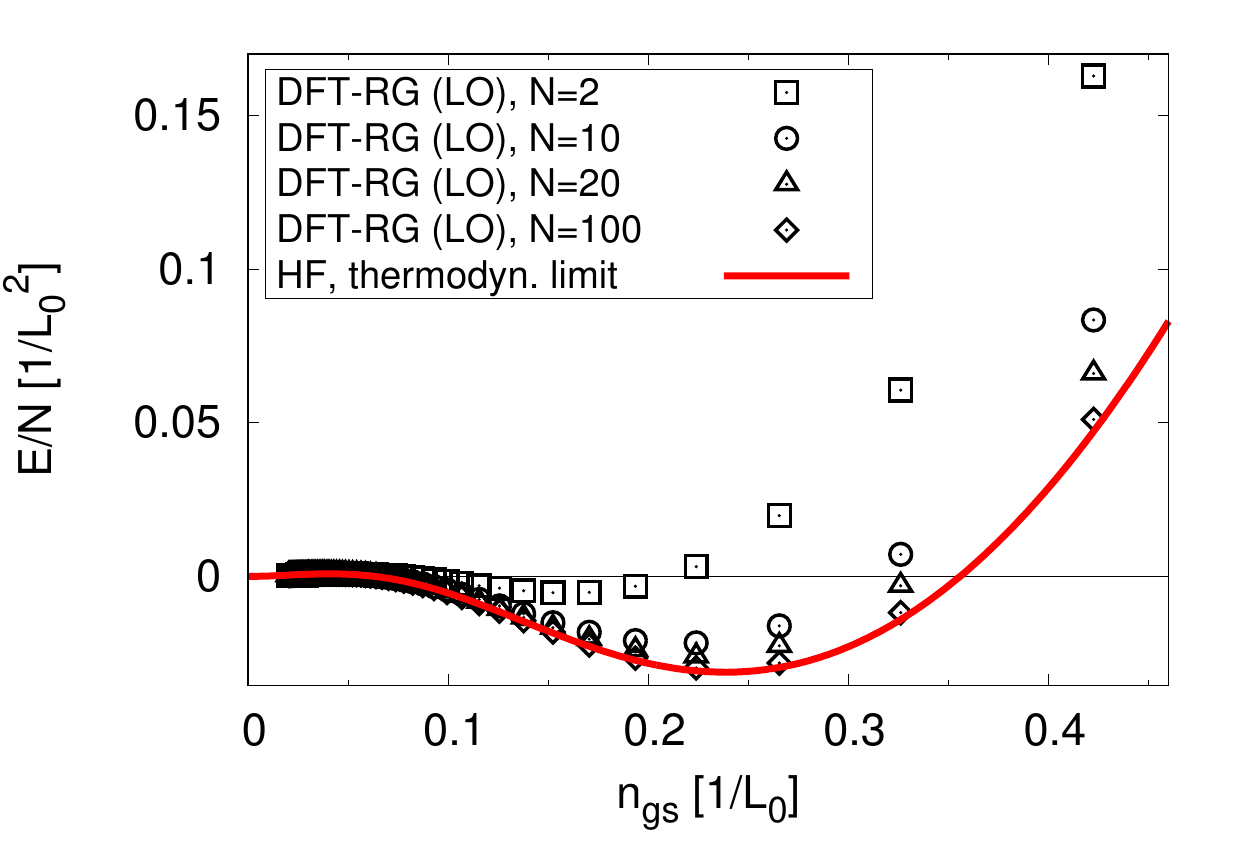}
\caption{\label{fig:hf} (color online) Energy per fermion $E/N$ for $N=2,10,20,100$ fermions as a function of $n_{\rm gs}=N/L$ 
as obtained from our DFT-RG approach at leading order (LO), {see Eq.~\eqref{eq:dfthf}.
Here, $L$ is measured} in units of $L_0$. 
For fixed~$n_{\rm gs}=N/L$, the 
leading-order DFT-RG results approach the Hartree-Fock (HF) approximation of the equation of state 
in the limit~$N\to\infty$, see Eq.~\eqref{eq:HFegs}.}
\end{figure}

We now aim at the regime where the mean interparticle distance is smaller than 
the range of the interaction. In the thermodynamic limit ($N/L=\text{const.}$ and~$L\to\infty$), the {\it Hartree-Fock} approximation is expected
to become reliable in this regime, at least for purely attractive interactions between the identical fermions~\cite{PhysRevA.63.033606,PhysRevA.77.061603,Deuretzbacher}. 
If the interaction is not only attractive but also short-range repulsive, the {\it Hartree-Fock} approximation may only yield reliable results in a certain range of densities
in which the mean interparticle distance is of the order of the range of the interaction or less but still sufficiently greater than the scale associated with the
short-range repulsive part. For the two-body potential~\eqref{eq:pot}, such a regime may be small if it exists at all. In any case, it is instructive from 
a field-theoretical point of view to study the relation of our DFT-RG approach to the {\it Hartree-Fock} approximation. To this end, we consider 
Eq.~\eqref{eq:diffeqegs1}, which {yields the leading-order correction to the ground-state energy, and use Eqs.~\eqref{eq:egs0} 
and~\eqref{eq:g20ic} to obtain:  
\be
\frac{1}{N}E_{\lambda=1}^{\text{LO}}
&=&\frac{1}{N}\sum_{n}\epsilon_n^{(I)} \theta(- \bar{\epsilon}_n^{\,(I)})
+\frac{g}{2N} \sum_{k,l} \theta(-\bar{\epsilon}_{k}^{\,(I)})\theta(-\bar{\epsilon}_{l}^{\,(I)})
\int_{x_1}\int_{x_2} (\phi_{k}^{(I)}(x_1))^{\ast}\phi_{k}^{(I)}(x_1){\mathfrak U}(x_1-x_2)(\phi_{l}^{(I)}(x_2))^{\ast} \phi_{l}^{(I)}(x_2) \nn\\
&& \quad - \frac{{g}}{2N}\sum_{k,l} \theta(-\bar{\epsilon}_{k}^{\,(I)})\theta(-\bar{\epsilon}_{l}^{\,(I)}) \int_{x_1}\int_{x_2} {\mathfrak U}(x_1-x_2)
(\phi_{k}^{(I)}(x_1))^{\ast} (\phi_{l}^{(I)}(x_2))^{\ast} \phi_{l}^{(I)}(x_1)\phi_{k}^{(I)}(x_2)
\,,\label{eq:dfthf}
\ee
where} we have used {that
\be
n_{{\rm gs},\lambda}(x)=\sum_k \theta(-\bar{\epsilon}_{k}^{\,(I)}) (\phi_{k}^{(I)}(x))^{\ast}\phi_{k}^{(I)}(x)\,. 
\ee
The} first term on the right-hand side of Eq.~\eqref{eq:dfthf} is associated with the kinetic energy of the fermions. 
The second term is the so-called {\it Hartree} term and the third
term is the so-called exchange or {\it Fock} term.
In other words, the expression~\eqref{eq:dfthf} is 
nothing but the {\it Hartree-Fock} energy as obtained from a {\it Slater} determinant defined by one-particle wave functions~$\phi_n^{(I)}$.
Thus, in our DFT-RG approach, the {\it Hartree-Fock} approximation is associated with 
the leading-order correction of the ground-state energy in an expansion in powers of the dimensionless parameter~$\bar{g}_N$.
In the following we shall indeed refer to Eq.~\eqref{eq:dfthf} as the leading-order (LO) DFT-RG approximation in accordance with
our discussion in Sec.~\ref{sec:mbpt}.\footnote{Note that Eq.~\eqref{eq:dfthf} is a {\it universal} result in
the sense of being independent of the choice of the regulator function~$\mathcal{R}_{\lambda}$ implicitly 
introduced in Eq.~\eqref{eq:action}. This follows immediately from the general properties of~$\mathcal{R}_{\lambda}$ 
specified in Eq.~\eqref{eq:regfctcond}.} 
Note that this LO approximation can also be obtained by simply setting~$\partial_{\lambda}G_{\lambda}^{(n)}=0$
for~$n\geq 1$.\footnote{This does not
imply that~$G^{(n)}_{\lambda}=0$ for~$n\geq 1$.}
In Fig.~\ref{fig:hf}, {we show $E/N\equiv E_{\lambda=1}^{\text{LO}}/N$} for~$N=2,10,20,100$ as a function of the ground-state density~$n_{\rm gs}$ as
obtained from Eq.~\eqref{eq:dfthf}. 

The {\it Hartree-Fock} approximation of the equation of state can be obtained from Eq.~\eqref{eq:dfthf} by taking the thermodynamic limit. In this limit,
the momenta~$p$ of the fermions become continuous and the associated one-particle wave-functions are given by
\be
\phi_p(x)=\frac{1}{\sqrt{L}}\,\E^{-\I px}\,,
\ee
where the volume-dependent prefactor normalizes these states and the limit~$L\to\infty$ is assumed to be taken in the end. 
The Fourier transform~$\mathfrak{U}(p)$ of the function~$\mathfrak{U}(x_1-x_2)$ is then defined as follows:
\be
\mathfrak{U}(x_1-x_2) = \sum_m \mathfrak{U}_{m,m}(\phi_{m}^{({\rm P})}(x_1))^{\ast}\phi_{m}^{({\rm P})}(x_2)
\stackrel{(L\to\infty)}{\longrightarrow}  \int_{-\infty}^{\infty}\frac{{\rm d}p}{2\pi}\, \mathfrak{U}(p) {\rm e}^{{\rm i}p(x_1-x_2)}\,,
\ee
where~$ \mathfrak{U}_{m,m}$ and~${\mathfrak{U}}(p)$ are both dimensionless. 
With these conventions at hand, we eventually obtain the {\it Hartree-Fock} energy~$E_{\rm HF}$ in the thermodynamic limit from Eq.~\eqref{eq:dfthf}. 
In agreement with the literature~\cite{Deuretzbacher}, we find
\be
\frac{1}{N}E_{\rm HF}=\frac{k_{\rm F}^2}{6}+\frac{\bar{g}_N}{2} \int_{-k_{\rm F}}^{k_{\rm F}}\frac{{\rm d}p_1}{2\pi} \int_{-k_{\rm F}}^{k_{\rm F}}\frac{{\rm d}p_2}{2\pi}
\left( {\mathfrak{U}}(0) - {\mathfrak{U}}(p_1-p_2) \right)\,,
\label{eq:HFegs}
\ee
where~$k_{\rm F}=\pi N/L$ is the {\it Fermi} momentum and $\bar{g}_{N}=gL/N=g\pi/k_{\rm F}$.
We note that the contribution 
from the {\it Hartree} term vanishes identically for the two-body potential~\eqref{eq:pot} in the thermodynamic limit, i.e.~${\mathfrak{U}}(p)\to 0$ for $p\to 0$. Moreover, 
we have~${\mathfrak{U}}(p)>0$ for any finite momentum~$p$ for this particular interaction. These are special features
of the two-body potential~\eqref{eq:pot} and, loosely speaking, they imply that there is no interaction between the fermions in the limit of vanishing momentum transfer.
For small momentum transfers, we find~${\mathfrak{U}}(p)\sim p^2$ for the two-body potential in Eq.~\eqref{eq:pot}. 
Thus, the exchange energy in Eq.~\eqref{eq:HFegs} scales as~$k_{\rm F}^3\sim (N/L)^3$ 
in the low-density limit and is therefore subleading compared to the kinetic energy, implying that~$E_{\rm HF}/N$ approaches
zero from above, see Fig.~\ref{fig:hf}.
In the high-density limit, the exchange energy is also subleading
compared to the kinetic energy but for a different reason: Here, the exchange energy becomes constant for~$k_{\rm F}\sim N/L\to\infty$.

\subsection{Variational Principle and Conventional DFT}\label{sec:var}

From our discussion of the relation of our DFT-RG approach and the {\it Hartree-Fock} approximation
it follows immediately that the ground-state energy $E$ as obtained from the leading-order DFT-RG approximation~\eqref{eq:dfthf}
is greater than or equal to the true ground-state energy. As stated above, Eq.~\eqref{eq:dfthf}
can also be obtained from an evaluation of the expectation value $\langle \hat{H}\rangle\equiv E$ with respect to
the ground-state wave function of the non-interacting $N$-body system in a box,\footnote{If not stated otherwise, the expectation value~$\langle \cdot\rangle$
is computed with respect to the ground-state wave function of the system under consideration.}
where~$\hat{H}=\hat{T}+\hat{U}$.
Here, $\hat{T}$ is the kinetic operator and the operator~$\hat{U}$ is associated {with a two-body interaction potential.}
From the {\it Rayleigh-Ritz} variational principle 
we then conclude that the estimate for the ground-state energy given in Eq.~\eqref{eq:dfthf} is bounded from below by the exact solution for the ground-state energy.

In order to discuss higher-order approximations in our DFT-RG framework,
let us consider again the flow equation for the ground-state energy 
as it follows {from Eq.~\eqref{eq:egsflow}:
\be
\partial_\lambda E_{\lambda} \equiv \partial_{\lambda}\langle \hat{H}\rangle_{\lambda} 
&=& \frac{1}{2}\int_{x_1}\int_{x_2} n_{{\rm gs},\lambda}(x_1)U(x_1-x_2) n_{{\rm gs},\lambda}(x_2)\nn\\
&& \qquad +\frac{1}{2}\int_{x_1}\int_{x_2}U(x_1-x_2)\left( G^{(2)}_{\lambda}(0,x_2,0,x_1) - n_{{\rm gs},\lambda}(x_2)\delta(x_2-x_1)\right)\,.
\ee 
Using} the relation~\eqref{eq:G2Pauli} between the density-density correlation function and the absolute square of the
ground-state wave function, we {obtain
\be
\partial_\lambda E_{\lambda} \equiv \partial_{\lambda}\langle \hat{H}\rangle_{\lambda} 
= \frac{1}{2}N(N-1)\int_{x_1}\int_{x_2} U(x_1,x_2) \int_{x_3}\cdots\int_{x_N} |\Psi_{{\rm gs},\lambda}(x_1,x_2,x_3,\dots,x_N)|^2
= \langle \hat{U}\rangle_{\lambda}
\,,\label{eq:dfthftheorem}
\ee
where} the subscript~$\lambda$ refers to the fact that the ground-state wave function in general depends on~$\lambda$.
Thus, we recover the {\it Hellmann-Feynman} theorem from the DFT-RG equation for the ground-state energy. 

An integration of Eq.~\eqref{eq:dfthftheorem}
with respect to $\lambda$ eventually yields the ground-state {energy:
\be
\langle \hat{H}\rangle 
=  \frac{\pi^2}{6}\left(\frac{N}{L}\right)^2\left(N-\frac{1}{N}\right) +
\frac{1}{2}N(N-1)\int_{x_1}\int_{x_2} U(x_1,x_2) \int_{0}^{1}{\rm d}\lambda \int_{x_3}\cdots\int_{x_N} |\Psi_{{\rm gs},\lambda}(x_1,x_2,x_3,\dots,x_N)|^2
\,,\label{eq:hfexact}
\ee
where} the first term on the right-hand side is the energy of $N$~non-interacting fermions in a box with (anti)periodic boundary conditions. 
Hence, the ground-state energy can either be calculated 
from~$|\Psi_{{\rm gs},\lambda}(x_1,\dots,x_N)|^2$ or, equivalently, from the reduced quantity~$\Phi_{N,\lambda}$ which follows
from $|\Psi_{{\rm gs},\lambda}(x_1,\dots,x_N)|^2$ by integration over the coordinates $x_i$ with $i=3,\dots,N$, see Eq.~\eqref{eq:G2Pauli}.
It is obvious that~$\langle\hat{H}\rangle$ on the left-hand side of Eq.~\eqref{eq:hfexact} should only be identified
with the exact ground-state energy of the system, i.e. the expectation value of~$\hat{H}$ with respect 
to the exact ground-state wave function, if~$\Phi_{N,\lambda}$
is exact, which implies that the infinite tower of {RG flow} equations for the correlation functions has been solved exactly.
As discussed in Sec.~\ref{sec:ddcf}, the change
of $|\Psi_{{\rm gs},\lambda}(x_1,\dots,x_N)|^2$ under a variation of $\lambda$ 
can be extracted from the flow equation of the $N$-density correlation function~$G^{(N)}_{\lambda}$. Now we recall that the computation of the 
full $\lambda$-dependence of the correlation function~$G^{(N)}_{\lambda}$ 
requires the full $\lambda$-dependence of the correlation functions~$G^{(m)}_{\lambda}$ with $1\leq m\leq N+2$.
Of course, the associated infinite tower of flow equations for the correlation functions 
can in general not be solved without relying on approximations. For example, one may consider
a truncation of this infinite tower to a finite set of flow equations, see also Sec.~\ref{sec:res} below. In any case, 
any truncation of this tower will in general yield only
an approximate solution~$\Phi_{N,\lambda}^{\rm (A, RG)}$ of~$\Phi_{N,\lambda}$.
We may now ask whether the ground-state energy obtained from such an {\it approximate} solution is always greater than or equal to the
exact ground-state energy. A necessary and sufficient condition for this to be the case is that an 
approximation~$\Psi_{{\rm gs},\lambda}^{\rm (A)}(x_1,\dots,x_N)$ of the exact ground-state wave function
{can be constructed} such that {
\be
\Phi_{N,\lambda}^{\rm (A, RG)}(x_1,x_2) = \int_{x_3}\cdots\int_{x_N} |\Psi_{{\rm gs},\lambda}^{\rm (A)}(x_1,\dots,x_N)|^2\quad\text{and}\quad
E_{\lambda}^{\rm (A, RG)} = \langle \Psi_{{\rm gs},\lambda}^{\rm (A)} | \hat{H} | \Psi_{{\rm gs},\lambda}^{\rm (A)}\rangle\,\label{eq:varcond}
\ee
for~$\lambda\in [0,1]$. Here,} $E_{\lambda}^{\rm (A, RG)}$ denotes the approximate DFT-RG result for the ground-state energy
as obtained {from~$\Phi_{N,\lambda}^{\rm (A, RG)}(x_1,x_2)$ where the} latter has been obtained from a given truncation of the 
underlying infinite tower of flow equations. Indeed, the condition~\eqref{eq:varcond} implies
\be
E_{\lambda}^{\rm (A, RG)} \geq  \langle \hat{H}\rangle \big|_{\text{exact}}
\ee
according to the {\it Rayleigh-Ritz} variational principle. In particular,  the condition~\eqref{eq:varcond} 
implies that the approximate wave function~$\Psi_{{\rm gs},\lambda}^{\rm (A)}(x_1,\dots,x_N)$
satisfies the {\it Hellmann-Feynman} theorem. 
The LO approximation within our DFT-RG framework represents a simple example for a case where such 
an approximate wave function can be {found.
As} discussed in Sec.~\ref{sec:hfa}, this approximation corresponds to the {\it Hartree-Fock} approximation. 
Hence, in this case, the approximate ground-state
wave function~$\Psi_{{\rm gs},\lambda}^{\rm (A)}$, which fulfills the condition~\eqref{eq:varcond}, is given 
by a {\it Slater} determinant built up from one-particle wave functions of the non-interacting theory. 

There also exists a simple example for 
an approximation scheme which potentially violates the condition~\eqref{eq:varcond}: Assume that an approximation of the DFT-RG flow equations
has been constructed such {that~$E_{\lambda=1}^{\rm (A, RG)}$ agrees} identically with
the second-order result from perturbation theory~(PT), $E^{\rm (PT)}=E^{(0)}+\bar{g}_N E^{(1)} + \frac{1}{2}\bar{g}_N^2 E^{(2)}$.
The perturbative calculation requires the wave function~$|\Psi_{{\rm gs}}^{\rm (PT)}\rangle$ in the leading-order approximation as input,
\be
|\Psi_{{\rm gs}}^{\rm (PT)}\rangle = |\Psi_{{\rm gs},0}\rangle+\bar{g}_N|\delta \Psi_{{\rm gs}}\rangle\,.
\ee
{Here, the index `0' refers to the non-interacting system.} Computing the expectation value of~$\hat{H}$ with respect to this wave function, we obtain
\be
\langle \Psi_{{\rm gs}}^{\rm (PT)} | \hat{H} |\Psi_{{\rm gs}}^{\rm (PT)}\rangle = \langle\Psi_{{\rm gs},0}| \hat{H} |\Psi_{{\rm gs},0}\rangle
+ 2 \bar{g}_N \langle \delta\Psi_{{\rm gs}}| \hat{H} |\Psi_{{\rm gs},0}\rangle
+ \bar{g}_N^2 \langle \delta\Psi_{{\rm gs}}| \hat{H} |\delta\Psi_{{\rm gs}}\rangle\,.
\ee
Apparently, $\langle \Psi_{{\rm gs}}^{\rm (PT)} | \hat{H} |\Psi_{{\rm gs}}^{\rm (PT)}\rangle$ includes a term of order~$\bar{g}_N^3$ and 
therefore~$\langle \Psi_{{\rm gs}}^{\rm (PT)} | \hat{H} |\Psi_{{\rm gs}}^{\rm (PT)}\rangle$ and~$E^{\rm (PT)}$ are not identical.\footnote{Note that a factor
of~$\bar{g}_N$ is included in the operator~$\hat{U}=\hat{H}-\hat{T}$.}
In particular, 
$\langle \Psi_{{\rm gs}}^{\rm (PT)} | \hat{H} |\Psi_{{\rm gs}}^{\rm (PT)}\rangle$ satisfies the {\it Rayleigh-Ritz} variational principle,  
whereas~$E^{\rm (PT)}$ may violate the associated variational bound, i.e.~$E^{\rm (PT)}$ is not necessarily greater than or equal to the exact ground-state energy.
Coming back to our DFT-RG framework, this implies that a truncation, which 
has been constructed such that~$E_{\lambda=1}^{\rm (A, RG)}$ agrees identically with the perturbative result at, e.g., second order, 
does not necessarily satisfy the {\it Rayleigh-Ritz} variational principle and therefore it is in general also not possible to construct 
an approximate wave function~$\Psi_{{\rm gs},\lambda}^{\rm (A)}$ in this case which 
satisfies the condition~\eqref{eq:varcond}. 

From these considerations we conclude that the ground-state energy obtained from a
 given truncation of our set of RG flow equations for the correlation functions 
 is not necessarily greater than or equal to the exact ground-state energy. In general, truncations 
 correspond to approximations of the energy density functional {(or, equivalently, the
 exchange-correlation functional in the terminology of conventional DFT)}. With respect to conventional
 ans\"atze for the energy density functional,
 however, note that any truncation of the vertex expansion~\eqref{eq:vertexexp}, in which 
 {we at least take into account the three-density correlation function, generates terms of 
 arbitrarily high orders in~$(\rho-\rho_{\rm gs})$
 when we solve the associated set of RG flow equations.
 This can be readily seen from the flow equation~\eqref{eq:gfloweq} for the density functional~$\Gamma[\rho]\sim E[\rho]$ which can be rewritten in the following {way:
 \be
\Gamma[\rho]&=& \Gamma_0[\rho] + \frac{1}{2}\int_{\chi_1}\int_{\chi_2} \rho (\chi_1)U_{2b}(\chi_1,\chi_2) \rho(\chi_2) \nn\\
&& \qquad\qquad 
+ \frac{1}{2}\int_0^{1}{\rm d}\lambda
\int_{\chi_1}\int_{\chi_2}U_{2b}(\chi_1,\chi_2)\left( \left(\Gamma^{(2)}_{\lambda}[\rho]\right)^{-1}\!\!(\chi_2,\chi_1) - \rho(\chi_2)\delta(\chi_2-\chi_1)\right)\,.
\label{eq:efunc}
 \ee
Here, the} first term on the right-hand side corresponds to the kinetic energy functional associated with the non-interacting initial system, e.g. fermions in a box
with (anti)periodic boundary conditions in our case. Since the inverse of the second functional derivative of~$\Gamma_{\lambda}[\rho]\sim E_{\lambda}[\rho]$
appears on the right-hand side of Eq.~\eqref{eq:efunc}, it indeed follows that terms of arbitrarily high orders in~$(\rho-\rho_{\rm gs})$ are generated
in the RG flow, even if a finite truncation of~$\Gamma_{\lambda}$ has been used initially.} We emphasize again that
the vertex expansion~\eqref{eq:vertexexp} should not be confused
with the local density approximation. In fact, the correlation functions associated with {the vertex expansion depend} on time-like and spatial coordinates and
can be recast into a derivative/gradient expansion, see our discussion of Eq.~\eqref{eq:gradexp}. 

{In conventional {\it Hohenberg-Kohn} (HK) DFT, the energy density functional is parametrized in terms of the time-{\it independent}
density~$n(x)$. For example, $E_{\rm HK}$ may be written as follows:\footnote{The first term on the right-hand side 
of Eq.~\eqref{eq:EKS} denotes the kinetic energy functional, $E_{\rm xc}[n]$
is the exchange-correlation functional and~$V$ is the external (fermion-confining) potential.} {
\be
E_{\rm HK}[n]= T[n] + \int_x V(x)n(x) + \frac{1}{2}\int_{x_1}\int_{x_2} n (x_1)U(x_1-x_2) n(x_2) + E_{\rm xc}[n]\,.\label{eq:EKS}
\ee
In} this work, on the other hand,
we consider a functional of the time-dependent density~$\rho(\tau,x)$.
Although the ground-state may be described by
a time-independent density~$n_{\rm gs}(x)$ in both cases, the two functionals are in general {\it not} identical. For example, the correlation functions
associated with an expansion of~$E_{\rm HK}[n]$ about the ground-state density do not depend on the imaginary time by construction, 
whereas the correlation functions associated with an expansion of~$\Gamma_{\lambda}[\rho]$ about the ground-state density are time-dependent and allow to
extract, e.g., energies of excited states, see also our discussion of Eq.~\eqref{eq:projrule} above.
In this sense, the functional~$\Gamma_{\lambda}[\rho]$ contains more information than 
the {\it Hohenberg-Kohn} energy density functional and should therefore be considered as a
generalization of it. From a comparison of Eq.~\eqref{eq:EKS} evaluated at the ground-state density~$n_{\rm gs}$
with Eq.~\eqref{eq:hfexact}, we deduce that, loosely speaking,
a particular ansatz for~$E_{\rm xc}[n_{\rm gs}]$ is associated with an approximation for~$\langle \hat{U}\rangle$ which involves 
the absolute square of the ground-state wave function~$|\Psi_{{\rm gs},\lambda}|^2$. In the limit of many fermions, an LDA-type
ansatz for~$E_{\rm xc}$ is often found to yield reasonable results for ground-state properties, even on a quantitative level. 
This may be traced back to the fact that the LDA is directly obtained from
the equation of state of the associated many-body problem (in the thermodynamic limit). For few-body systems, on the other hand, an LDA ansatz
for~$E_{\rm xc}$ is in general not sufficient for an accurate computation of ground-state properties and more involved
ans\"atze for~$E_{\rm xc}$ are required. Within our DFT-RG approach, the problem of finding an appropriate ansatz for~$E_{\rm xc}$ is replaced
by the computation of the density correlation functions.}

We close this discussion with a comment on the relation of our approach to the {\it Hellmann-Feynman} theorem. First of all, we note that the {\it Hellmann-Feynman} theorem
does not provide a recipe for the computation of the $\lambda$-dependent wave function. On the other hand, 
our RG approach provides us with a recipe for the computation of the $\lambda$-dependence of the absolute square of the 
ground-state wave function via its relation to the density correlation functions. In this sense, our DFT-RG approach may be viewed as an extension of the {\it Hellmann-Feynman} theorem
which allows for a systematic computation of the $\lambda$-dependent absolute square of the ground-state wave function
entering the computation of the ground-state energy.

In conventional DFT studies (see, e.g., Ref.~\cite{EngelDreizler} for an introduction), 
the {\it Hellmann-Feynman} theorem serves as a starting point for the derivation of so-called coupling-constant integration methods which in principle allow to 
compute a representation of the exchange-correlation functional in terms of {\it Kohn-Sham} orbitals and eigenvalues. Our DFT-RG approach does not rely
on {the computation of the latter. It} rather relies on a hierarchy of correlation functions computed from a non-interacting but confined system 
{which defines} the starting point of the associated RG flow. 

\subsection{Fermion Self-Interactions}

In conventional DFT, an infamous problem in the construction of density functionals is the potential appearance of spurious fermion self-interactions which
need to be removed by, e.g., ``counter terms" in order to remove the associated self-interaction energy spoiling the predictions for ground-state energies.
For example, the {\it Hartree} term yields a finite contribution to the energy even for a system containing only a single fermion.

Our DFT-RG approach relies on expanding the density functional about the ground state rather than making a global ansatz of this functional. As we have
shown above, this expansion about the ground state can even be related to many-body perturbation theory. 
{We shall now see} that this
also allows us to keep the problem of spurious self-interactions systematically under control. Indeed, we already deduce from Eq.~\eqref{eq:dfthftheorem} that
$\partial_{\lambda}E_{\lambda}=0$ for~$N=1$. Thus, no fermion self-interactions are present if we solve the infinite tower of RG flow equations exactly. To be more 
precise, the derivation of Eq.~\eqref{eq:dfthftheorem} relies on Eq.~\eqref{eq:G2Pauli} which, for~$N=1$, reduces to the following ``sum rule":
\be
G^{(2)}_{\lambda}(0,x_1,0,x_2)-n_{{\rm gs},\lambda}(x_1)\delta(x_1-x_2) + n_{{\rm gs},\lambda}(x_2)n_{{\rm gs},\lambda}(x_2)=0\,.
\label{eq:fsisr}
\ee
In practice, a solution of the flow equations for the ground-state density and the density-density correlation function requires a truncation of the infinite coupled
set of flow equations associated with the density correlation functions. Any given truncation of this set may result in a violation of the 
sum rule~\eqref{eq:fsisr}. In our explicit numerical studies below, we estimate
the strength of the violation of this sum rule by computing the ground-state energy of the one-fermion system which may be considered as a quality measure of the given
truncation. 

We emphasize that the density-density correlation function of the non-interacting system satisfies 
this sum rule, see Eq.~\eqref{eq:g20ic}. As a consequence, the predictions 
from the leading-order approximation within our DFT-RG framework do not suffer from the fermion self-interaction problem.
Considering the perturbative second-order energy correction~$E_{\lambda=1}^{(2)}$ (see Eq.~\eqref{eq:egsnlo}), there is also no contribution
to the ground-state energy in case of a single-fermion system as it should be, i.e.~$E_{\lambda=1}^{(2)}=0$ for~$N=1$.
Indeed, it is not surprising that these spurious self-interaction contributions are not present in our DFT-RG approach if we consider the solution
of the flow equations order by order in the parameter~$\bar{g}_N$. As discussed above, this is equivalent to conventional
many-body perturbation theory where we do not encounter contributions from spurious fermion self-interactions to the ground-state energy. 
{This observation helps to guide the construction of truncations within our present framework:
If a truncation is constructed in this spirit such that it reproduces many-body perturbation theory exactly up
to a given order, then the contributions from fermion self-interactions generated in the full RG flow on the level of this trunction 
also approach zero 
in the perturbatively accessible weak-coupling limit. Moreover, the truncation can be systematically improved
by including higher-order corrections of the perturbative series. In contrast to such a construction of truncations of the
energy density functional, the size of contributions from fermion self-interactions is {\it a priori} completely uncontrolled
when we simply ``guess" a truncation of the energy density functional and the systematic cancellation of these spurious contributions 
may be~difficult.}

\section{Results}\label{sec:res}

In this section we now employ our DFT-RG approach to compute ground-state properties of systems of $N$ identical fermions in a box
interacting via the two-body potential given in Eq.~\eqref{eq:pot}.

\subsection{DFT-RG Flow Equations}\label{sec:restrunc}

Before we discuss our numerical results, we specify the flow equations underlying our studies. To this end, we make use of the findings
from our discussion of the relation between the DFT-RG approach and many-body perturbation theory.
Using Eqs.~\eqref{eq:egsflow} and~\eqref{eq:egslo}, we find the following differential equation for the ground-state energy:
\be
\frac{1}{N}\partial_{\lambda}E_{\lambda}= \frac{\bar{g}_N}{2}\mathfrak{U}_{0,0}\left(\frac{N}{L}\right)^2
- \frac{\bar{g}_N}{2L^2}\sum_{k,l} 
\theta(-\bar{\epsilon}_{k}^{(I)})\theta(-\bar{\epsilon}_{l}^{(I)})\,\mathfrak{U}_{k-l,k-l}
+\frac{\bar{g}_N}{2L}\int_{-\infty}^{\infty}\frac{{\rm d}\omega}{2\pi}\sum_a \mathfrak{U}_{a,a}\Delta \tilde{G}^{(2)}_{\lambda,a,a}(\omega)\,,
\label{eq:egsF}
\ee
where we have tacitly introduced the Fourier transform of the density-density correlation function~$G^{(2)}_{\lambda}$,
\be
G^{(2)}_{\lambda}(\tau_1,x_1,\tau_2,x_2)=\int_{-\infty}^{\infty}\frac{{\rm d}\omega}{2\pi}\sum_{a,b}\, 
\tilde{G}^{(2)}_{\lambda,a,b}(\omega) (\phi_{a}^{({\rm P})}(x_1))^{\ast}\phi_{b}^{({\rm P})}(x_2){\rm e}^{-{\rm i}\omega(\tau_1-\tau_2)}\,.
\ee
{In Eq.~\eqref{eq:egsF}, we have decomposed this function} into the density-density correlation function of the non-interacting system ($\lambda=0$)
and its modification~$\Delta G^{(2)}_{\lambda}$ in the presence of interactions between the fermions: 
\be
\tilde{G}^{(2)}_{\lambda,a,b}(\omega)  = \tilde{G}^{(2)}_{\lambda=0,a,b}(\omega) + \Delta \tilde{G}^{(2)}_{\lambda,a,b}(\omega)\,.
\label{eq:G2split}
\ee
{This decomposition is convenient from} a numerical point of view as it allows
us to treat those terms on the right-hand side of the flow equation for~$\tilde{G}^{(2)}_{\lambda,a,b}$ analytically which contain contributions of the form~$\sim \delta(x_i-x_j)$,
see Eq.~\eqref{eq:G2eqFC} below.
In any case, the initial condition for the energy is given by the energy of the non-interacting $N$-fermion system, see Eq.~\eqref{eq:egs0}. 

The right-hand side of the flow equation for the ground-state density~$n_{\rm gs}$ vanishes identically as discussed above. 
However, we stress again that~$n_{\rm gs}$
should not be confused with the intrinsic density of the system which is a non-trivial function 
and can be extracted from the density-density correlation function, see Eq.~\eqref{eq:G2Pauli},
at least for the two-body system. In general, the $N$-density correlation function~$G^{(N)}_{\lambda}$ is required to compute the  
intrinsic density of the $N$-fermion system.

For the density-density correlation function, the exact flow equation reads:
\be
\partial_{\lambda}G^{(2)}_{\lambda}(\tau_1,x_1,\tau_2,x_2)
&=& - \int_{\tau_3}\int_{x_3}\int_{x_4} G^{(2)}_{\lambda}(\tau_1,x_1,\tau_3,x_3)U(x_3-x_4)G^{(2)}_{\lambda}(\tau_3,x_4,\tau_2,x_2)\nn\\
&& \quad -\frac{1}{2} \int_{\tau_3}\int_{x_3}\int_{x_4} U(x_3-x_4) G^{(4)}_{\lambda}(\tau_3,x_4,\tau_3,x_3,\tau_1,x_1,\tau_2,x_2)\,.
\label{eq:deG2}
\ee
In our numerical studies, we shall neglect the flow of all correlation functions~$G^{(n)}_{\lambda}$ with $n>2$. Note that 
this does not imply that, e.g., we drop~$G^{(4)}_{\lambda}$ completely in the flow equation~\eqref{eq:deG2}. It only
implies that we replace~$G^{(4)}_{\lambda}$ by~$G^{(4)}_{\lambda=0}$. Recall that the $n$-density correlation functions of a non-interacting system are in general
also finite. In Sec.~\ref{sec:mbpt}, we have discussed that the results for the ground-state energy
obtained from the flow equations~\eqref{eq:egsF} and~\eqref{eq:deG2} agree identically with those from second-order perturbation theory
when we also replace~$G^{(2)}_{\lambda}$ by $G^{(2)}_{\lambda=0}$ in Eq.~\eqref{eq:deG2}.
If we go beyond second-order perturbation theory
by solving the differential equation~\eqref{eq:deG2} without replacing ~$G^{(2)}_{\lambda}$ by $G^{(2)}_{\lambda=0}$, 
we indeed take into account arbitrarily high orders in~$\bar{g}_N$. However, we also find 
that the quantity~$\Phi_{N,\lambda}(x_1,x_2)$ defined in Eq.~\eqref{eq:G2Pauli} is finite {for~$x_1\to x_2$.
Thus, this} truncation of our infinite tower of flow equations {is not consistent with the} {\it Pauli} exclusion principle. 

In order to obtain a flow equation for the density-density correlation function which manifestly respects 
the {\it Pauli} exclusion principle, we either have to include the complete infinite
set of flow equations or improve our truncation such that the {\it Pauli} exclusion principle is respected for any value of~$\lambda$. 
To find such an improved truncation, we first note that
the condition~$\Phi_{N,\lambda}(x,x)=0$ for $\lambda \in [0,1]$ (associated with the {\it Pauli} exclusion principle)
is equivalent to the following condition:
\be
\Delta G^{(2)}_{\lambda} (0,x,0,x) = 0\,.\label{eq:G2Pexclcond}
\ee
This follows from the decomposition of~$G^{(2)}_{\lambda}$ into~$G^{(2)}_{\lambda=0}$ 
and~$\Delta G^{(2)}_{\lambda}$ together with Eqs.~\eqref{eq:G2Pauli} and~\eqref{eq:g20ic}. For~$\lambda=0$, this condition is trivially satisfied. For any finite~$\lambda$,
a sufficient criterion to satisfy this condition is obtained by requiring that the flow of $\Delta G^{(2)}_{\lambda}$ vanishes identically for~$\tau_1=\tau_2=0$ and $x_1=x_2=x$:
\be
\partial_{\lambda}\Delta G^{(2)}_{\lambda} (0,x,0,x) =0\,. \label{eq:dG2cond}
\ee
If we now insist on the above truncation scheme, i.e. we neglect the flow of all correlation functions~$G^{(n)}_{\lambda}$ with $n>2$
and replace those functions by their initial conditions~$G^{(n)}_{\lambda=0}$, a simple but non-trivial prescription to satisfy 
the condition~\eqref{eq:dG2cond} is given by
replacing~$G^{(4)}_{\lambda=0}$ in the flow equation~\eqref{eq:deG2} as follows: 
\be
G^{(4)}_{\lambda=0}(\chi_1,\chi_2,\chi_3,\chi_4) \to f_{\mathcal{P}}(\lambda)G^{(4)}_{\lambda=0}(\chi_1,\chi_2,\chi_3,\chi_4)\,.
\ee
Here, the ``{\it Pauli}-blocking function" $f_{\mathcal P}$ does not depend on the spatial and the time-like coordinates but only on the flow parameter~$\lambda$.
For~$\tau_1=\tau_2=0$ and $x_1=x_2=0$,\footnote{Without loss of generality, we have set~$x_1=x_2=0$ {since the density-density correlation function does only} depend
on $|x_1-x_2|$ in our present study. For other confining geometries and interaction potentials, this prescription yields a function~$f_{\mathcal P}$ which depends
on the spatial coordinates.} the flow equation for the density-density correlation function can then be solved for the function~$f_{\mathcal P}$:
\be
&&f_{\mathcal P}(\lambda)=\\
&&\;\; -2\left(\int_{\tau_3}\int_{x_3}\int_{x_4} G^{(2)}_{\lambda}(0,0,\tau_3,x_3)U(x_3\!-\!x_4)G^{(2)}_{\lambda}(\tau_3,x_4,0,0)\right)
\left( \int_{\tau_3}\int_{x_3}\int_{x_4} U(x_3\!-\!x_4) G^{(4)}_{\lambda=0}(\tau_3,x_4,\tau_3,x_3,0,0,0,0)
\right)^{-1}\!\!\!\,.
\nn
\ee
For~$\lambda \to 0$, we have~$f_{\mathcal P}(\lambda)\to 1$. Within the present truncation, this follows directly from the fact that the right-hand side
of Eq.~\eqref{eq:deG2} approaches the leading order result for~$G^{(2)}_{\lambda}$
in the perturbative expansion, see also our discussion in Sec.~\ref{sec:mbpt}. Taking
the function~$f_{\mathcal P}$ into account, we obtain the following 
flow equations for the Fourier coefficients~$\Delta \tilde{G}^{(2)}_{\lambda,a,a}(\omega)$: {
\be
\partial_{\lambda}\Delta \tilde{G}^{(2)}_{\lambda,a,a}(\omega)
= -  g\Delta \tilde{G}^{(2)}_{\lambda,a,a}(\omega)\mathfrak{U}_{a,a}\Delta \tilde{G}^{(2)}_{\lambda,a,a}(\omega)
-2g\tilde{G}_{0,a,a}^{(2)}\mathfrak{U}_{a,a}\Delta \tilde{G}^{(2)}_{\lambda,a,a}(\omega)
-\mathcal{G}_{a,a}^{(2,1)}(\omega) - f_{\mathcal P}(\lambda)\mathcal{G}_{a,a}^{(4,1)}(\omega),
\label{eq:G2eqFC}
\ee
where}
\be
\tilde{G}_{0,a,a}^{(2)}(\omega)=\frac{2}{L}\sum_{k} \theta(\bar{\epsilon}_k^{(I)})\theta(-\bar{\epsilon}_{k-a}^{(I)})\frac{|\epsilon_k^{(I)} - \epsilon_{k-a}^{(I)}|}{\omega^2 + |\epsilon_k^{(I)} - \epsilon_{k-a}^{(I)}|^2}
\ee
is the $a$-th Fourier coefficient of the density-density correlation {function of the non-interacting system. The quantities~$\mathcal{G}_{a,a}^{(2,1)}$
and~$\mathcal{G}_{a,a}^{(4,1)}$ are defined as} {
\be
\mathcal{G}_{a,a}^{(2,1)}(\omega)&=& \frac{2}{L^2}\, g\mathfrak{U}_{a,a}\bigg\{\bigg(
\sum_{k} \theta(-\bar{\epsilon}_{k}^{(I)})\theta(\bar{\epsilon}_{k+a}^{(I)}) \frac{|\epsilon_{k+a}^{(I)}-\epsilon_{k}^{(I)}|^2 - \omega^2}{(\omega^2 + |\epsilon_{k+a}^{(I)}-\epsilon_{k}^{(I)}|^2 )^2}
\nn\\
&&\quad  -2\sum_{k\neq l}\, 
\frac{1}{\epsilon_{k+a}^{(I)}-\epsilon_{l+a}^{(I)}+\epsilon_{l}^{(I)}-\epsilon_{k}^{(I)}}  
\theta(\bar{\epsilon}_{k+a}^{(I)})\theta(-\bar{\epsilon}_{k}^{(I)})\left(\theta(-\epsilon_{l}^{(I)})-\theta(-\epsilon_{l+a}^{(I)})\right)
\frac{|\epsilon_{k+a}^{(I)}-\epsilon_{k}^{(I)}|}{\omega^2 + |\epsilon_{k+a}^{(I)}-\epsilon_{k}^{(I)}|^2}
 \bigg\}\,,
\\
\mathcal{G}_{a,a}^{(4,1)}(\omega) &=& -\frac{2}{L^2} \sum_{k}\sum_{m\neq 0}  g\mathfrak{U}_{m,m} \bigg\{ 
\theta(-\bar{\epsilon}_{k}^{(I)})\left(\theta(-\bar{\epsilon}_{k-a+m}^{(I)})\!-\!\theta(-\bar{\epsilon}_{k+m}^{(I)}) \right) 
\frac{|\epsilon_{k+m}^{(I)}-\epsilon_{k-a+m}^{(I)}|^2 - \omega^2}{(\omega^2 + |\epsilon_{k+m}^{(I)}-\epsilon_{k-a+m}^{(I)}|^2)^2}
\nn\\
&&\;\; +\frac{2}{\epsilon_{k+m}^{(I)}\!-\!\epsilon_{k-a+m}^{(I)}\!+\!\epsilon_{k-a}^{(I)}\!-\!\epsilon_{k}^{(I)}} 
\theta(-\epsilon_{k}^{(I)})\theta(\epsilon_{k-a}^{(I)})\left(\theta(-\bar{\epsilon}_{k-a+m}^{(I)})\!-\!\theta(-\bar{\epsilon}_{k+m}^{(I)}) \right)
\frac{|{\epsilon}_{k}^{(I)}-{\epsilon}_{k-a}^{(I)}|}{\omega^2 +|{\epsilon}_{k}^{(I)}-{\epsilon}_{k-a}^{(I)}|^2}
\bigg\}.
\ee
The} initial condition for the differential equations~\eqref{eq:G2eqFC} are~$\Delta \tilde{G}^{(2)}_{\lambda,a,a}(\omega)=0$ for $a\in\mathbb{Z}$ and~$\omega\in \mathbb{R}$.
Note that the function~$f_{\mathcal P}$ depends implicitly on all coefficients~$\Delta \tilde{G}^{(2)}_{\lambda,a,a}(\omega)$. Therefore, the flow equations 
for these quantities are coupled whereas
they are decoupled if we set~$f_{\mathcal P}(\lambda)= 1$ for~\mbox{$\lambda\in [0,1]$}. 
In the following we refer to the set of equations~\eqref{eq:egsF} and~\eqref{eq:G2eqFC} as the next-to-leading order (NLO)
approximation within our DFT-RG framework, as it appears to be the most 
natural extension of the leading-order approximation which we have introduced and
discussed in Sec.~\ref{sec:hfa}.

In order to solve the set of flow equations~\eqref{eq:egsF} and~\eqref{eq:G2eqFC} numerically, we have introduced a cutoff~$\Lambda_{\rm F}$ for the Fourier modes and checked the convergence of our results
as a function of the cutoff. To be more specific, we have used values for~$\Lambda_{\rm F}$ up to $\Lambda_{\rm F}= 400$ for~$N=10$ in the volume range considered
in this work and found that, e.g.,
the results for the ground-state energy for~$N=10$ for~$\Lambda_{\rm F}= 300$ and~$\Lambda_{\rm F}= 400$ 
only deviate on the sub per mille level, provided the cutoff associated with the integrations over $\omega$
on the right-hand side of the flow equations has been chosen sufficiently large.
To perform these integrations over~$\omega$ numerically, we have first projected the interval~$(-\infty,\infty)$ onto the compact interval~$[-1,1]$:
\be
\bar{\omega}=\frac{2}{\pi}\arctan\left(s_{\omega} \omega L^2\right)\,.\label{eq:ChebyhsevDF}
\ee
Here, $s_{\omega}$ is a dimensionless scaling factor at our disposal. Standard {\it Chebyshev-Gauss} quadrature can then be employed to 
perform the integrations over~$\omega$,
where~$s_{\omega}$ can be used to improve the convergence. 
We have checked that our results are converged as a function of the maximal number~$\Lambda_{\rm C}$ of {\it Chebyshev} nodes~$\bar{\omega}_i$,
\be
\bar{\omega}_i=\cos\left(\frac{2i-1}{2\Lambda_{\rm C}}\pi\right).
\ee
To be more specific, in our numerical studies discussed below, we have used values for~$\Lambda_{\rm C}$ up to $\Lambda_{\rm C}= 280$ for~$N=10$.
{Note that the ground-state energy appears to converge faster as a function of the cutoff~$\Lambda_{\rm C}$ in our numerical studies. The high values
of~$\Lambda_{\rm C}$ are required to fulfill the condition~\eqref{eq:G2Pexclcond} (i.e. the {\it Pauli} exclusion principle) in the RG flow. In 
fact, the latter requires that also the tail of the density-density correlation function in the large $\omega$-limit~($\sim 1/\omega^2$) is resolved
accurately. In future studies, an improvement of the convergence of our results with respect to~$\Lambda_{\rm C}$ may be achieved
by improving the distribution of the {\it Chebyshev} nodes which is partially controlled by the function~\eqref{eq:ChebyhsevDF} in our present
studies.}

\subsection{Ground-State Properties of Bound States}

\begin{figure}[t]
\includegraphics[width=0.6\columnwidth]{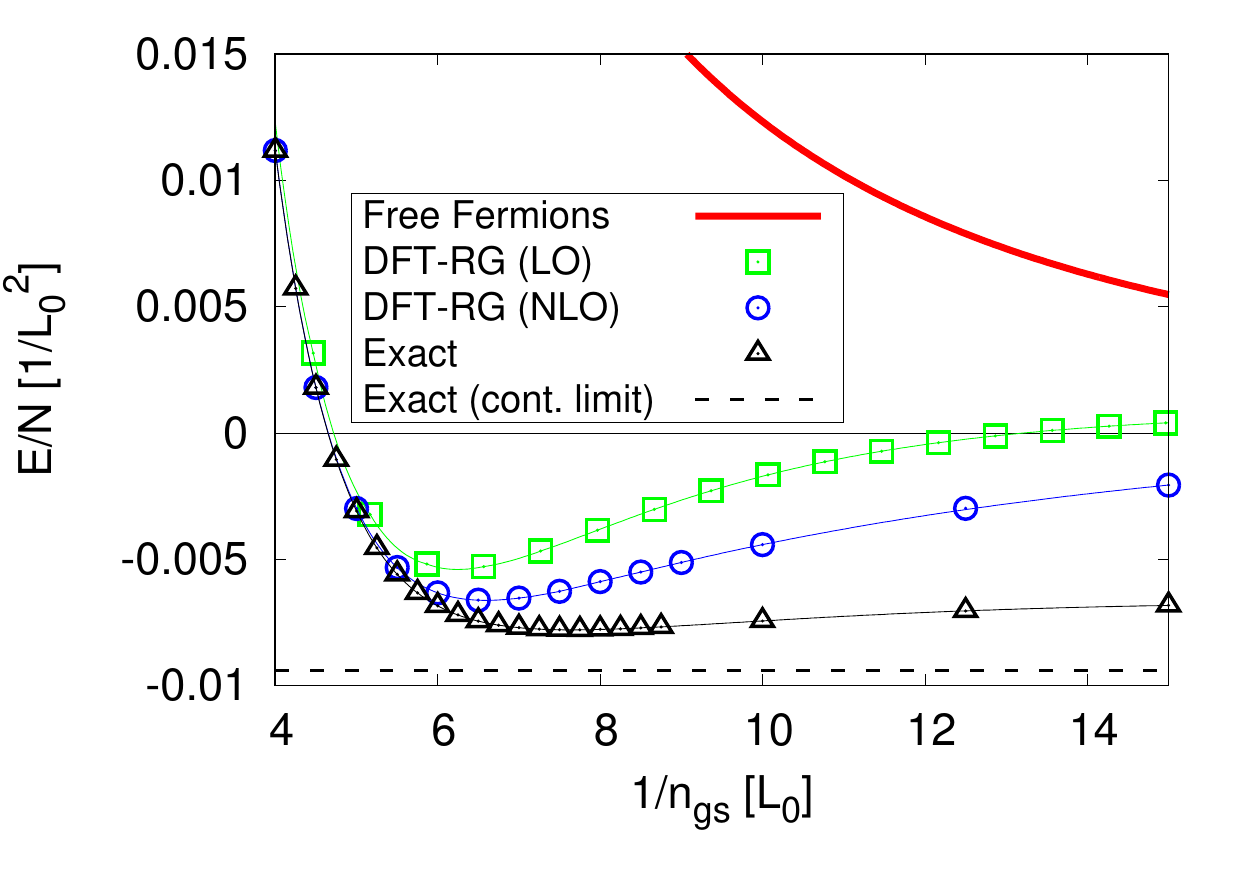}
\caption{\label{fig:2b} (color online) Ground-state energy of the two-body problem as a function of the inverse ground-state density 
as obtained from different studies. Note that the ground-state density should not be confused with
the intrinsic density of the system, see main text for details.}
\end{figure}

\begin{figure}[t]
\includegraphics[width=0.45\columnwidth]{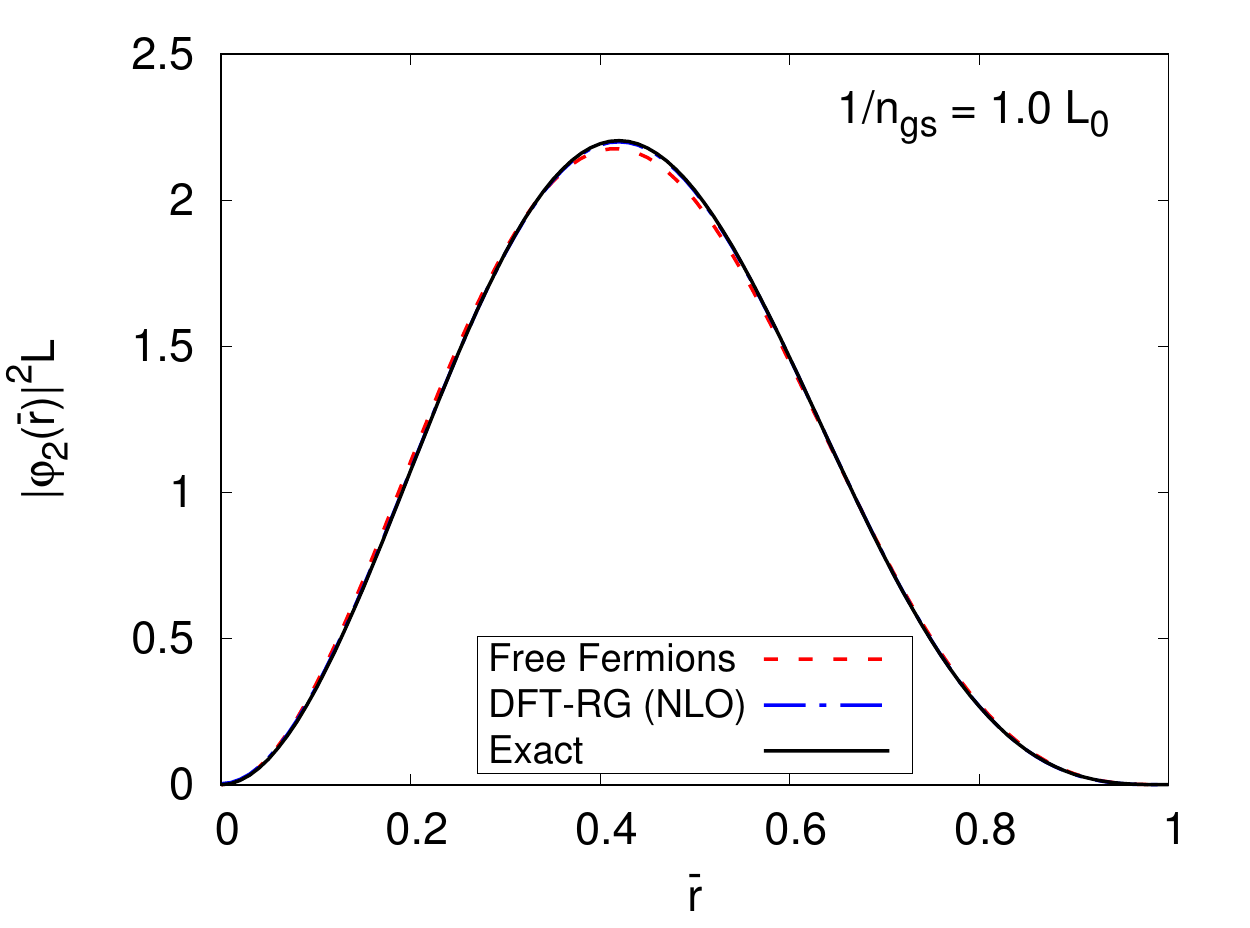}
\includegraphics[width=0.45\columnwidth]{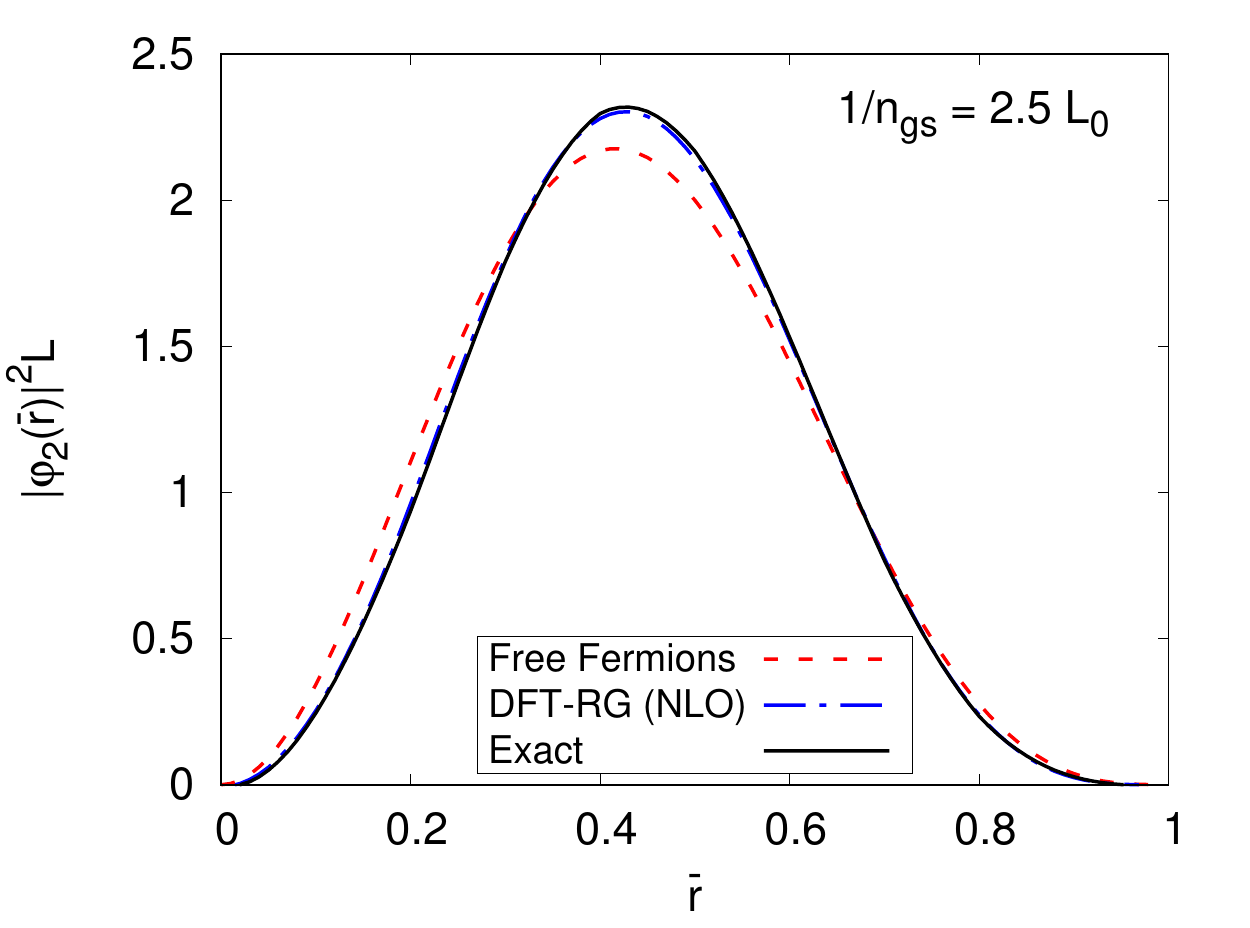}
\includegraphics[width=0.45\columnwidth]{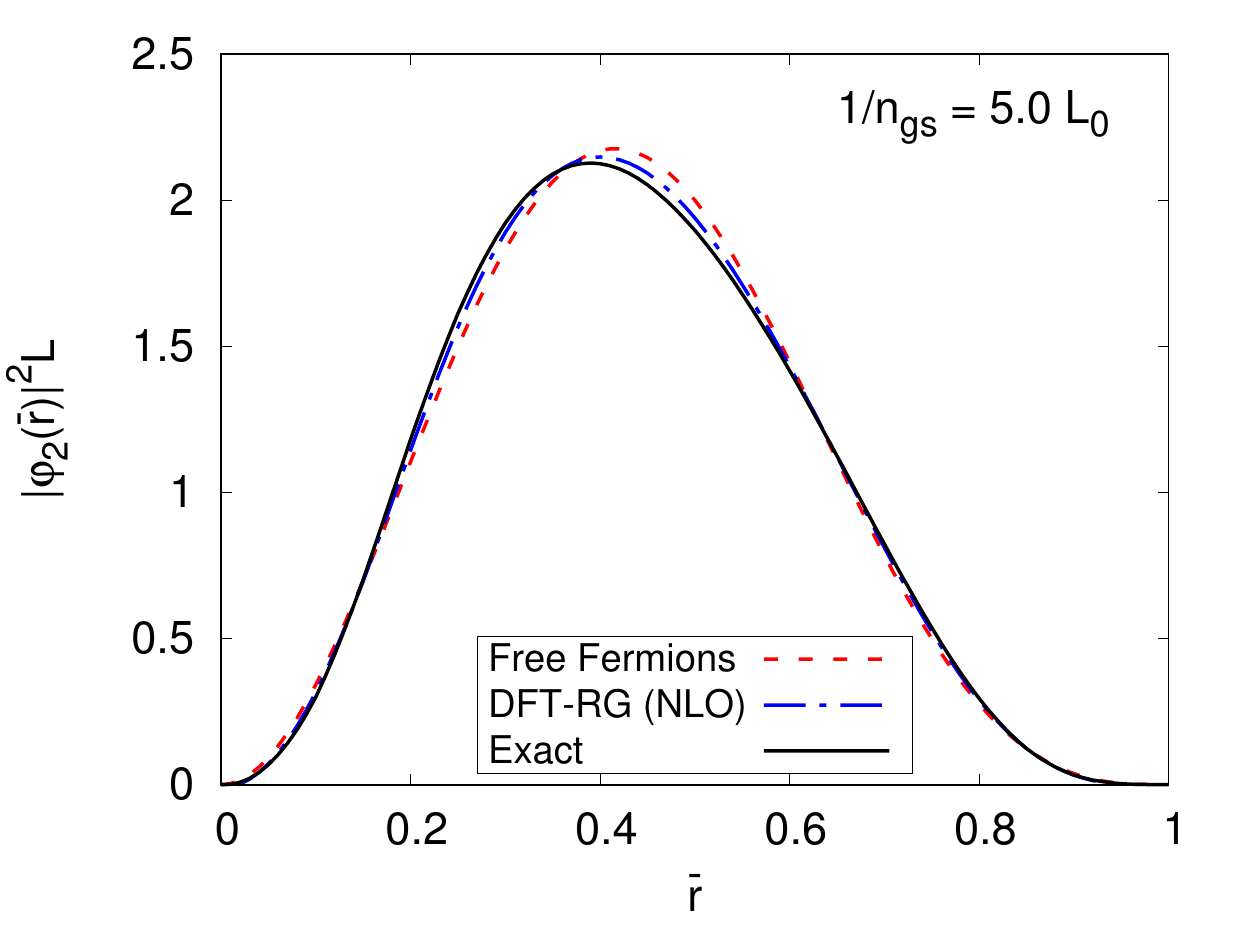}
\includegraphics[width=0.45\columnwidth]{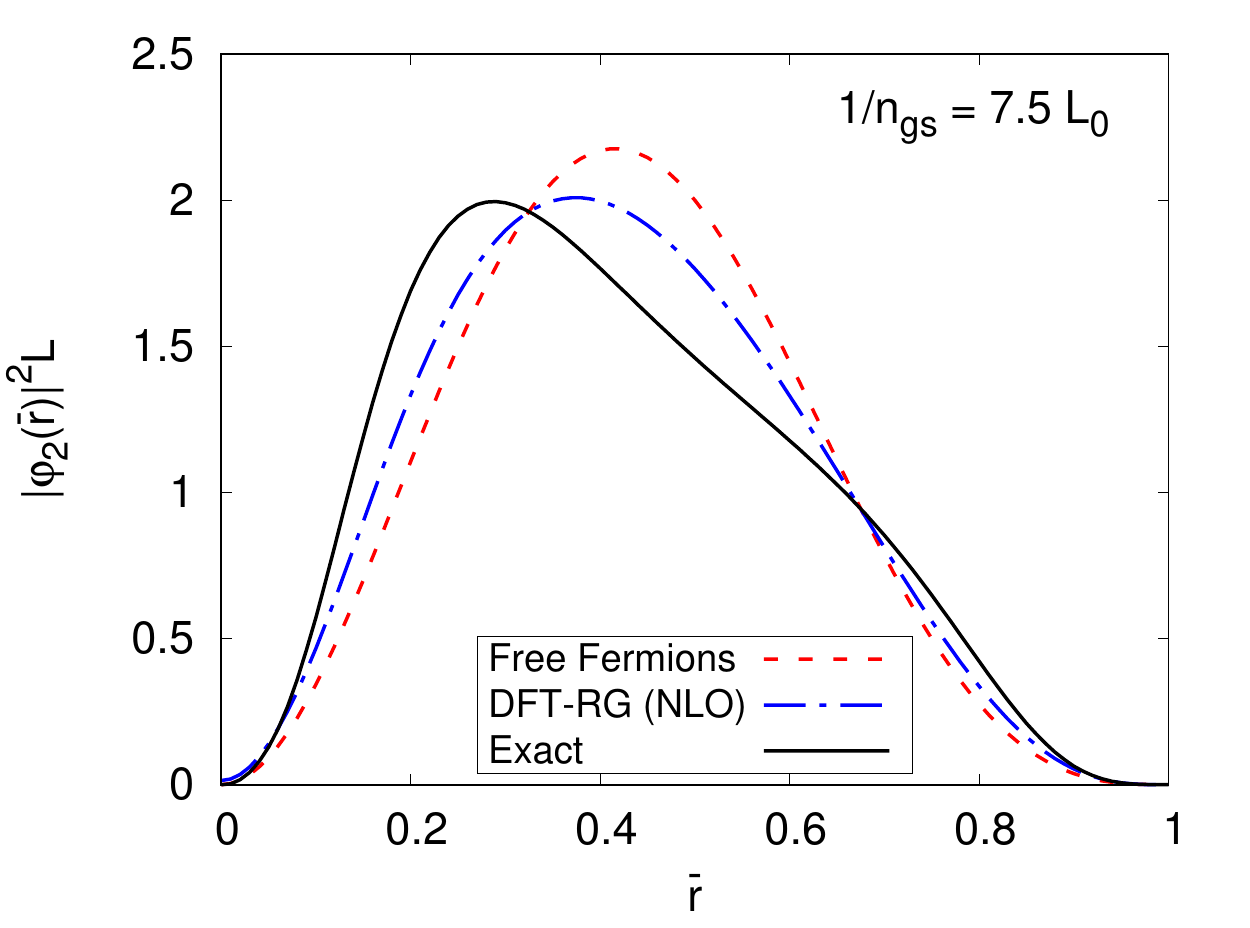}
\caption{\label{fig:2bwf} (color online) Dimensionless absolute square of the ground-state wave function  $|\varphi_{2}(\bar{r})|^2 L$  in the center-of-mass frame
as a function of $\bar{r}=r/L$ as obtained from our DFT-RG approach at NLO in comparison with the exact solution and the result for the non-interacting two-body system. 
}
\end{figure}

Let us finally discuss the results from the numerical solution of the DFT-RG flow equations.
In Fig.~\ref{fig:2b}, we show the ground-state energy per fermion~$E/N$ for $N=2$ as a function of the inverse ground-state density~$n_{\rm gs}$ as obtained from
different calculations, namely the non-interacting two-fermion system, 
DFT-RG in the leading-order (LO) approximation as given by Eq.~\eqref{eq:dfthf},
DFT-RG in the next-to-leading order (NLO) approximation as discussed in Sec.~\ref{sec:restrunc}, and the exact result as obtained from a diagonalization
of the Hamilton operator in a sufficiently large subspace spanned by the two-particle wave functions of the non-interacting
system in the {box.
For comparison, we also show the result for the ground-state energy of the
two-body problem in the continuum limit which has been calculated by solving the {\it Schr\"odinger} equation in the center-of-mass frame in
a sufficiently large subspace spanned by harmonic oscillator eigenfunctions. Again, we emphasize that~$n_{\rm gs}=N/L$ in Fig.~\ref{fig:2b} should not be confused with the intrinsic density
of the system. For~$N=2$, the latter can be extracted from the density-density correlation function, see Eqs.~\eqref{eq:G2Pauli} and~\eqref{eq:iddef} as well {as 
Fig.~\ref{fig:2bwf}.}}

We observe that the results for the ground-state energy from our DFT-RG study approach
the exact solution from above. To be more specific, we find that our DFT-RG results at LO and NLO
agree very well with the exact solution for small~$1/n_{\rm gs}$, i.e. small volumes or high densities. In fact, for $1/n_{\rm gs} \lesssim 4\,L_0$, our DFT-RG results 
already agree identically with the exact solution on the scale of the plot. For even smaller volumes,~$1/n_{\rm gs} \lesssim 2.9\,L_0$, 
we then find that the energy of the interacting two-fermion
system becomes larger than the energy of the non-interacting system. This can be traced back to the fact that the repulsive core of the two-body interaction contributes predominantly
for such small volumes whereas the attractive (long-range) tail of the interaction potential is cut off. Thus, the observed increase of the 
energy~$E/N$ for small volumes originates from both
an increase of the kinetic energy~$\sim 1/L^2$ and an increase of the potential energy associated with the repulsive core of the interaction. 
For~$1/n_{\rm gs} \gtrsim 4\,L_0$, on the other hand,
the results from the DFT-RG study at LO start to deviate from the results at NLO and the exact solution. For~$1/n_{\rm gs} \gtrsim 6\,L_0$, 
we then observe that the results from the DFT-RG study at NLO also start to deviate significantly from the exact results. 

For the DFT-RG study at LO and NLO, we find that the results for~$E/N$ assume a minimum at a finite value of~$1/n_{\rm gs}$, 
whereas the exact solution appears to ``flatten out" already around~$1/n_{\rm gs} \sim 6 \,L_0$ and approaches the 
value of the ground-state energy in the continuum limit, i.e. for~$1/n_{\rm gs}\to\infty$. 
Hence, our present truncation is not capable to reproduce the correct large-volume scaling behavior of the 
density-density correlation function which is required to recover the exact result in the continuum limit. 
In this low-density {regime, density correlation functions of higher order become important 
and their flows can no longer be neglected.}

In Fig.~\ref{fig:2b}, we also observe that the exact solution for the ground-state energy
in the finite box approaches the continuum-limit value from above. We shall exploit this observation below. However, note that the exact solution assumes 
a local maximum $E/N \approx -0.0068\, (1/L_0^2)$ at~$1/n_{\rm gs} \approx 17.5\, L_0$, decreases again for larger volumes and
{converges slowly  
to the continuum-limit} result~$E/N|_{\text{cont.}} \approx -0.0094\, (1/L_0^2)$.
Indeed, for~$1/n_{\rm gs} \approx 80\, L_0$, the exact solution in the box 
still deviates from the continuum-limit result by~$\sim 10\%$. 
This slow convergence behavior may
be considered as an indication for the importance of the low-momentum modes 
which are cut off by the finite volume but required for an 
accurate description of the dynamics in the continuum limit.

At this point, a comment on the energy associated with the
center-of-mass motion is in order. Formally, we may decompose the total energy~$E$ into the binding energy~$E_{\rm B}$ 
and the center-of-mass energy~$E_{\rm cm}$ associated with the free motion of the center of mass of the $N$ fermion system,
$E=E_{\rm cm}+E_{\rm B}$. Assuming that $E$ is the ground-state energy, we have~$E_{\rm cm}=0$ for fermions in a box
with (anti)periodic boundary conditions and we are left with~$E=E_{\rm B}$. Note that this holds for any number of fermions.
Thus, our results for the energy are not spoilt by contributions associated with the center-of-mass motion of the system.

Before we continue with the discussion of the absolute square of the ground-state 
wave function, we would like to discuss briefly the above-mentioned spurious fermion self interactions.
In case of periodic boundary conditions for the fermions, the ground-state energy of a single-fermion system 
is zero. 
We have proven above that our DFT-RG framework also yields a vanishing ground-state energy for the
one-fermion system as it should be, 
provided that we do not truncate the infinite set of flow equations. For any truncation, we therefore expect that our results for the ground-state energies are 
contaminated by contributions from spurious fermion self-interactions. To test the quality of our DFT-RG results at NLO,
we have computed the ground-state energy
of the one-fermion system. At NLO, we find that~$EL_0^2\lesssim {\mathcal O}(10^{-4})$ for $1/(n_{\rm gs}L_0)\gtrsim 6$ which corresponds
to about~$1\%$ of the exact ground-state energy of the two-body problem in this range.\footnote{At leading order, 
$E/N$ vanishes identically for~$N=1$ within our DFT-RG framework.} 
We consider this value for~$E L_0^2$ for~$N=1$
to be reasonably small given the fact that it has been obtained with the simplest truncation taking into account the flow of the density-density correlation function.

Let us now turn to the discussion of our results for the absolute square of the ground-state wave function $|\varphi_{2}(\bar{r})|^2$ in the center-of-mass frame 
which have been extracted from the density-density correlation function, see Eqs.~\eqref{eq:G2Pauli} and~\eqref{eq:iddef}. 
In Fig.~\ref{fig:2bwf}, we show our results for the dimensionless wave function $|\varphi_{2}(\bar{r})|^2 L$
as a function of $\bar{r}=r/L=|x_1-x_2|/L$ as obtained from our DFT-RG study at NLO, in comparison with the exact solution 
and the non-interacting system.
Whereas the ground-state density is constant as a consequence of the use of antiperiodic boundary
conditions for~$N=2$, the ground-state wave function $|\varphi_{2}(\bar{r})|^2$ for~$N=2$
in the center-of-mass frame, i.e. the so-called intrinsic density, exhibits 
a non-trivial dependence on the spatial coordinate. 
For the non-interacting theory,~$|\varphi_{2}(\bar{r})|^2 L$ as a function of~$\bar{r}$ is universal, i.e. independent of the ground-state density~$n_{\rm gs}$.
{In contrast to the} non-interacting case, we observe 
that the exact solution as well as our DFT-RG results depend on~$n_{\rm gs}$. In particular, {for~$1/n_{\rm gs} \gtrsim 5\,L_0$, we} 
observe that the
position of the maximum of~$|\varphi_{2}(\bar{r})|^2 L$ is shifted to smaller values in terms of the {dimensionless quantity~$\bar{r}$ when} the volume is increased.
For sufficiently large volumes,
the exact solution~$|\varphi_{2}(\bar{r})|^2$ in a finite box then approaches the solution in the continuum limit. In particular for the maximum of~$|\varphi_{2}(\bar{r})|^2 L$, 
we have~$\bar{r}_{\text{max}}(L)\sim r_{\text{max}}(\infty)/L$ for the exact solution, 
where the constant~$r_{\text{max}}(\infty)>0$ is the position of the maximum in the
continuum limit and roughly coincides with the position of the minimum of the interaction potential, see Fig.~\ref{fig:pot}. 
In accordance with our discussion of the ground-state energy, we observe that our DFT-RG results for~$|\varphi_{2}(\bar{r})|^2$ are in very good
agreement with the exact solution for~$1/n_{\rm gs}\lesssim 5\,L_0$. For~$1/n_{\rm gs}\gtrsim 5\,L_0$, our DFT-RG results then 
start to deviate from the exact solution.
For~$1/n_{\rm gs}= 7.5\,L_0$, there is indeed already a significant difference between our DFT-RG 
result at NLO and the exact result which is also reflected in the
corresponding results for the ground-state energy, see Fig.~\ref{fig:2b}.

\begin{figure}[t]
\includegraphics[width=0.6\columnwidth]{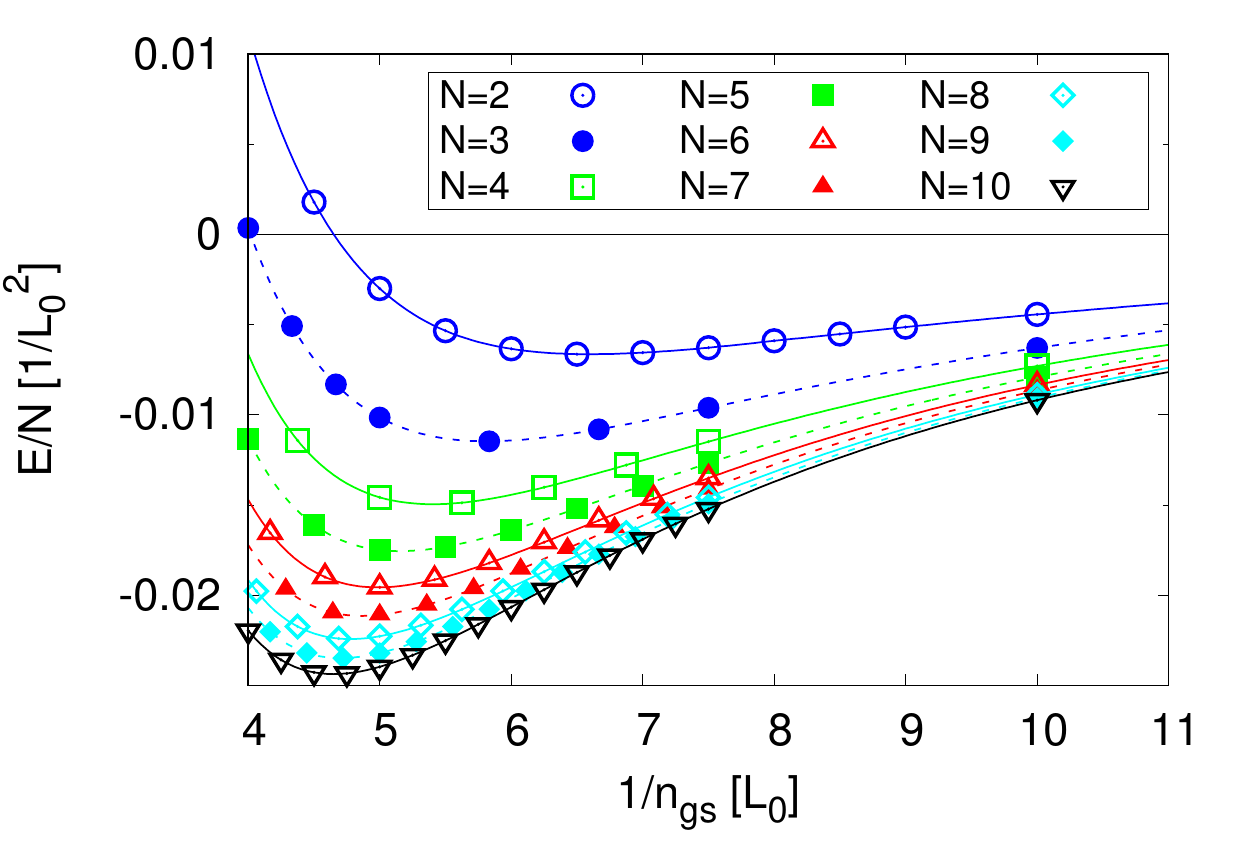}
\caption{\label{fig:fb} (color online) Ground-state energy of various $N$-fermion systems 
as a function of the inverse ground-state density. Note that the latter should not be confused with the intrinsic density of the system, see main text for details.}
\end{figure}

Finally we discuss our results for systems with more than two fermions. In Fig.~\ref{fig:fb}, 
we show our results for the ground-state energy per fermion~$E/N$ as a function
of the inverse ground-state density~$1/n_{\rm gs}$ for~$N=2,3,\dots,10$ fermions. We observe that the results for different fermion numbers are similar on a qualitative 
level: Starting from small values of~$1/n_{\rm gs}$, we find 
that~$E/N$ first decreases, reaches a minimum, then increases again and appears to tend 
to zero for small densities (i.e. large volumes). Moreover,
we observe that the position of the minimum is monotonously shifted to smaller values of~$1/n_{\rm gs}$ for increasing~$N$. 

\begin{figure}[t]
\includegraphics[width=0.6\columnwidth]{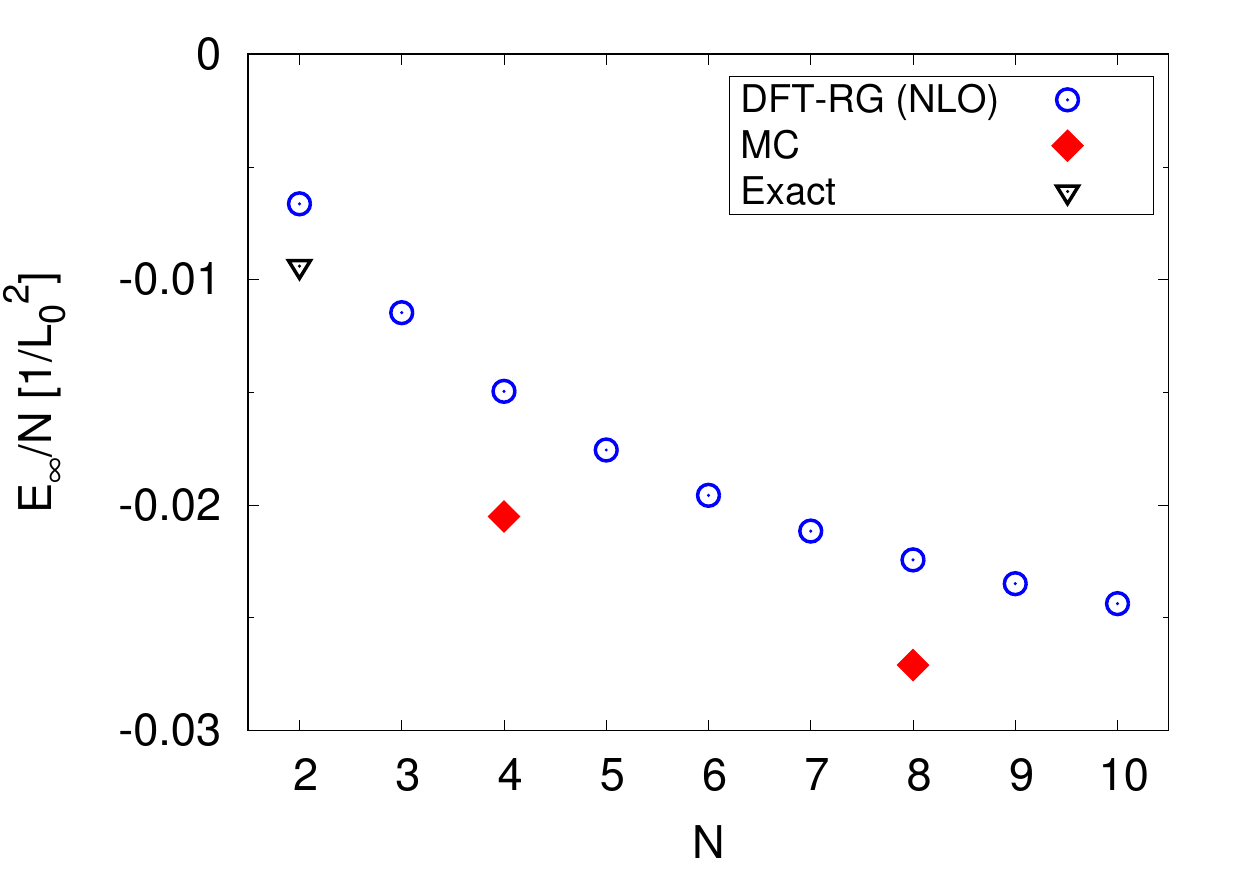}
\caption{\label{fig:gs} (color online) Estimates for the ground-state energy per fermion~$E/N$ for systems with~$N=2,\dots,10$ fermions in the continuum limit 
compared to the exact result for~$N=2$ and MC results~\cite{Alexandrou:1988jg} for~$N=4$ and~$N=8$. Note that an MC result
for~$N=2$ in the continuum limit is not given in Ref.~\cite{Alexandrou:1988jg} but only for the finite-volume case which we find to be in
disagreement with our exact solution. For $N=4$ and $N=8$, the MC results for the infinite-volume limit
are found to be in good agreement with Hartree-Fock calculations in the infinite-volume limit~\cite{Alexandrou:1988jg}.}
\end{figure}

The comparison of our DFT-RG results with 
the exact solution for~$N=2$ suggests that {our DFT-RG results are reliable for values of~$1/n_{\rm gs}$ up to about the value where the exact solution assumes a local minimum. For} 
larger values of~$1/n_{\rm gs}$, on the other hand, our present truncation is not capable to recover the correct scaling behavior of the ground-state energy. However, 
an estimate for the ground-state energy in the continuum limit can still be obtained from our present truncation. To this end, 
we exploit the fact that the results for the ground-state energy for the two-body problem from our DFT-RG studies at LO and NLO 
approach the exact result from above for a given
value of the box size~$L$. Moreover, we have found that 
the exact solution in the box also approaches the exact continuum-limit value from above. Assuming that these observations also hold for~$N>2$, we can estimate the ground-state
energy of the $N$-body system in the continuum limit from a minimization of~$E/N$ with respect to~$L$:
\be
E_{\infty}=\inf_{L}  E(L)\,.
\ee
In Fig.~\ref{fig:gs} we compare our DFT-RG results for~$E_{\infty}/N$ as a function of~$N$ with 
the exact result for~$N=2$ and results from Monte Carlo (MC) studies~\cite{Alexandrou:1988jg} for~$N=4$ and~$N=8$. 
The error bars of the MC results are smaller than the size of the symbols in Fig.~\ref{fig:gs}.
In Ref.~\cite{Alexandrou:1988jg}, the ground-state energies in the continuum limit have also been computed in the {\it Hartree-Fock} 
approximation for~$N=4,8,12$ and found to agree well with the MC data.
The continuum-limit result for the ground-state energy of the two-fermion system, which we use as a benchmark for our DFT-RG studies, 
is neither given for the MC calculation nor the {\it Hartree-Fock} calculation. However,
the MC result presented for~$N=2$ in the presence of a finite volume  
appears to be in disagreement with our exact solution in the finite box,\footnote{This discrepancy may be traced back to differences in the implementation
of the two-body interaction in the finite volume. We have studied two possible definitions, see our discussion of Eq.~\eqref{eq:Udecomp}, and found 
that the discrepancy between the MC result and our exact solution is present for both implementations of the two-body interaction.} 
even on a qualitative level (i.e. the energy from the MC study is found to be 
negative {for~$1/n_{\rm gs}\approx 4.3\,L_0$ rather} than positive). In any case, we find that our present best estimate for the 
ground-state energy of the two-body system underestimates
the exact value by about~$30\%$. Moreover, 
we observe that our results for the ground-state energy per fermion 
agree on a qualitative level with the available MC results for the continuum limit but
the energies from the DFT-RG studies are found to be consistently greater than those from the MC calculations.

%
\section{Conclusions and Outlook}\label{sec:conc}

In the present work we have discussed an RG approach to DFT which allows to directly compute the energy density functional from the 
microscopic interactions of a given theory. In particular, it does not only open up the possibility to compute ground-state energies but
also excited states. Moreover, it gives access to the absolute square of the ground-state wave function from which the intrinsic density
can be computed. From a conceptional point of view, our approach does not rely on a (global) parametrization 
of the density functional
but on an expansion about the ground-state density. This expansion  
is essentially determined by the density correlation functions being the fundamental building blocks of our approach.
From these correlation functions, we can extract physical observables such
as the energies of the ground state {and excited states as well as the intrinsic density. 
In this respect, we have also discussed in Sec.~\ref{sec:var} that the density functional associated with 
the $2$PPI effective action underlying our present work
should be considered as a generalization of the conventional {\it Hohnberg-Kohn} energy density functional
since it contains more information than the latter.} 

Our approach allows to derive systematically differential equations for the density correlation functions. The initial conditions 
{for the associated in general infinite} tower of equations are derived from a given confined but non-interacting $N$-fermion system. 
The confining geometry, which is necessary
to localize the non-interacting system and fix the particle number, is at our disposal. In our explicit calculations in this work, we have
used a box with (anti)periodic boundary conditions as confining geometry which we expect to be most convenient for our studies aiming
at the computation of ground-state properties of selfbound systems in the continuum {limit. In fact,
we have discussed that our choice to use a box with (anti)periodic boundary conditions together with 
our implementation of the two-body interaction {in such a box} allows us to preserve translation invariance of the density 
correlation functions in our calculations and to keep the center-of-mass energy under control to a large extent,
which are clearly attractive features for our present studies aiming at properties of selfbound systems.
Depending on the problem under consideration, however, other} confining geometries, such as a
harmonic trap or a box with ``hard walls", are possible as well and 
may be of interest in future studies of, e.g., trapped low-dimensional mass- and spin-imbalanced 
ultracold Fermi gases which potentially allow to study the transition from few- to many-body {physics in a clean and controllable fashion 
paralleled by experimental studies~\cite{2013Sci...342..457W,2013PhRvL.111q5302Z}.}
For future studies of selfbound systems it may be worthwhile to explore the {dependence of the results from a given truncation 
on different regulator functions as well as the use} 
of confining geometries which depend on the flow parameter
themselves and are adapted dynamically in the RG flow. 
In any case, the infinite tower of flow equations for the density correlation functions underlying our RG approach needs to be truncated
for explicit calculations. Following Ref.~\cite{Kemler:2013yka}, we have
shown that such truncations of the infinite set of DFT-RG equations can be directly related to many-body perturbation theory. This is particularly useful 
to find meaningful truncation schemes and 
construct the energy density functional in a controlled and systematic manner. 

In addition to our conceptional developments, we have applied our DFT-RG framework to a one-dimensional nuclear model
and found that our estimates for the continuum-limit values of the ground-state energy of various $N$-fermion systems
are already in reasonable agreement with results from the exact two-body solution and MC calculations for the four- and eight-body system.
{For the two-body problem, we have explicitly found
that our DFT-RG results approach the exact value from above when we go from the LO
to the NLO approximation within our framework. Since our result for 
the ground-state energy for the two-body problem at LO differs from the exact value by more than~$40\%$
and still by about~$30\%$ at NLO, we expect significant contributions 
to the ground-state energy to come from terms associated with N${}^2$LO and even N${}^3$LO corrections to the density functional, at least
in case of the two-body problem.
We add that the LO result does not depend on our choice for the regulator function. However,
the use of other regulator functions 
beyond the {\it Callan-Symanzik}-type function underlying our present work should be exploited in future studies beyond the LO approximation
in order to minimize the deviation from the exact result at a given order within our DFT-RG approach.
At this point, we would like to emphasize again that} our calculations do not rely on fitting parameters of the functional but 
only uses the fundamental interaction potential as input to define the many-body problem. 
This is promising also with respect to studies of the formation of bound states in three dimensions which are now in reach given
the developments of the present work. With respect to nuclear physics, it may indeed be an attractive feature of our DFT-RG approach that
it allows to compute the energy density functional from the microscopic interactions of the theory as it  
opens up a new direction to {derive} the energy density functional {from, e.g., chiral} effective field theory 
interactions~\cite{Epelbaum:2002ji,Epelbaum:2002vt,Epelbaum:2005pn,Navratil:2007we,Epelbaum:2008ga}.
Apart from such ambitious explicit (quantitative) computations of nuclear 
energy density functionals, our DFT-RG approach may be 
viewed as a tool to gain a deeper insight into the general structure of these functionals. In fact, we have demonstrated that our DFT-RG approach
is related to {many-body} perturbation theory in a simple and systematic fashion. Moreover, we have discussed that our expansion about the ground-state density of the system
can be mapped on the more conventional gradient/derivative expansion of the energy density functional. These connections of our DFT-RG framework
to well-known and established frameworks, 
{such as many-body perturbation theory, gradient/derivative expansions and expansions in terms of density matrices, may} be considered as
one of the most attractive features of our approach and may help to guide the development of microscopic 
energy density functionals for future DFT studies of nuclei.

{\it Acknowledgments.--~} The authors thank J.~E.~Drut, R.~J.~Furnstahl, H.-W.~Hammer, K.~Hebeler, F.~Karbstein, J.~Polonyi, 
D.~Roscher, and A.~Schwenk for useful discussions. 
Moreover, the authors are very grateful to J.~E.~Drut, R.~J.~Furnstahl, and A.~Schwenk for comments on the manuscript.
J.B. acknowledges support by HIC for FAIR within the LOEWE program of the State of Hesse. Moreover,
S.K. and J.B. {acknowledge support by the Deutsche Forschungsgemeinschaft (DFG) through Grant BR 4005/3-1, and M.P. and J.B. 
by the DFG through grant SFB 1245.}

\bibliographystyle{apsrev}
\bibliography{bib_source}

\end{document}